\documentclass[%
superscriptaddress,
 reprint,
 nofootinbib,
 amsmath,amssymb,
 aps, prx
]{revtex4-2}
\usepackage{bbm}
\usepackage{color}
\usepackage{braket}
\usepackage{graphicx}
\usepackage{dcolumn}
\usepackage{bm}
\newcommand{\Tr}{\text{Tr}}
\newcommand{\be}{\begin{equation}}
\newcommand{\ee}{\end{equation}}

\begin{document}

\preprint{APS/123-QED}

\title{Entanglement of Local Operators and the Butterfly Effect}%

\author{Jonah Kudler-Flam}
\email{jkudlerflam@uchicago.edu}
\affiliation{Kadanoff Center for Theoretical Physics, University of Chicago, IL~60637, USA}%

\author{Masahiro Nozaki}
\email{mnozaki@yukawa.kyoto-u.ac.jp}
\affiliation{
Berkeley Center for Theoretical Physics, Berkeley, CA 94720, USA
}%
\affiliation{
 iTHEMS Program, RIKEN, Wako, Saitama 351-0198, Japan
}%
\author{Shinsei Ryu}
\email{ryuu@uchicago.edu}
\affiliation{Kadanoff Center for Theoretical Physics, University of Chicago, IL~60637, USA}%

\author{Mao Tian Tan}
\email{mtan1@uchicago.edu}
\affiliation{Kadanoff Center for Theoretical Physics, University of Chicago, IL~60637, USA}%

\date{\today}

\begin{abstract}
  We study the robustness of quantum and classical information to perturbations implemented by local operator insertions.
  We do this by computing multipartite entanglement measures in the Hilbert space of local operators in the Heisenberg picture.
  The sensitivity to initial conditions that we explore is an illuminating manifestation
  of the butterfly effect in quantum many-body systems. We present a ``membrane theory'' in Haar random unitary circuits to compute the mutual information, logarithmic negativity, and reflected entropy in the local operator state by mapping to a classical statistical mechanics problem and find that
  any local operator insertion delocalizes information as
  fast as is allowed by causality. Identical behavior is found for
  conformal field theories admitting holographic duals where the bulk geometry is described by the eternal black hole with a local object situated at the horizon.
  In contrast to these maximal scramblers, only an $O(1)$ amount of information is found to be delocalized by local
  operators in integrable systems
  such as free fermions and Clifford circuits.
\end{abstract}

\maketitle

\tableofcontents

Chaos in classical systems is described by the sensitivity of phase space trajectories to initial conditions. Systems displaying chaos will generically have nearby trajectories diverge exponentially at early times, characterized by a Lyapunov exponent. One can think of this sensitivity to initial conditions as a manifestation of the butterfly effect; a small change, such as a butterfly flapping its wings, can have extraordinary consequences on the state of the system at later times. 

Quantum chaos is an old topic with many developments (see e.g.~\cite{stockmann_1999, 2016AdPhy..65..239D}) that addresses the connection between classically chaotic systems and their underlying quantum mechanics; how do highly nonlinear classical dynamics emerge from the linear unitary evolution of the Schr{\"o}dinger equation? Recently, there has been considerable excitement across multiple fields of physics due to a quantum manifestation of the butterfly effect \cite{2014JHEP...03..067S,PhysRevLett.115.131603,2016JHEP...08..106M} characterized by out-of-time-ordered four-point correlation functions (OTOCs)
\begin{align}
    C_{\beta}(x,t) \equiv \frac{\langle V^{\dagger}W^{\dagger}(x,t) V W(x,t)\rangle_{\beta}}{\langle V^{\dagger}V\rangle_{\beta}\langle W^{\dagger} W \rangle_{\beta}},
    \label{OTOC}
\end{align}
where $V,W$ are local operators in the Heisenberg picture and $\langle \cdots \rangle_{\beta}$ specifies that the correlator is evaluated in a finite temperature state of inverse temperature $\beta$.
This directly probes the spreading of a local operator's spatial support. In analogy to classical chaos, (\ref{OTOC}) can exhibit a quantum Lyapunov exponent, $\lambda_L$, at early times 
\begin{align}
    C_{\beta}(x,t) \sim 1 - e^{\lambda_{L}(t-t_*-x)} + \cdots
\end{align}
and tends to zero at late-times, describing the ``scrambling'' of the quantum information of the initial state. Importantly, this exponential behavior of the OTOC is \textit{not} a generic feature of quantum chaotic systems. In fact, it has only been found for large-$N$ theories \cite{2014JHEP...03..067S,2016JHEP...08..106M,2015JHEP...03..051R,2015JHEP...05..132S,PhysRevLett.115.131603,2017JHEP...05..125G,2016PhRvL.117k1601J,2016PhRvD..94j6002M} and does not generically occur in realistic (finite-$N$) quantum chaotic systems\footnote{Our working definition of quantum chaos is an energy spectrum whose statistics mimic random matrix theory.} (see e.g.~\cite{2017JHEP...10..138H,2019arXiv190808059C}).

The OTOC also leaves certain information-theoretic questions open about chaos. How close are the quantum states of subsystems with and without the perturbation? How much and how fast is information delocalized (scrambled) by the perturbation? To what extent does the choice of local operator influence the scrambling process? In this paper, we address these questions by studying the \textit{local operator entanglement}, quantum correlations of local operators\footnote{Here, we take the opportunity to draw the reader's attention to what, to our knowledge, is the earliest work on operator entanglement \cite{2001PhRvA..63d0304Z}.}. We can study the local operators directly by computing correlation measures not in the original Hilbert space, $\mathcal{H}$, but the doubled Hilbert space of endomorphisms $\mbox{End}(\mathcal{H})\simeq \mathcal{H}_1 \otimes \mathcal{H}_2^*$
\begin{align}
    \mathcal{O}(x,t) \equiv e^{iHt}\mathcal{O}(x) e^{-iHt} \rightarrow \ket{\mathcal{O}(x,t)}.
    \label{loc_op_state}
\end{align}
In practice, this is done by ``flipping the bra vector to a ket''
\begin{align} \label{dual_to_local_operator}
    \ket{\mathcal{O}(x,t)} \equiv \mathcal{N} \sum_{m,n}  e^{i (E_n-E_m) t} \mathcal{O}_{nm}(x) \ket{n}_1\ket{m^*}_2,
\end{align}
where we have expanded in an energy eigenbasis and imposed an appropriate normalization. Entanglement in the Hilbert space of local operators has been considered previously in Refs.~\cite{2007PhRvA..76c2316P,2009PhRvB..79r4416P,2017JPhA...50w4001D,2018arXiv180300089J,2019PhRvL.122y0603A,2019arXiv190907407B,2019arXiv190907410B}.

\begin{figure}
    \centering
    \includegraphics[width =.23\textwidth]{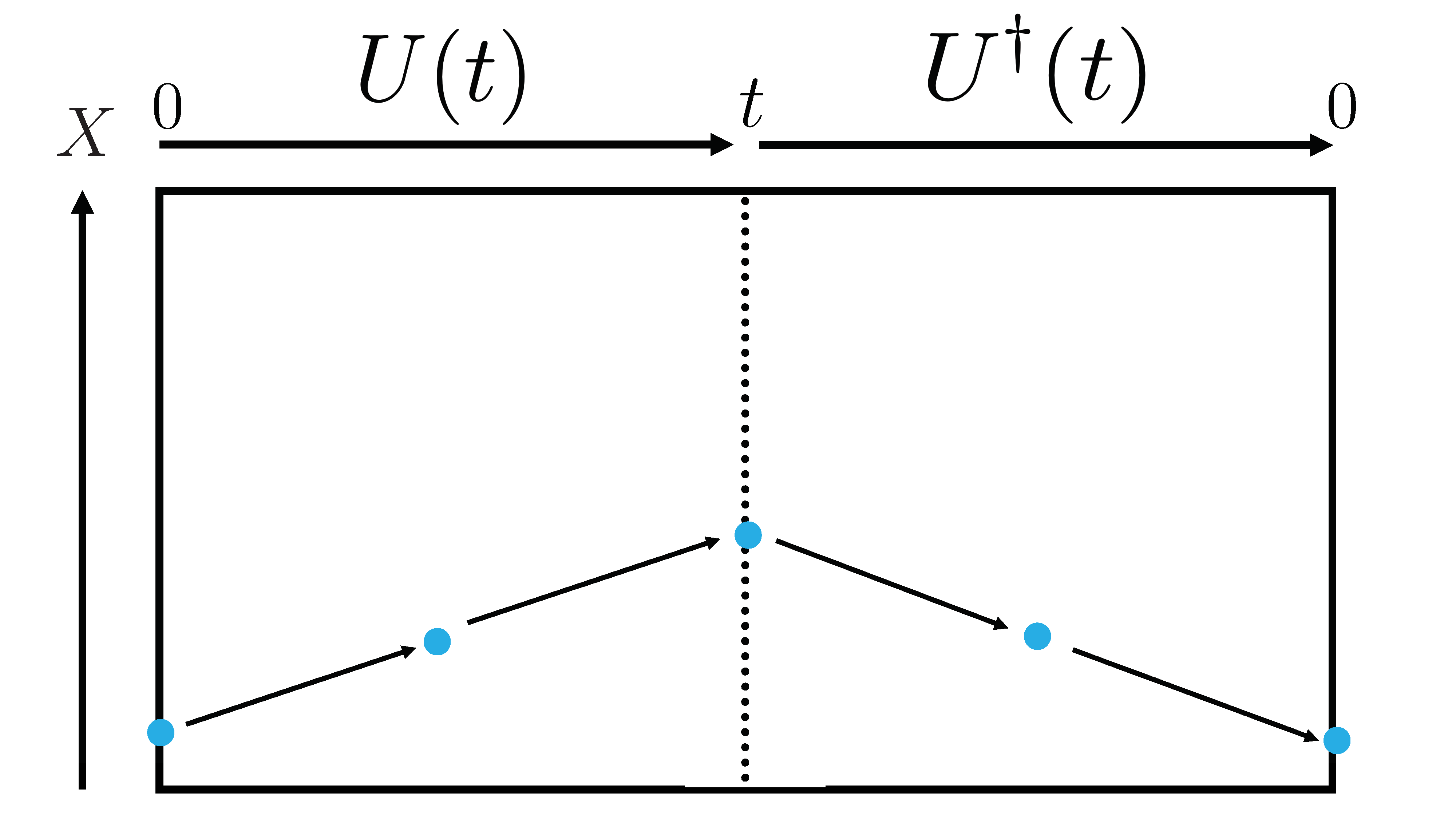}
    \includegraphics[width =.23\textwidth]{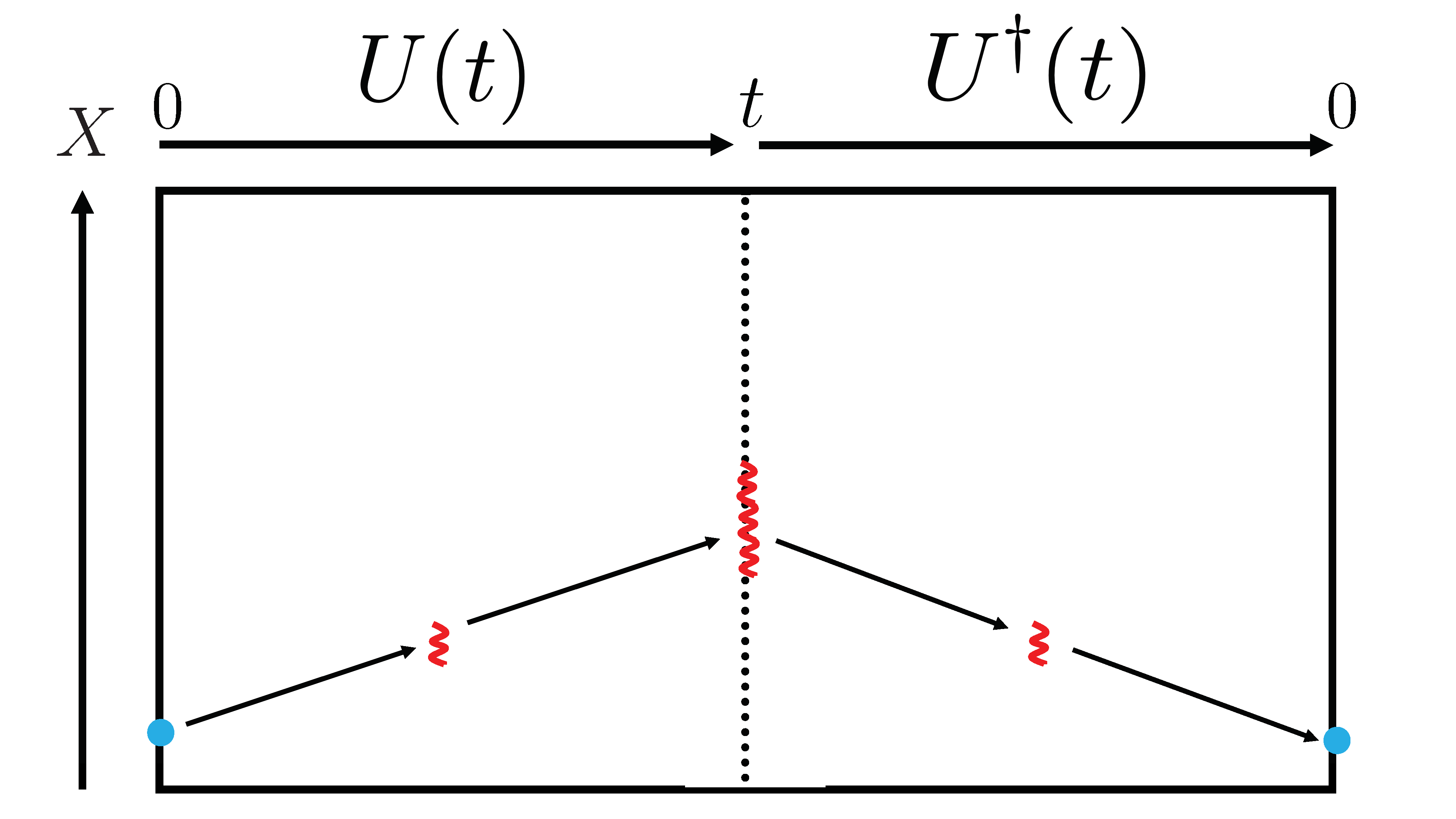}
    \caption{We represent the local operator as a quantum circuit. Left: localized quantum information (blue) remains local under time evolution by integrable channels and returns to where it began when evolved backwards because nothing ($\mathcal{O} = \mathbb{I}$) happens at time $t$. Right: localized information spreads out (red) under time evolution by non-integrable Hamiltonians but recoheres when evolved backwards to $t =0$.}
    \label{identity_cartoon}
\end{figure}

Let us now try to understand how correlations in this state, \eqref{dual_to_local_operator}, characterize the butterfly effect by showing how information flows under this time evolution. Consider the trivial case where the operator, $\mathcal{O}$, is the identity, $\mathbbm{1}$. The identity should have no effect on the state. This is shown in Fig.~\ref{identity_cartoon} where the time evolution operator moves around quantum information. For integrable systems, information that is initially localized will remained localized at time $t$ in the sense that it will not mix in an ergodic manner in the local Hilbert space. This is due to the constraints imposed by integrability. This lack of ``local ergodicity'' does not exclude the possibility of spreading in space as is seen in integrable system exhibiting diffusion. For chaotic systems, the initially localized information becomes spread out at time $t$ and mixed in the local Hilbert space without the severe constraints imposed by integrability. This is when the operator (identity) is inserted. Because the identity acts trivially, the backwards time evolution brings the information back into a localized packet whether or not the system is integrable\footnote{This forward and backward evolution is reminiscent of other quantum chaos diagnostics such as the Loschmidt echo and OTOC. It would be interesting to further explore the connections between these quantities.}. This means that the mutual information between subregions is simply proportional to their overlap in the spatial direction.

\begin{figure}
    \centering
    \includegraphics[width =.23\textwidth]{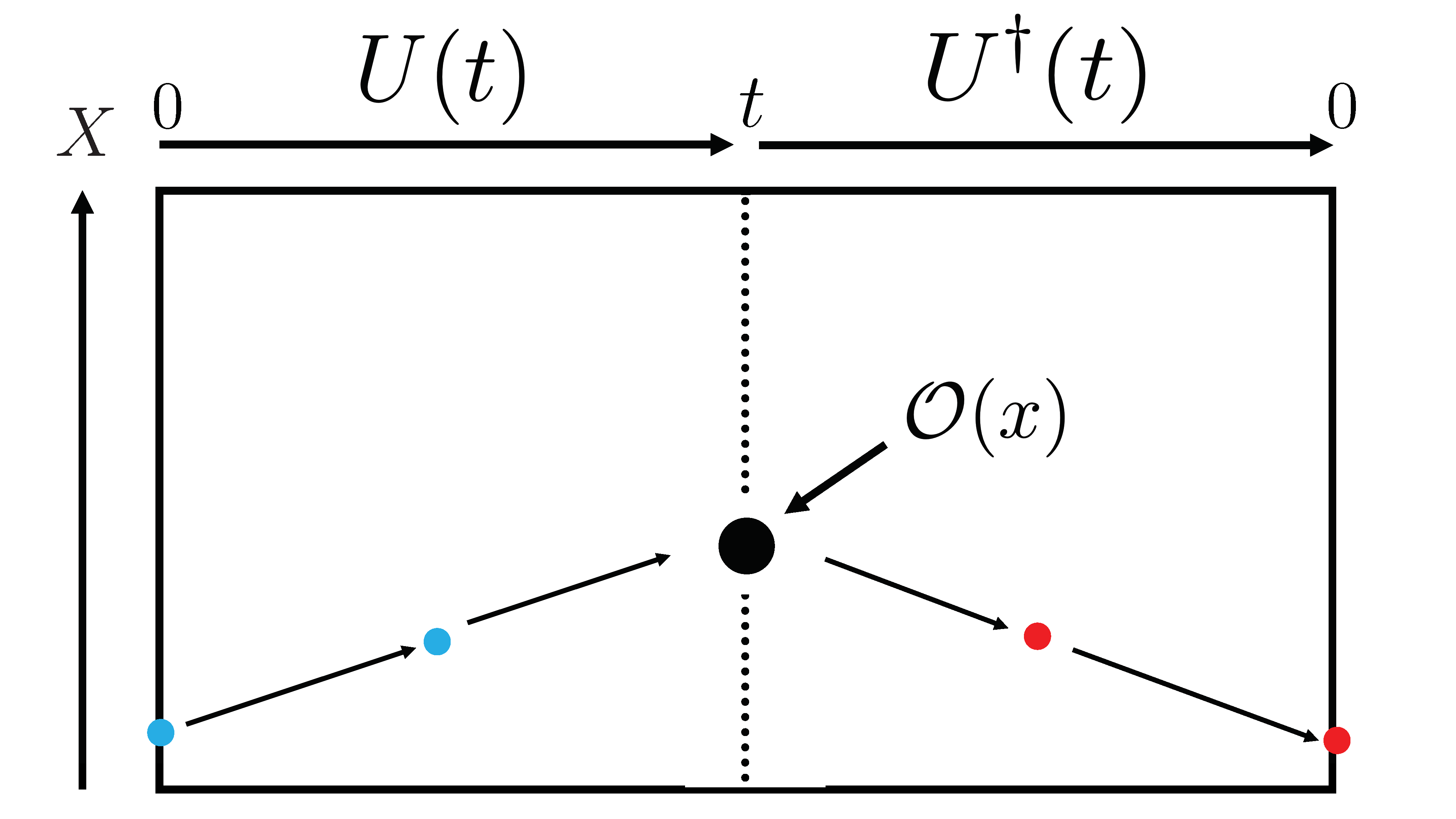}
    \includegraphics[width =.23\textwidth]{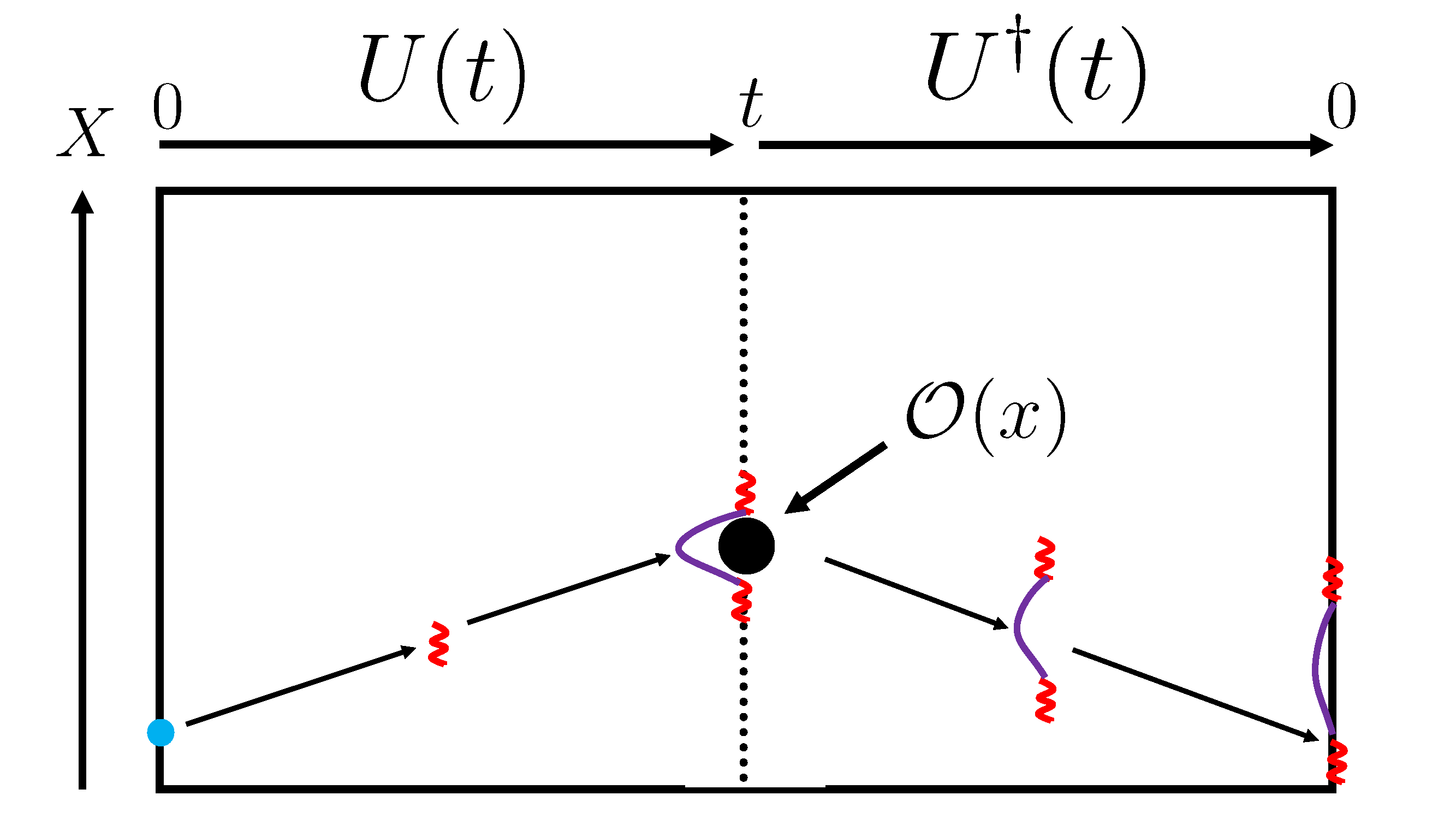}
    \caption{The Heisenberg evolution of a (nontrivial) local operator is shown. For integrable channels (left), initially localized information remains largely localized even after the perturbation by operator $\mathcal{O}$. In contrast, for chaotic channels (right), local information delocalizes in time. Then, the perturbation by operator $\mathcal{O}$ effects the state such that this information cannot recohere under backwards time evolution, but rather continues to grow (decoheres).}
    \label{op_intro_cartoon}
\end{figure}

We progress to nontrivial operators. When these operators are inserted, they may scatter the information (see Fig.~\ref{op_intro_cartoon}). For integrable systems, the localized quantum information will remain localized but it may be transmitted to a different location than it started at when it is evolved back to $t=0$. For a chaotic system, the butterfly effect implies that the local perturbation created by the operator may ruin its coherence, hence it remains delocalized after the backwards time evolution. Given long enough times, the information will be spread out over the entire system.

\begin{figure}
    \centering
    \includegraphics[width  = .46\textwidth]{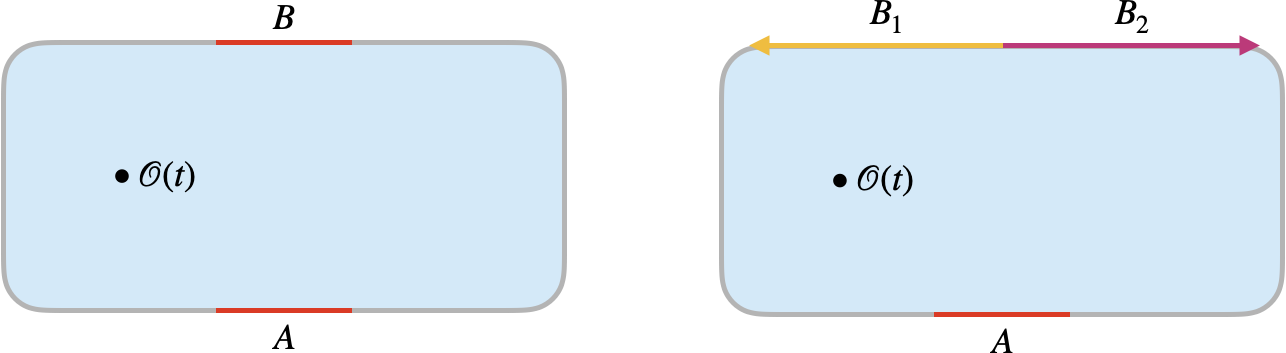}
    \caption{We show the partitioning used throughout the paper. Left: the BOMI configuration for symmetric intervals $A$ and $B$. Right: the TOMI configuration for finite interval $A$ and semi-infinite intervals $B_1$ and $B_2$.}
    \label{config_cartoon}
\end{figure}

We propose that an illuminating diagnostic of the amount of quantum information initially in region $A$ that is scrambled by operator $\mathcal{O}$ is the tripartite mutual information (TOMI)\footnote{This quantity was studied for the non-local time evolution operator in Ref.~\cite{2016JHEP...02..004H}.}, defined as
\begin{align}
    I_3(A, B_1, B_2) = I(A, B_1) + I(A, B_2)-I(A,B),
    \label{I3_def}
\end{align}
where $B = B_1 \cup B_2$ is the entire output Hilbert space, $\mathcal{H}_2$. This characterizes how much total (classical + quantum) information from $A$ is lost unless the entire output system is measured. The local operator entanglement allows us to understand how different operators scramble information. Analogously, we also study tripartite operator logarithmic negativity (TOLN) to characterize the purely quantum information that is scrambled. This is defined by replacing the bipartite operator mutual informations (BOMI) on the right hand side of \eqref{I3_def} with logarithmic negativities (BOLN). We show the generic setup in Fig.~\ref{config_cartoon}. While the operator choice for OTOC may be seen as a disadvantage because it can be misleading (e.g.~spin-spin OTOC in the Ising model \cite{PhysRevLett.115.131603}), it should be seen as an advantage for local operator entanglement because the mutual information probes correlations of all operators; not all butterflies have the same effect.

\subsection{Summary of results}

\begin{figure}
    \centering
    \includegraphics[width = .48\textwidth]{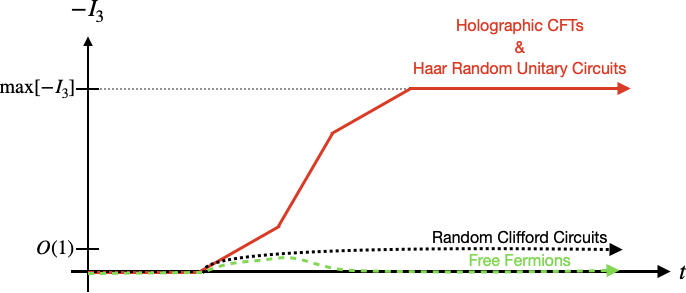}
    \caption{Here, we show the behavior of the tripartite local operator mutual information for the four systems that we study in this paper. $-I_3$ characterizes how much information has been scrambled. The red line represents holographic CFTs and Haar random unitary circuits which saturate the physical bound. In contrast, random Clifford circuits (black dotted line) saturate to an operator dependent $O(1)$ value and free fermions (green dashed line) have vanishing late-time $I_3$.}
    \label{summary_cartoon}
\end{figure}

In the rest of the paper, we have many technical results that the casual reader may not
wish to sift through. Here, we summarize our central findings. We also present a cartoon summarizing results for $I_3$ in Fig.~\ref{summary_cartoon}.

\paragraph{Random Unitary Circuits}

Random unitary circuits are tractable toy models of local Hamiltonians displaying chaotic phenomena. In Section \ref{RandU_sec}, we put forward an effective description of the entanglement dynamics of local operators in terms of the free energy of a membrane in spacetime. This effective description is discussed in further detail for R\'enyi entropy, logarithmic negativity, and reflected entropy by mapping the random unitary circuit to a classical statistical mechanics problem in Appendix \ref{app_stat_mech}. 

We find for Haar random unitary channels that local information is entirely delocalized by the local operator, regardless of the operator chosen. This manifests by the tripartite information increasing in magnitude as fast as is allowed by causality, ultimately saturating to the lower bound on all quantum systems which is proportional to the number of degrees of freedom in subsystem $A$.

In contrast, when the quantum channel is composed of random unitary elements from the Clifford group instead of the full unitary group, we find that a very small amount of information is scrambled; this value is independent of system sizes but dependent on operator choice. This is notably different than the observed maximal scrambling behavior of Clifford circuits for the unitary time-evolution operator \cite{2020JHEP...01..031K}. 
We explain this discrepancy by emphasizing the importance of Clifford gates being unitary 3-designs. Moreover, we find that depending on the operator, the mutual information and logarithmic negativity behave differently. This demonstrates how quantum and classical information may be scrambled in different ways in quantum channels.

\paragraph{Integrable System}
We study free fermions as an example of an integrable system. In particular, we consider the tight-binding Hamiltonian for simplicity. The local operator is taken to be a fermion parity operator acting on a single site. The resulting local operator state is Gaussian, allowing us to employ the correlator method 
to compute the local operator entanglement entropy.

When the input and output subsystems are spatially identical, the mutual information starts of at a maximal value before beginning to dip at some time determined by causality. After the wave front of the operator leaves the subsystems, the BOMI begins to relax back to its initial value, though we do not have a proof that the BOMI fully relaxes back to its original value due to finite size effects. This $O(1)$ change (not extensive with system size) in the BOMI is a signature of the (trivially) integrable nature of free fermions. Very little (if any) information is scattered or delocalized. The TOMI is quite similar. It is initially zero but decreases once the operator is within the subregion. Eventually, the TOMI attains its most negative value before it slowly relaxing back to zero. The late-time behaviour of TOMI indicates the lack of scrambling from operators in the free fermion system.

\paragraph{Chaotic CFTs}

The local operator entanglement of holographic 2D CFTs are studied in section
\ref{HolographicCFTs}.
These are conformal field theories with large central charge and sparse low lying spectra, and are considered \textit{maximally chaotic} due to their early time exponential behavior in the OTOC \cite{2014JHEP...03..067S, PhysRevLett.115.131603,2016JHEP...08..106M}. Another way in which they saturate the fundamental bounds of quantum information scrambling is the decay of the tripartite entanglement of the time-evolution operator \cite{2018arXiv181200013N,2020JHEP...01..031K,2020JHEP...01..175K}.

When the input and output subsystems are symmetric, the BOMI for the local
operator begins at its maximum value.
After the operator has had time to reach the intervals, it begins to decrease
linearly at the maximum rate allowed by causality.
Unlike the free fermion BOMI, the BOMI for holographic CFTs decreases all the way to zero.
This tells us that the local operator eventually delocalizes the information completely and is consistent with the expectation that these conformal field theories are maximally chaotic. 
The TOMI for holographic CFTs is also found to decrease from zero to a maximally negative value at the maximal rate 
\begin{equation}
    \lim_{t\rightarrow\infty}I_3(A,B_1,B_2)=-2S_A^\text{reg.},
    \label{I3qft_reg}
\end{equation}
where $S_A^\text{reg.}$ is the UV finite thermodynamic entropy of subregion $A$ at a temperature determined by a regulator i.e.~it does not contain the standard UV divergence of von Neumann entropy in continuum theories due to short distance modes near the entangling surface.
We stress that these results are significantly stronger than analyses of operator entanglement in the past because this maximal scrambling of information occurs regardless of any details about the operator. The smallest perturbation entirely destroys the quantum information of the state.

An additional notable phenomenon is that the BOMI and TOMI have step function discontinuities associated to when the local operator enters and leaves the associated subregions. The magnitude of these step functions is determined by the conformal weight of the operator. Only for heavy operators ($\Delta \sim c$) are they discontinuities macroscopic.

Finally, we note that these findings precisely match with the results for the
Haar random unitary circuits with bond dimension $q$
in Section \ref{RandU_sec}
once identifying the bond dimension with the Cardy density of states
\begin{align}
    \label{cardy_bond_dim}
    q = e^{\frac{\pi c}{3 \beta}}
 \end{align}
where $c$ is the central charge and $\beta$ is the effective temperature
which is just a regulator for us.
One caveat is that the membrane computation for the random unitary circuits does not have the discontinuities previously mentioned. This discrepancy may either show a difference between the two theories or the analogy may be restored once we account for $O(1)$ contributions in the membrane theory.

\paragraph{Holography}

In Section \ref{holography_sec}, we identify the geometry dual to the local operator state \eqref{loc_op_state}. Because this state lives in two copies of the original Hilbert space, it is natural that the dual geometry has two identical asymptotic boundaries. This is the eternal black hole dual to the thermofield double state with the temperature playing the role of the cutoff. The local operator perturbs the eternal black hole in a similar manner to Refs.~\cite{2014JHEP...03..067S,2015JHEP...08..011C}. It is a massive particle that backreacts on the geometry. We are able to compute the operator mutual information directly from the geometry using the Ryu-Takayanagi formula which precisely matches the CFT calculation.

\section{Random unitary circuits and 
membrane theory}
\label{RandU_sec}

In this section, we motivate intuition by comparing two effective theories of entanglement dynamics, the quasi-particle picture and the membrane theory, which model integrable and chaotic dynamics respectively.

The quasi-particle picture has been proposed as a universal description of entanglement dynamics in integrable theories \cite{2005JSMTE..04..010C,2017PNAS..114.7947A,2018ScPP....4...17A}. This posits that when an integrable system is sufficiently excited above its ground state, the entanglement between subsystems may be entirely accounted for by quasi-particle pairs that carry entanglement content that travel at known speeds. These dynamical inputs may be fixed by thermodynamic Bethe ansatz techniques. The entanglement in inherently bipartite by construction because only Bell pair-like correlations are accounted for. This description largely matches our results for free fermions as the local operator state is an excitation above the ground state of the Hamiltonian $H_1\otimes \mathbb{I}_2 - \mathbb{I}_1 \otimes H_2$.

Severe breakdowns of the quasi-particle picture occur for non-integrable systems because multipartite entanglement becomes increasingly important (see e.g.~Refs.~\cite{2013JHEP...05..080N,2015JHEP...09..110A,2015JHEP...02..171A,2019JHEP...08..063K,2019JHEP...01..025K,2020JHEP...02..017K,2020arXiv200105501K,2020arXiv200811266K}). Thus, the information about entanglement can no longer be carried by quasi-local objects. Recently, a compelling case has been made that, for quantum chaotic systems, the entanglement dynamics are captured by a codimension-one membrane in spacetime, a manifestly non-local object \cite{2017PhRvX...7c1016N,2018PhRvX...8b1013V}. The dynamical input into this \textit{membrane theory} is the tension of the membrane which may be explicitly computed in certain cases. A particular instance where this may be computed is for Haar random unitary circuits. In this section, we will study these circuits and adapt the membrane theory to local operator entanglement.

\subsection{Haar random unitary circuits}

We begin with a simpler problem of computing just the late-time behavior of local operator entanglement by modeling the random unitary circuit as one big Haar random operator\footnote{It has been shown that local random unitary circuits are approximate $k$-designs at a circuit depth scaling as $O(N k)$, where $N$ is the total number of qudits \cite{2019arXiv190512053H}. In this subsection, we will need at most $k=4$, so this quantifies what we mean by ``late-time.''}. In the following sections, we will refine these results in order to understand early-time behavior and the membrane theory.

The advantage of modeling chaotic dynamics with Haar random unitary circuits is that analytic results are tractable due to well known results from random matrix theory.
In general, we will only need the Weingarten formula which computes the integral of monomials of unitary operators with the Haar measure \cite{2002math.ph...5010C}
\begin{align}
    &\int \left[ dU \right] U_{i_1, j_1}U_{i_2, j_2} \dots U^*_{i_1', j_1'}U^*_{i_2', j_2'}\dots 
    \nonumber 
    \\
    &= \sum_{\sigma, \tau \in S_n} \delta_{i_1i_{\sigma(1)}'}\dots\delta_{i_ni_{\sigma(n)}'}\delta_{j_1j_{\tau(1)}'}\dots\delta_{j_nj_{\tau(n)}'}\mbox{Wg}(d, \sigma \tau^{-1}),
\end{align}
where $d$ is the rank of the unitary.
The sum is over elements of the permutation group and $\mbox{Wg}$ is the Weingarten function. We will consider the large system size limit such that the term with $\sigma \tau^{-1} = e$ (the identity) is dominant and approximately
\begin{align}
    \mbox{Wg}(d, e) = \frac{1}{d^n}+ \mathcal{O}(d^{-n-2}).
\end{align}
leading to 
\begin{align}
    &\int \left[ dU \right] U_{i_1, j_1}U_{i_2, j_2} \dots U^*_{i_1', j_1'}U^*_{i_2', j_2'}\dots 
    \nonumber
    \\
    &\simeq \frac{1}{d^n} \sum_{\sigma\in S_n} \delta_{i_1i_{\sigma(1)}'}\dots\delta_{i_ni_{\sigma(n)}'}\delta_{j_1j_{\sigma(1)}'}\dots\delta_{j_nj_{\sigma(n)}'}.
    \label{Wg_approx}
\end{align}

\begin{figure}
    \centering
    \includegraphics[height = 4.0cm]{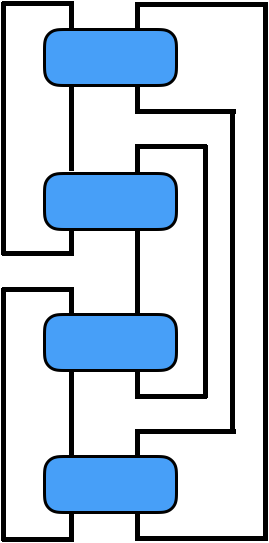} \qquad\quad
    \includegraphics[height = 4.0cm]{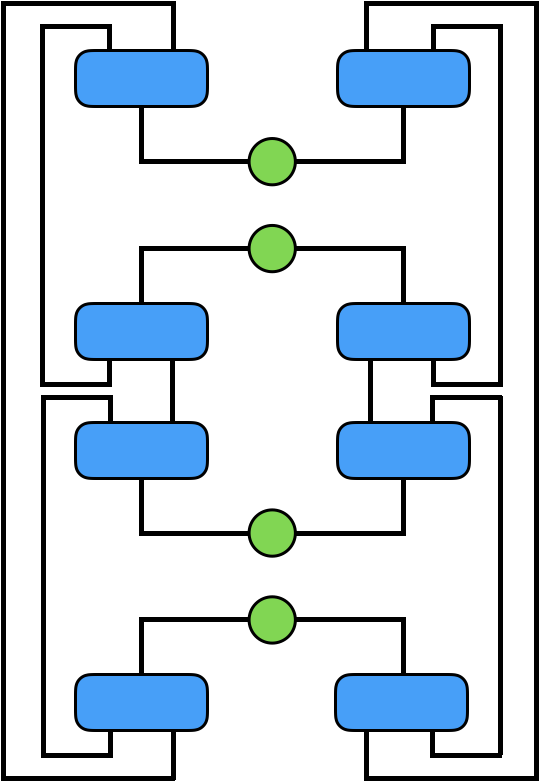}
    \caption{Left: The network that computes the late-time value of the second R\'enyi operator mutual information for the time evolution operator. Right: The network that computes the second R\'enyi local operator mutual information. The blue rectangles are random unitary operators and the green circles are local operators.}
    \label{haardiagrams}
\end{figure}

In Fig.~\ref{haardiagrams}, we show the diagrams that compute the average of $\Tr \rho^2$ (the purity) for the states $\ket{U(t)}$ and $\ket{\mathcal{O}(x,t)}$. Assuming that the average and the logarithm approximately commute for large system sizes, this computes the average second R\'enyi entropy. While it is in principle possible to compute all the ``average R\'enyi entropies'', which we denote
\begin{align}
    \tilde{S}_n = \frac{1}{1-n}\log \overline{\Tr \rho^n},
\end{align}
and analytically continue to $n=1$, we will only consider $n=2$ for simplicity. An interesting fact is that while, in general, $\tilde{S}_n \neq \overline{S_n}$, in the von Neumann limit\footnote{We first became aware of this fact from Ref.~\cite{2011PhRvB..83d5110F} though are unaware of its origins.}, they are equal, allowing one to properly find the average entanglement entropy. In fact, such an analytic continuation was computed for operator entanglement of the time evolution operator in Ref.~\cite{2017PhRvB..95i4206Z}.

We warm up with the time-evolution operator and then proceed to local operators. For a $q^L \times q^L$ time-evolution operator, the corresponding state is normalized as
\begin{align}
    \ket{U(t) }= \frac{1}{q^{L/2}} \sum_{i \tilde{j}}U_{i, \bar{j}} \ket{i}\otimes \ket{\tilde{j}} \equiv \frac{1}{q^{L/2}} U_{i, \bar{j}}.
\end{align}
Here, the two indices represent the input and output Hilbert spaces respectively as the total state is an element of the Hilbert space of $U(q^L)$.
The density matrix is then
\begin{align}
    \rho(t) = \frac{1}{q^{L}} U_{i, \bar{j}}U^*_{ i',\bar{j}'}.
\end{align}
We can bipartition both the input and output systems arbitrarily
\begin{align}
    \rho(t) = \frac{1}{q^{L}} U_{A B C D }U^*_{ A' B'C' D'}.
\end{align}
The reduced density matrix on $AC$ is
\begin{align}
    \rho_{AC}(t) = \frac{1}{q^{L}} U_{A B C D }U^*_{A' B C' D }
\end{align}
where summation over repeated indices is implied. The average purity is then
\begin{align}
    &\overline{\Tr \rho_{AC}^2(t)} 
    \nonumber
    \\&= \frac{1}{q^{2L}} \int \left[dU \right]U_{A' B_1 C' D_1 }U^*_{A B_1 C D_1 }U_{A B_2 C D_2 }U^*_{A' B_2 C' D_2 } \nonumber
    \\
    &= \frac{1}{q^{2L}} \int \left[dU \right]U_{A_1 B_1 C_1 D_1 }U^*_{A_1' B_1' C_1' D_1' }U_{A_2 B_2 C_2 D_2 }U^*_{A_2' B_2' C_2' D_2' }
    \nonumber \\
    &
    \quad 
    \times \left(\delta_{A_1 A_2'}\delta_{B_1B_1'}\delta_{C_1 C_2'}\delta_{D_1D_1'}\delta_{A_1'A_2}\delta_{C_1'C_2}\delta_{B_2 B_2'}\delta_{D_2 D_2'} \right) .
\end{align}
The Wiengarten formula involving only four unitaries needed for the above is simple enough that we may write it out explicitly in terms of Kronecker deltas 
\begin{align}
    &\int \left[dU \right]U_{i_1 j_1 }U^*_{i_1' j_1' }U_{i_2 j_2 }U^*_{i_2' j_2' } 
    \nonumber
    \\
    &= \frac{1}{q^{2L} - 1}\left(\delta_{i_1 i_1'}\delta_{i_2 i_2'}\delta_{j_1 j_1'}\delta_{j_2 j_2'} + \delta_{i_1 i_2'} \delta_{i_2 i_1'}\delta_{j_1 j_2'}\delta_{j_2 j_1'}   \right) \nonumber
    \\
    &-\frac{1}{q^L(q^{2L}-1)}\left(\delta_{i_1 i_1'}\delta_{i_2 i_2'}\delta_{j_1 j_2'}\delta_{j_2 j_1'} + \delta_{i_1 i_2'} \delta_{i_2 i_1'}\delta_{j_1 j_1'}\delta_{j_2 j_2'}   \right).
\end{align}
This leads to
\begin{align}
    \overline{\Tr \rho_{AC}^2(t)} 
    &= \frac{1}{q^{4L}-q^{2L}}(q^a q^{2b} q^c q^{2d}+q^{2a} q^{b} q^{2c} q^{d})
    \nonumber \\
    &\quad 
    -\frac{1}{q^{3L}(q^{2L}-1)}(q^{a} q^{2b} q^{2c} q^{d}+q^{2a} q^{b} q^{c} q^{2d})
    \nonumber \\
    &=  \frac{1}{q^{2L}-1}(q^bq^d +q^aq^c)
    \nonumber \\
    &\quad  - \frac{1}{q^L(q^{2L}-1)}(q^bq^c+q^aq^d).
\end{align}
Thus
\begin{align}
    \tilde{S}^{(2)}_{AC} = -\log \left[\frac{1}{q^{2L}-1}(q^bq^d +q^aq^c) - \frac{1}{q^L(q^{2L}-1)}(q^bq^c+q^aq^d)\right].
\end{align}
Here, $a$, $b$, $c$, and $d$ are the number of qudits in subsystems $A$, $B$, $C$, and $D$ respectively. Let's look at the tripartite mutual information, 
taking $A$ to be $O(1)$ and $B_1$ and $B_2$ to be semi-infinite (scale as e.g.~$L/2)$. Then,
\begin{align}
\tilde{I}^{(2)}_{AB_1},\tilde{I}^{(2)}_{AB_2} 
&\sim \log \Big[\frac{1}{q^{2L}}(q^{L-a}q^{L/2} +q^a q^{L/2}) 
\nonumber \\
&\qquad - \frac{1}{q^L(q^{2L})}(q^{L-a}q^{L/2}+q^aq^{L/2})\Big]  
\nonumber \\
&\qquad
+\left(\frac{L}{2}+a\right)\log{q}  
     \rightarrow 0,
 \\
\tilde{I}^{(2)}_{AB} &= 2 a \log q, 
\end{align}
where $a$ is the length of the subsystem. Therefore, TOMI tends to $-2a \log q$. If we had taken the output subsystem to be size $L-\epsilon$, then
\begin{align}
    \tilde{I}^{(2)}_{AB_{\epsilon}} &= -\log \Big[\frac{1}{q^{2L}-1}(q^{L-a}q^{\epsilon} +q^aq^{L-\epsilon}) 
    \nonumber \\
    &\quad - \frac{1}{q^L(q^{2L}-1)}(q^{L-a}q^{L-\epsilon}+q^aq^{\epsilon})\Big] 
    \nonumber
    \\&\quad 
    -(L-\epsilon+a)\log{q} 
    \nonumber \\
    &
    \rightarrow  \log \Big[q^{\epsilon-a} \nonumber
    + q^{a-\epsilon} + q^{-\epsilon-a}\Big]+(\epsilon-a)\log{q} .
\end{align}
Thus, the condition for nontrivial mutual information is 
\begin{align}
    \tilde{I}^{(2)}_{AB_{\epsilon}} \simeq \begin{cases} 0 & \epsilon > a \\
    2(a-\epsilon) \log q & \epsilon < a 
    \end{cases}.
\end{align}
This means that at late times in a chaotic quantum channel, one needs at least $(L-a)/L$ of the system to recover any information from $A$.

We now progress to local operator entanglement. As shown in Fig.~\ref{haardiagrams}, we have twice the number of unitaries to worry about. In the limit of large Hilbert space dimension such that we can make the approximation of (\ref{Wg_approx}), this is still tractable by brute force. Our state associated to the local operator is
\begin{align}
    \ket{\mathcal{O}(t)} = \frac{1}{\sqrt{\langle \mathcal{O}^{\dagger}\mathcal{O}\rangle}} U_{i, \bar{j}}\mathcal{O}_{\bar{j},i'}U^*_{ i',\bar{j}'},
\end{align}
so the density matrix is
\begin{align}
    \rho^{\mathcal{O}}(t) = \frac{1}{\langle \mathcal{O}^{\dagger}\mathcal{O}\rangle}U_{i_1, \bar{j}_1}\mathcal{O}_{\bar{j}_1,i_1'}U^*_{ i_1',\bar{j}_1'}U^*_{i_2, \bar{j}_2}\mathcal{O}^{\dagger}_{\bar{j}_2,i_2'}U_{ i_2',\bar{j}_2'}.
\end{align}
Again, we bipartition the input and output Hilbert spaces
\begin{align}
    \rho^{\mathcal{O}}(t) &= \frac{1}{\langle \mathcal{O}^{\dagger}\mathcal{O}\rangle}U_{A_1B_1 C_1D_1}\mathcal{O}_{A_1'B_1'C_1D_1}
    \nonumber
    \\& U^*_{ A_1' B_1'C_1'D_1'}U^*_{A_2B_2C_2D_2}\mathcal{O}^{\dagger}_{A_2'B_2'C_2D_2}U_{ A_2'B_2'C_2'D_2'},
\end{align}
so that we may define the reduced density matrix on $AC$
\begin{align}
    \rho^{\mathcal{O}}_{AC} &= \frac{1}{\langle \mathcal{O}^{\dagger}\mathcal{O}\rangle}U_{A_1B C_1D_1}\mathcal{O}_{A_1'B_1'C_1D_1}U^*_{ A_1' B_1'CD_1'}
    \nonumber
    \\
    &U^*_{A_2BC_2D_2}\mathcal{O}^{\dagger}_{A_2'B_2'C_2D_2}U_{ A_2'B_2'CD_2'}.
\end{align}
The Haar averaged purity is
\begin{align}
    &\overline{\Tr \left(\rho^{\mathcal{O}}_{AC}\right)^2 }\nonumber
    \\
    &= \frac{1}{\langle \mathcal{O}^{\dagger}\mathcal{O}\rangle^2}\left(\mathcal{O}^{\dagger}_{A_2B_2C_2D_2}\mathcal{O}_{A_3B_3C_3D_3}\mathcal{O}^{\dagger}_{A_6B_6C_6D_6}\mathcal{O}_{A_7B_7C_7D_7}\right)\nonumber
    \\
    &\times \Big(\delta_{D_1D_4'}\delta_{C_1C_8'}\delta_{A_1'A_8}\delta_{B_1'B_4} \delta_{B_1B_2}\delta_{A_1A_2} \delta_{D_1'D_2} \delta_{C_1'C2}\nonumber
    \\
    &\times \delta_{A_3A_4'} \delta_{B_3B_4'} \delta_{C_3C_4} \delta_{D_3D_4} \delta_{C_4'C_5} \delta_{A_4A_5'} \delta_{D_5D_8'} \delta_{B_5'B_8} \delta_{B_5B_6} \nonumber 
    \\
    &\times \delta_{A_5A_6} \delta_{D_5'D_6} \delta_{C_5'C_6} \delta_{A_7A_8'} \delta_{B_7B_8'} \delta_{C_7 C_8} \delta_{D_7D_8}\Big)
    \nonumber
    \\
    &\times\int \left[dU\right]\Big[ U_{A_1B_1 C_1D_1}U^*_{ A_1' B_1'C_1'D_1'} U_{A_4B_4C_4D_4}U^*_{ A_4'B_4'C_4'D_4'}\nonumber
    \\
    &\times U_{A_5B_5 C_5D_5}U^*_{ A_5' B_5'C_5'D_5'}U_{A_8B_8C_8D_8}U^*_{ A_8'B_8'C_8'D_8'}\Big].
\end{align}
We can see from (\ref{Wg_approx}) that this integral will involve 24 terms, even after the approximation. After the contraction of many Kronecker delta's, one finds at leading order
\begin{align}
    \overline{\Tr \left(\rho^{\mathcal{O}}_{AC}\right)^2 }&\simeq \frac{q^aq^c + q^bq^d}{q^{2L}}  + \frac{q^aq^d + q^bq^c}{q^{2L}} \frac{\langle \mathcal{O}^{\dagger} \mathcal{O}\mathcal{O}^{\dagger} \mathcal{O}\rangle}{\langle \mathcal{O}^{\dagger}\mathcal{O}\rangle^2}.
\end{align}
In general, the second term will be subleading. The immediate consequence is that the answer is operator independent. We then find
\begin{align}
    \tilde{S}^{(2)}_{AC} \simeq -\log \left[\frac{q^{a+c} + q^{2L-a-c}}{q^{2L}} \right].
\end{align}
The mutual information is
\begin{align}
  \tilde{I}^{(2)}_{AC} &\simeq 2L\log q - \log \left[\frac{(q^{c} + q^{2L-c})(q^{a} + q^{2L-a})}{q^{a+c} + q^{2L-a-c}} \right]
  \nonumber
  \\
    &\simeq - \log \left[\frac{q^{c-a} + q^{2L-c-a}}{q^{a+c} + q^{2L-a-c}} \right].
\end{align}
To find the TOMI, we take $c,d = L/2$
\begin{align}
    \tilde{I}_3^{(2)} \simeq  -2a\log q + \log 2.
\end{align}
The second term is subleading in the scaling limit. Because this saturates the bound on tripartite mutual information, we know the state must be approximately maximally entangled, so all of the R\'enyi entropies are approximately equivalent. Thus, we find 
\begin{align}
    \tilde{I}_3^{(n)} \simeq -2a \log q, \quad \forall n.
    \label{Haar_I3}
\end{align}
This is the lower bound on $I_3$ allowed by quantum mechanics. Given that this is an important point that will continue come up in this work, we now show the simple derivation. $I_3$ is composed of three individual mutual informations. The mutual information is positive semi-definite, so
\begin{align}
    I_3(A, B_1, B_2) \geq -I(A, B_1 \cup B_2).
\end{align}
The mutual information between subregion $A$ and the entire output $B_1 \cup B_2$ is a time independent quantity in finite-dimensional systems. The input and outputs are maximally entangled by construction, so the mutual information is twice the logarithm of the Hilbert space dimension
\begin{align}
    I_3(A, B_1, B_2) \geq -2 \log \left|\mathcal{H}_A \right| .
\end{align}
This is only well-defined for subsystems in discrete models. When we move on to quantum field theories, we must regulate the Hilbert space with \eqref{I3qft_reg} as the analog. This will lead to strange effects such as the regularized dimension of the Hilbert space associated to a finite region being time-dependent.

\subsection{Random Clifford circuits}
\begin{figure*}
    \centering
    \includegraphics[width = .43\textwidth]{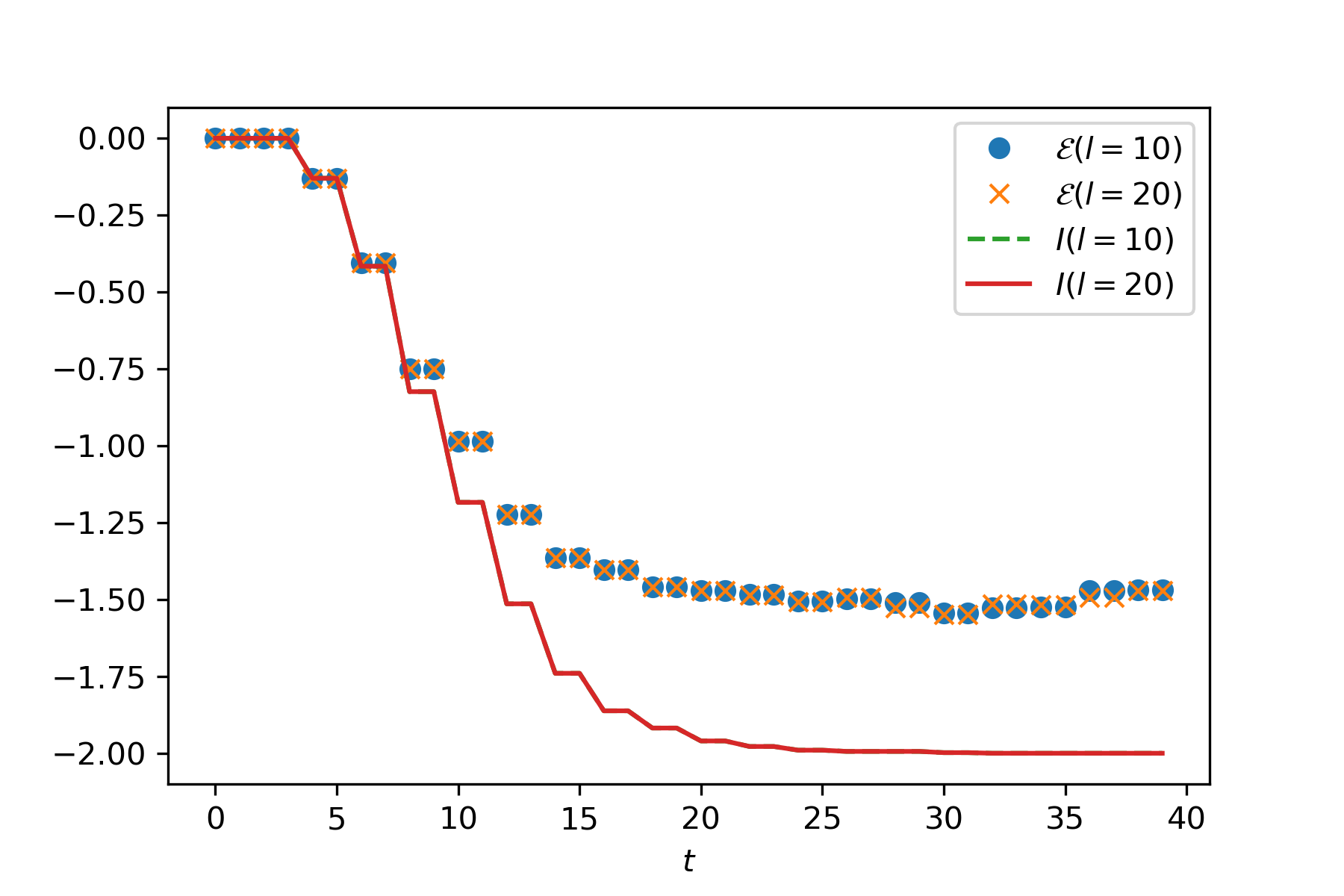}
    \includegraphics[width = .43\textwidth]{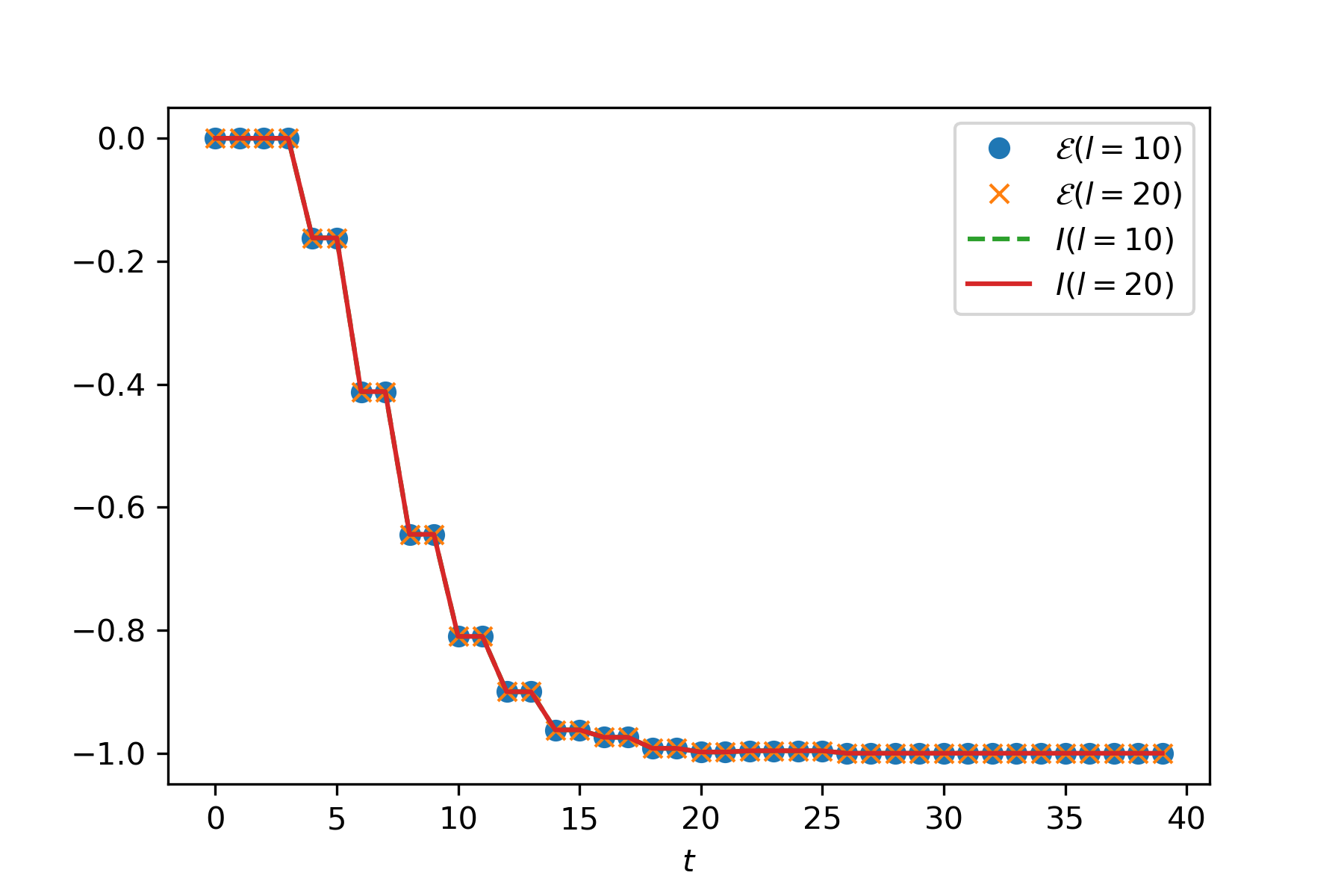}
    \caption{The change in the bipartite mutual information and logarithmic negativity for the CNOT gate (left) and Hadamard gate (right) under random Clifford evolution. The intervals are symmetric and of lengths $l = \{10,20\}$ with the local operator inserted $5$ lattice sites to the left of the intervals.  Note that, after averaging (500 realizations), the size of the system plays no role. Many of the plots are lie directly on one another. The change in negativity and mutual information are different for the CNOT gate but the same for the Hadamard.}
    \label{BOMI_clif}
\end{figure*}

The random unitary calculations above effectively capture the late-time behavior of chaotic channels. However, to study interesting early-time behavior, we have  a couple of options.
One option for modeling strongly-interacting dynamics for large systems sizes is random Clifford circuits. These are random unitary circuits that are composed of the Clifford group: phase, Hadamard, and CNOT gates. Using the stabilizer formalism, measuring entanglement in these circuits is tractable with computational times scaling polynomially. For entanglement of the unitary evolution operator, it was shown that these circuits maximally scramble and behave extremely similarly to holographic quantum channels \cite{2020JHEP...01..031K}. However, it is also known that these circuits have pathological OTOC \cite{2018PhRvX...8b1013V}. 
Even though the late-time average OTOC is zero, the variance is order 1. This is due to Clifford gates being unitary 3-designs. As is clear from Fig.~\ref{haardiagrams}, one must take the fourth moment of the unitary group in order to compute $\tilde{S}^{(2)}$ for local operators. Because Cliffords are 3-designs, a priori, they may have distinct behavior from the Haar random unitaries and chaotic channels in general.

\begin{figure*}
    \centering
    \includegraphics[width = .43\textwidth]{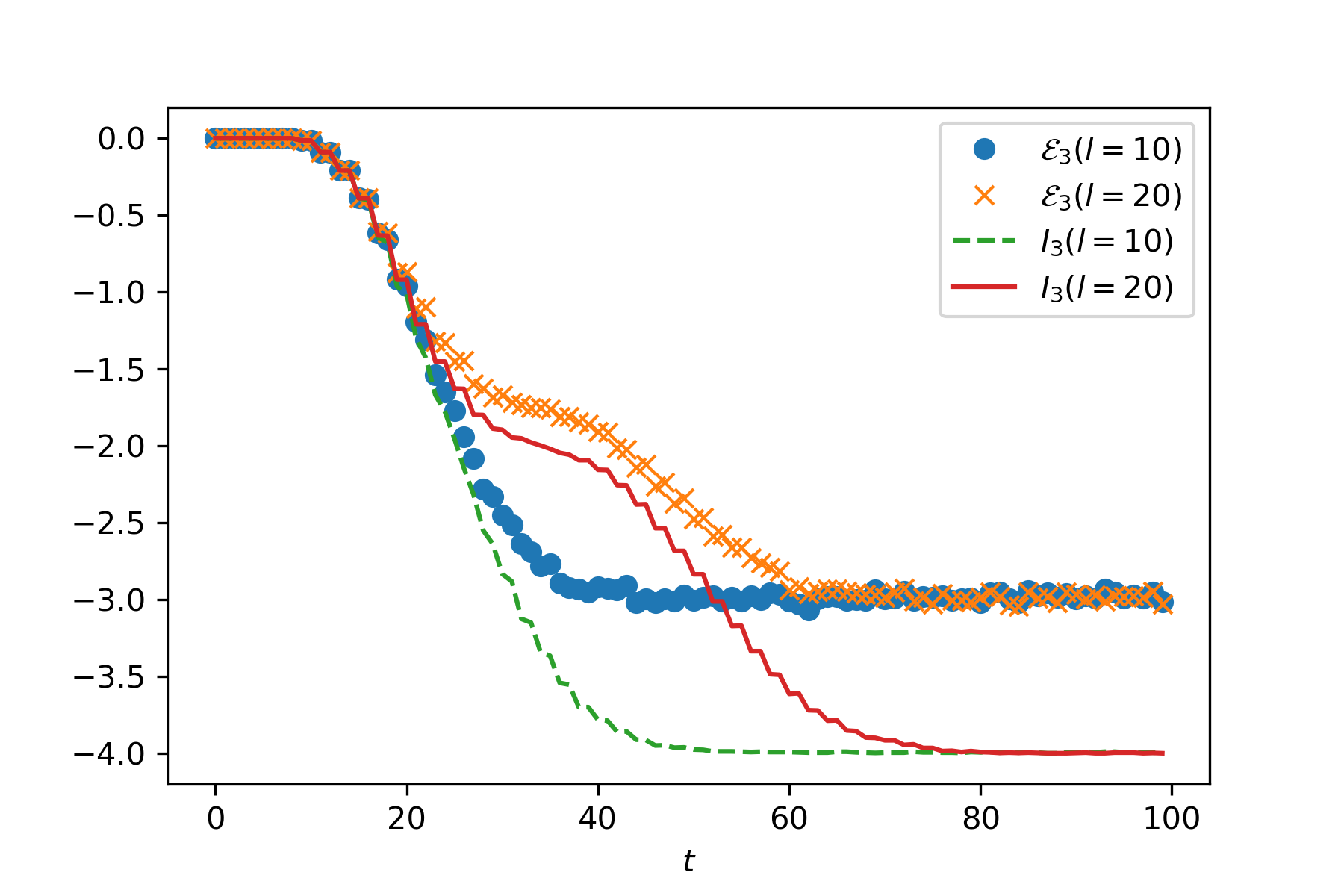}
    \includegraphics[width = .43\textwidth]{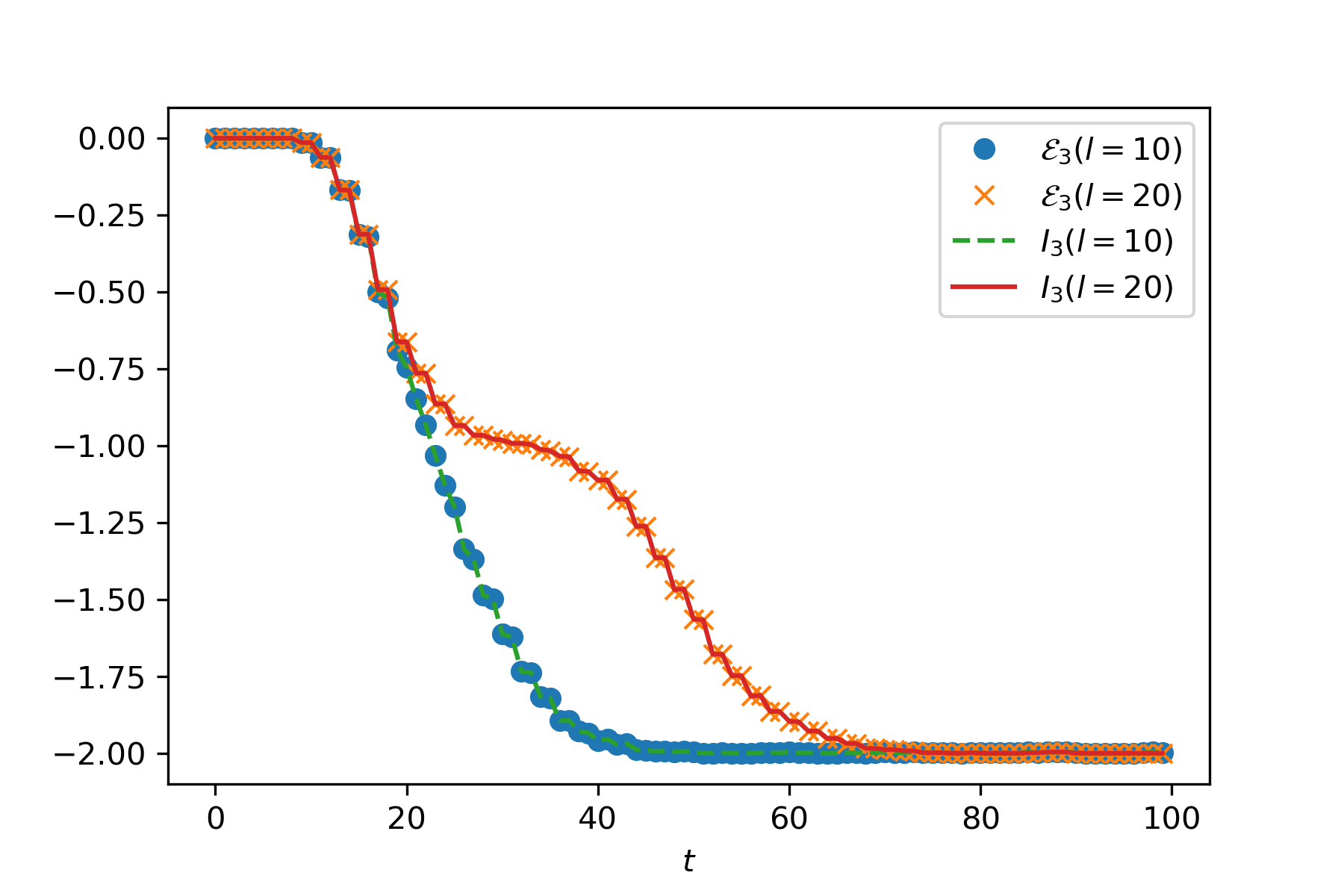}
    \caption{500 realizations for CNOT gate (left) and Hadamard gate (right). Notably, the CNOT gate scrambles both classical and quantum information while the Hadamard gate only scrambles quantum information. This can be seen by the fact that $\mathcal{E}_3$ is larger than $I_3$ for the CNOT gate, but identical to $I_3$ for the Hadamard. Again, the operator is inserted $5$ sites to the left of the symmetric intervals.}
    \label{TOMI_clif}
\end{figure*}

Indeed, this is what we find. The late-time value of the TOMI is a constant, independent on the size of subregion or total system size. However, it does depend on which operator we are evolving. Moreover, the operator has left and right-moving components, so given a configuration where the input subregion spatially overlaps with the partition of the output Hilbert space, twice the information will be scrambled compared to if the operator is initially outside of the interval overlaps. For simplicity, we have used the three local unitary operators that generate the Clifford group as our local operators. This behavior is reminiscent of integrable theories, however we find that there are no recurrences, even for the finite system, a feature of the stochastic time evolution.

In Fig.~\ref{BOMI_clif}, we show the time evolution of the operator mutual information and operator logarithmic negativity\footnote{We use the CNFP algorithm of Ref.~\cite{2005NJPh....7..170A} to compute the negativity or equivalently the total number of Bell pairs shared between regions.} for symmetric intervals. At early times, when the operator has not yet reached the intervals, the correlations are maximal, proportional to the area of the intervals. However, once the operator has time to reach the intervals, the correlations decrease because some information is being scattered as in Fig.~\ref{op_intro_cartoon}. We observe the following interesting features that distinguish this quantum channel from chaotic channels, particularly the Haar random unitaries that we have studied. (1) The amount of information scattered is independent of the size of the subregions. (2) The amount of information scattered is operator dependent. In particular, the CNOT gate scatters more than the Hadamard gate. (3) The quantum and classical information delocalize differently. For certain operators, the saturation value of $I_3$ is equivalent to $\mathcal{E}_3$, while for others the saturation value of $I_3$ has greater magnitude than $\mathcal{E}_3$ (see Fig.~\ref{TOMI_clif}). The latter indicates that some purely classical information has been scrambled.

\subsection{Membrane theory}

While the quasi-particle picture is an effective description of entanglement propagation for all integrable systems, it fails to capture the qualitative features of entanglement production in chaotic systems. It is highly desirable to obtain an analogous universal description of entanglement dynamics for chaotic quantum systems. Recently, it has been proposed that these chaotic theories have effective hydrodynamical descriptions where the von Neumann and R\'enyi entropies may be computed by the area of a spacetime codimension-one brane, $\mathcal{M}$, which is characterized by its tension, $\mathcal{T}$, \cite{2017PhRvX...7c1016N,2018arXiv180300089J,2018PhRvD..98j6025M,2018PhRvX...8b1013V}
\begin{align}
    S^{(n)}(A) = \int_{\mathcal{M}_A} dt\,  \mathcal{T}^{(n)}({v}, {x},t).
    \label{entropy_membrane_eq}
\end{align}
where $x$ is the position of the membrane and $v$ is the space-time velocity of the membrane ($dx/dt$). The membrane $\mathcal{M}_A$ is the extremal surface with respect to the integrand of \eqref{entropy_membrane_eq} that is homologous to subregion $A$. Though derived from finite-dimensional quantum circuits, there are strong parallels of this construction to the holographic description of von Neumann and R\'enyi entropies in the $AdS/CFT$ correspondence \cite{2006PhRvL..96r1602R,2006JHEP...08..045R,2007JHEP...07..062H,2016NatCo...712472D}. In essence, both prescriptions require finding the area of an extremal surface that is homologous to the subregions of interest. 

With motivations from the holographic description of 
reflected entropy \cite{2019arXiv190500577D}, it was later proposed that this mixed state entanglement measures may also be computed by the area of a different codimension-one membrane 
\cite{2020arXiv200105501K}
\begin{align}
    S_R^{(n)}(A,B) &= 2\int_{E_W(A,B)} dt\,  \mathcal{T}^{(n)}({v}, {x},t).
    \label{ln_line_tension}
\end{align}
We denote this membrane $E_W$ because in the language of AdS/CFT, this surface is the entanglement wedge cross section, a natural geometric object in the bulk that generalizes the Ryu-Takayanagi surface \cite{2018NatPh..14..573U}. 
In the membrane theory, the entanglement wedge of $A\cup B$ is the codimension-one spacetime region whose boundary is $\mathcal{M}_{A\cup B} \cup A \cup B$.
$E_W(A,B)$ is then defined as the extremal surface separating subregions $A$ and $B$ within the entanglement wedge.
We note that for random unitary circuits with large local bond dimension $q$, the spectrum is effectively flat and the line tension is thus equal for all R\'enyi's. While these membrane descriptions were well motivated by analysis of random unitary circuits for states after global quenches and for operator states of the unitary evolution operator, for local operator states, only a highly symmetric case has been analyzed \cite{2018arXiv180300089J}; we show that it is straightforward to generalize to generic configurations, giving an intuitive explanation for our late-time result (\ref{Haar_I3}) from the previous section. The line-tension for the local operator entanglement is dependent on space and time, not just velocity. The line tension is the same as it was for unitary operator entanglement within the light cone of the local operator
\begin{align}
\label{linetension_eq}
    \mathcal{T}(v, x,t) =
    \begin{cases}
\log q & v < 1 , \\
v \log q & v > 1.
\end{cases}
\end{align}
while outside the light cone
\begin{align}
\label{linetension_eq2}
    \mathcal{T}(v, x,t) = v \log q , \quad \forall v.
\end{align}
This may be quickly seen by considering the minimal cut through the quantum circuit displayed in Fig.~\ref{circuit_cartoons}. A more sophisticated derivation is explained in Appendix \ref{app_stat_mech} by mapping the random unitary circuit to a classical spin model. In the large $q$ limit, the number of bonds cut is asymptotically equal to the R\'enyi entropy. This becomes \eqref{linetension_eq} \& \eqref{linetension_eq2} in the scaling limit. 

\begin{figure}
    \centering
    \includegraphics[height = .22\textwidth]{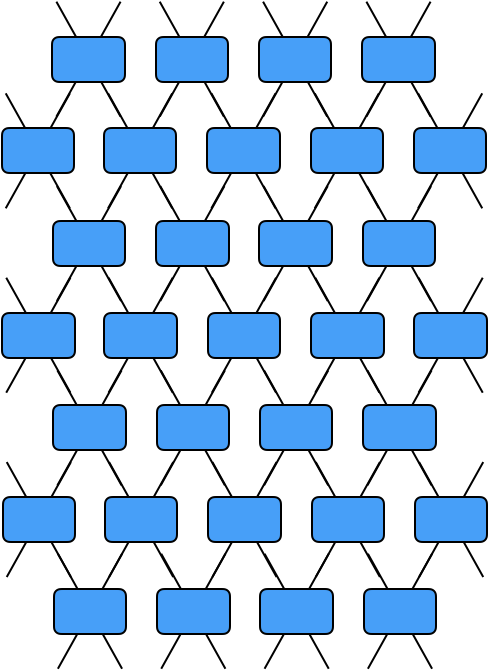} \quad 
    \includegraphics[height = .22\textwidth]{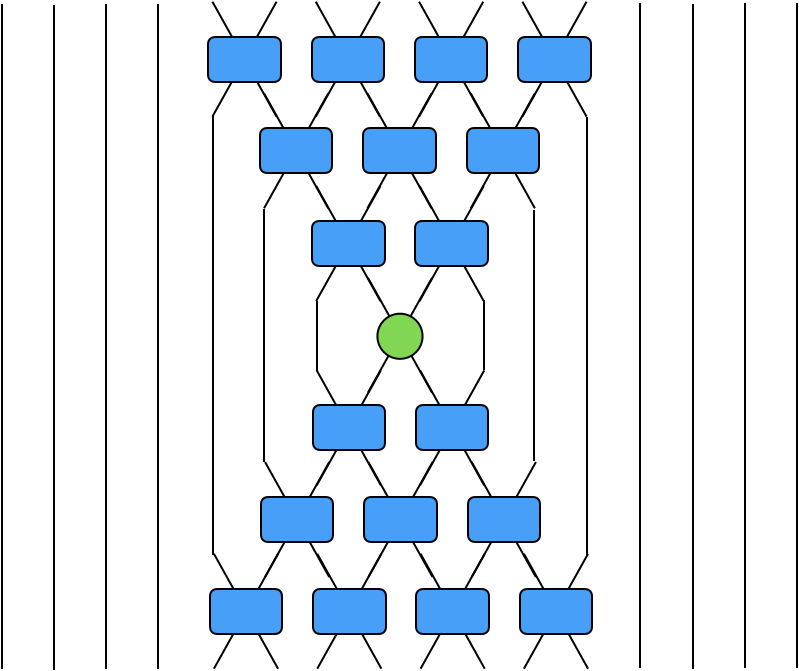}
    \caption{(left) cartoon of a random unitary circuit that would be analyzed for unitary time-evolution operator entanglement for example. (right) the random unitary circuit for the local operator is somewhat different because the $U$ and $U^{\dagger}$ cancel each other outside of the light cone of the local operator leaving maximally entangled Bell pairs.}
    \label{circuit_cartoons}
\end{figure}

\begin{figure*}
    \centering
    \includegraphics[width = .66\textwidth]{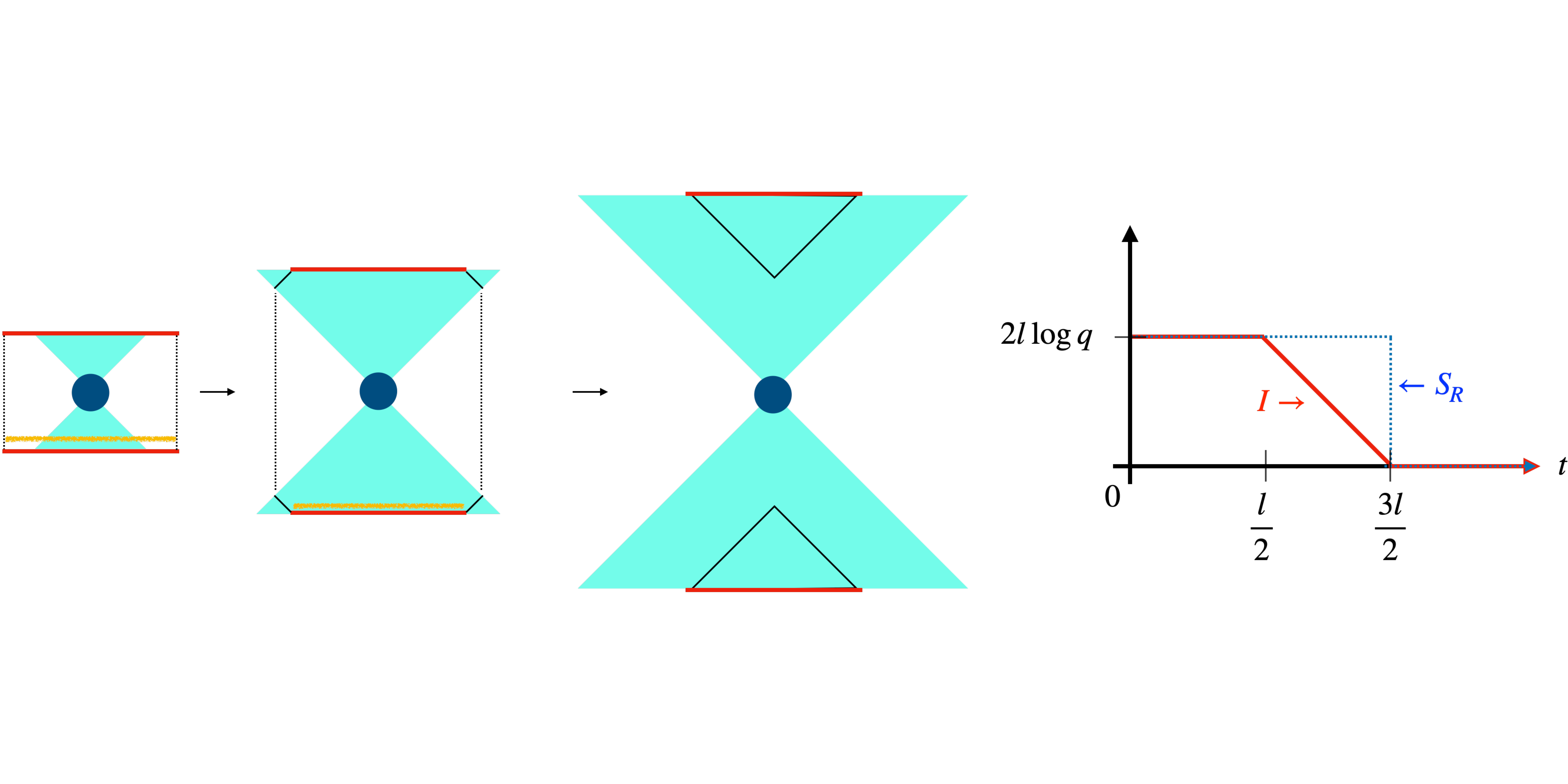}
    \caption{The membrane picture for the local operator is shown. The operator gains nontrivial support along light cones determined by the butterfly velocity (cyan). We show the minimal membranes for symmetric intervals. At early times (left), $I~ \& ~S_R$ are maximal. The black lines show the minimal membrane for $S_{A\cup B}$ and the orange lines show the minimal membrane for $S_R$. After $t = \frac{l}{2}$ (middle), the mutual information decreases while $S_R$ remains constant. Finally, when the minimal membrane becomes disconnected after $t = \frac{3l}{2}$ (right), the mutual information reaches zero and $S_R$ discontinuously jump to zero.}
    \label{loc_op_sym_cartoon}
\end{figure*}

It is instructive to work out a couple examples. 
We show three time steps for the entanglement entropy of symmetric intervals
of length $l$
in Fig.~\ref{loc_op_sym_cartoon}. Initially, the two intervals
$A$ and $B$ in the input and output Hilbert spaces, respectively,
are maximally entangled with one another, so their total entropy is zero. Using (\ref{linetension_eq2}), the corresponding minimal membrane is of zero area (left) because it has $v =0 $. Once the light cone of the operator reaches outside of the intervals, it can break the entanglement between them and entangle them with the rest of the system. This is seen in the intermediate time step where the area of the minimal membrane grows linearly. At sufficiently late times, the entropy saturates to its maximum value, $2 l \log q$, which is described by the disconnected regime on the right in Fig.~\ref{loc_op_sym_cartoon}. In summary, we find 
\begin{align}
    S_{A\cup B} = \begin{cases}
    0 & t < l/2 \\
    2 (t-l/2) \log q  & l/2 <  t < 3l/2 \\
    2 l \log q & t > 3l/2
    \end{cases}.
\end{align}
Because the individual entropies of the intervals are constant in time, we find the mutual information is 
\begin{align}
    I(A, B) = \begin{cases}
    2 l \log q & t < l/2 \\
    2 (3l/2-t) \log q  & l/2 <  t < 3l/2 \\
    0 & t > 3l/2
    \end{cases}.
\end{align}

\begin{figure*}
    \centering
    \includegraphics[width = .8\textwidth ]{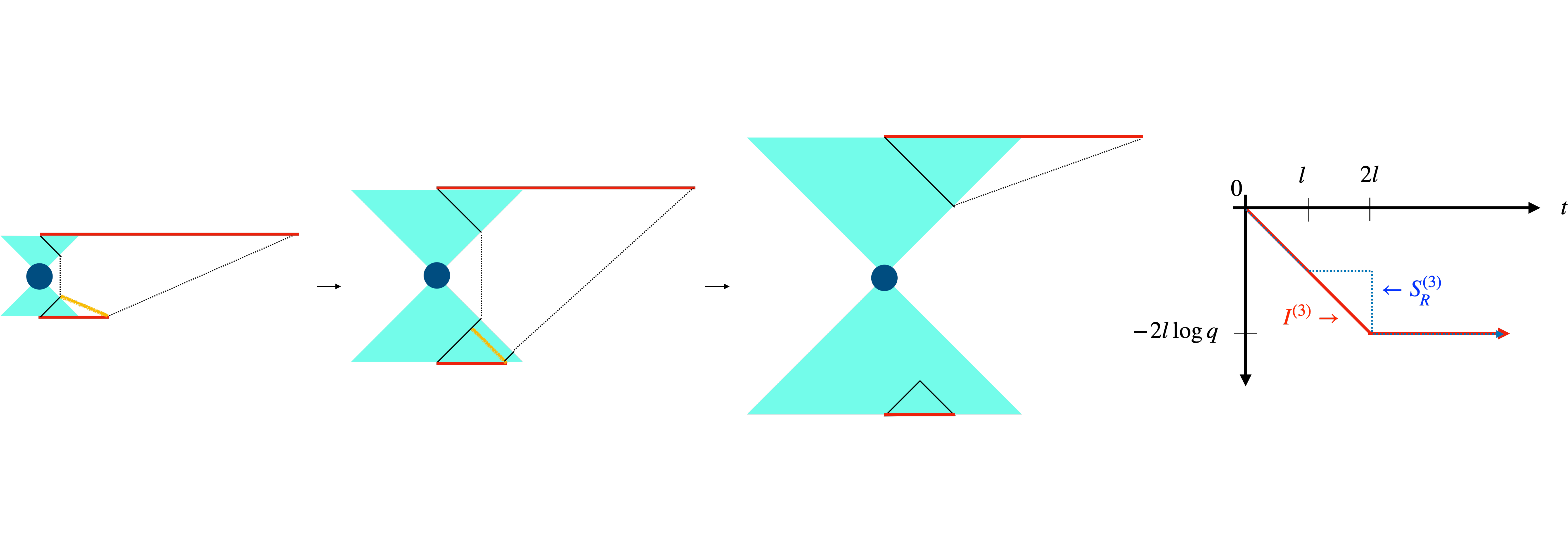}
    \caption{We show the membrane configurations for the only time-dependent term in the tripartite mutual information and reflected entropy. At early times (left), $E_W$ (orange) decreases linearly and the membrane for $S_{A\cup B}$ (black) increases linearly. At intermediate times (middle), $E_W$ stops decreasing but $S(A\cup B)$ continues to increase. At late times (right), the membrane is disconnected, so $I = S_R = 0$.}
    \label{loc_op_tripartite_cartoon}
\end{figure*}

We can also compute the full time dependence of the tripartite operator entanglement shown in Fig.~\ref{loc_op_tripartite_cartoon}. We take $A = (0,l),~ B_1 = (-\infty, 0), ~B_2 = (0,\infty)$ for simplicity, but the late-time value will be universal. We find
\begin{align}
    I(A,B_1) &= 0, \nonumber \\ 
    I(A,B_2) &= \max\left[0, 2  \log q(l - t) \right],
    \nonumber \\ 
    I(A,B) &= 2l \log q ,
\end{align}
which leads to a tripartite mutual information of 
\begin{align}
    I_3 = \max\left[-2l \log q, -2  t\log q   \right].
\end{align}
The saturation value is identical to the late-time result of the previous section (\ref{Haar_I3}) and is of maximum magnitude.

We can play the same game for reflected entropy. However, the relevant membrane is now given by the extremal cross section of the codimension-zero region bounded by $\mathcal{M}$ and the spacetime boundary. 
For the symmetric case shown in Fig.~\ref{loc_op_sym_cartoon}, we find step function behavior
\begin{align}
    S_R = \begin{cases}
    2 l \log q ,& t < \frac{3l}{2},
    \\
    0 , & t>  \frac{3l}{2}.
    \end{cases}
\end{align}
This is dramatically different than the mutual information. Similarly extreme differences were found between the mutual information and reflected entropy for irrational CFTs and random unitary circuits following a global quantum quench \cite{2020arXiv200105501K} and in operator entanglement of the reduced density matrix \cite{2019arXiv190709581W}. Interestingly, this discrepancy has never been observed for integrable theories. We would like to better understand this physically because its information theoretic implications are somewhat puzzling as discussed in Ref.~\cite{2020arXiv200105501K}. Multipartite entanglement must play a significant role, but the problem certainly deserves further attention.

For the semi-infinite configuration shown in Fig.~\ref{loc_op_tripartite_cartoon}, we have
\begin{align}
    S_R 
    = \begin{cases}
    (2 l - t) \log q ,& t < l,
    \\
    l\log q , &l < t <  2l
    \\
    0 , & 2l < t .
    \end{cases}
\end{align}
Because the other terms in the tripartite quantities are constant in time for the given configuration, we find
\begin{align}
    S_R^{(3)} 
    = \begin{cases}
     - t \log q ,& t < 2l,
    \\
    -l\log q , &l < t <  2l
    \\
    -2l\log q  , & 2l < t 
    \end{cases}.
\end{align}

Some of the CFT techniques that we use in subsequent sections are specific to von Neumann entropy and will not apply to the negativity and reflected entropy, so we do not evaluate these in CFT. While these calculations seem tractable, we leave this to future work and assume that the random unitary circuit analysis precisely describes the CFT computations once identifying the bond dimension with the Cardy density of states
\eqref{cardy_bond_dim},
$q = e^{\frac{\pi c}{3 \beta}}$.

\section{Free fermion system}

In this section, we will compute the local operator entanglement for a
(1+1)-dimensional lattice free fermion system
described by a quadratic Hamiltonian, $H = \sum_{x,y} c_x^\dagger \mathcal{H}_{xy} c_y$,
where $c_x/c_x^\dagger$ are the real space fermion annihilation/creation operators at site $x$
on the lattice.
Specifically, we will take the nearest-neighbor tight-binding Hamiltonian to be our free fermion Hamiltonian.
\begin{equation}
    H = -\tilde{t}\sum_x c_x^\dagger c^{\ }_{x+1} +\text{h.c.}
\end{equation}
This Hamiltonian is diagonalised by a Fourier transform $ A_{xk} = e^{i k x}/\sqrt{L} $, 
$A_{kx}^{-1}= e^{-ikx}/\sqrt{L}$,
where $L$ is the total length of the system.
The tight-binding dispersion relation is $E_k = -2\tilde{t}\cos{k}$.
As the computation of local operator entanglement for generic RCFTs is rather
involved, we will study the free fermion system numerically instead. 

In order to compare (match) free fermion numerics with field theory results,
UV regulators will have to be introduced in the operator state and taken to be
much larger than the lattice spacing in order to suppress lattice effects
\cite{2018arXiv181200013N,2020JHEP...01..031K}.
However, the introduction of UV regulators in the local operator state will
greatly complicate the expressions so we refrain from doing so.
We thus consider
the local operator state with no regulators
\begin{equation}
    |\mathcal{O}(t)\rangle = \mathcal{N} e^{-i H t}\mathcal{O}_A |\Omega \rangle
\end{equation}
where $H = H_B-H_A$, $\mathcal{O}_A = \mathcal{O}\otimes \mathbb{I}$ and $|\Omega \rangle$ is the infinite temperature thermofield double state.
The numerical results are not expected to agree precisely with field theory
calculations although they should capture the overall qualitative behaviour.
Here, the maximally entangled state can be written in terms of real space fermions as
\begin{equation}
    | \Omega \rangle = \prod_m \left( \frac{1+c_{Am}^\dagger c_{Bm}^\dagger}{\sqrt{2}}\right) |0 \rangle.
\end{equation}
As an local operator, we choose to work with
the single site fermion parity operator at some arbitrary site $z$,
\begin{equation}
    \mathcal{O} = 1-2 c_z^\dagger c_z = (-1)^{c_z^\dagger c_z}.
\end{equation}
This operator is the exponential of a quadratic fermion operator, so it is a Gaussian operator. 
Hence, we can utilize the correlator method to compute operator entanglement entropies.
Since the parity operator squares to one, the state is already normalized.
The initial local operator state is then given by the maximally entangled state with a sign flip at site $z$
\begin{equation}
    \mathcal{O}_A |\Omega \rangle = \prod_m \left( \frac{1+(-1)^{\delta_{mz}}c_{Am}^\dagger c_{Bm}^\dagger}{\sqrt{2}}\right) |0 \rangle.
\end{equation}
Noting 
that the time-evolution of the fermion operators under $H_B-H_A$ is given by 
\begin{align}\label{CommuteHamiltonianPastFermion02}
  e^{i H t}c_{Ix}^\dagger e^{-iHt}
  &= \frac{1}{L} \sum_{ka}e^{i(ka-kx+(-1)^ItE_k)}c_{Ia}^\dagger,
    \nonumber \\ 
    e^{ i Ht}c_{Ix} e^{-iHt}
  &= \frac{1}{L} \sum_{ka}e^{i(kx-ka-(-1)^ItE_k)}c_{Ia},
\end{align}
where $(-1)^A = -1$ and $(-1)^B=1$,
the correlation matrices are given by
\begin{align}\label{cdaggercmatrix}
C_{Ix, Jy} &= \langle \mathcal{O}(t)|c_{Ix}^\dagger c_{Jy} | \mathcal{O}(t)                \rangle 
            = \frac{1}{2}\delta_{IJ}\delta_{x,y},
            \nonumber\\ 
  F_{Ix,Jy} &= \langle \mathcal{O}(t)| c_{Ix}^\dagger c_{Jy}^\dagger |\mathcal{O}(t) \rangle
= \frac{1}{2}\epsilon_{IJ}\delta_{x,y}
\nonumber
\\
&-\frac{\epsilon_{IJ}}{L^2}\sum_{k,p} e^{i(kz-kx+(-1)^ItE_k+pz-py+(-1)^JtE_p)}.
\end{align}
The relevant operator entanglement entropies can then be computed by
diagonalizing subblocks of the correlation matrix.

\begin{figure*}
    \centering
    \includegraphics[width = .35\textwidth]{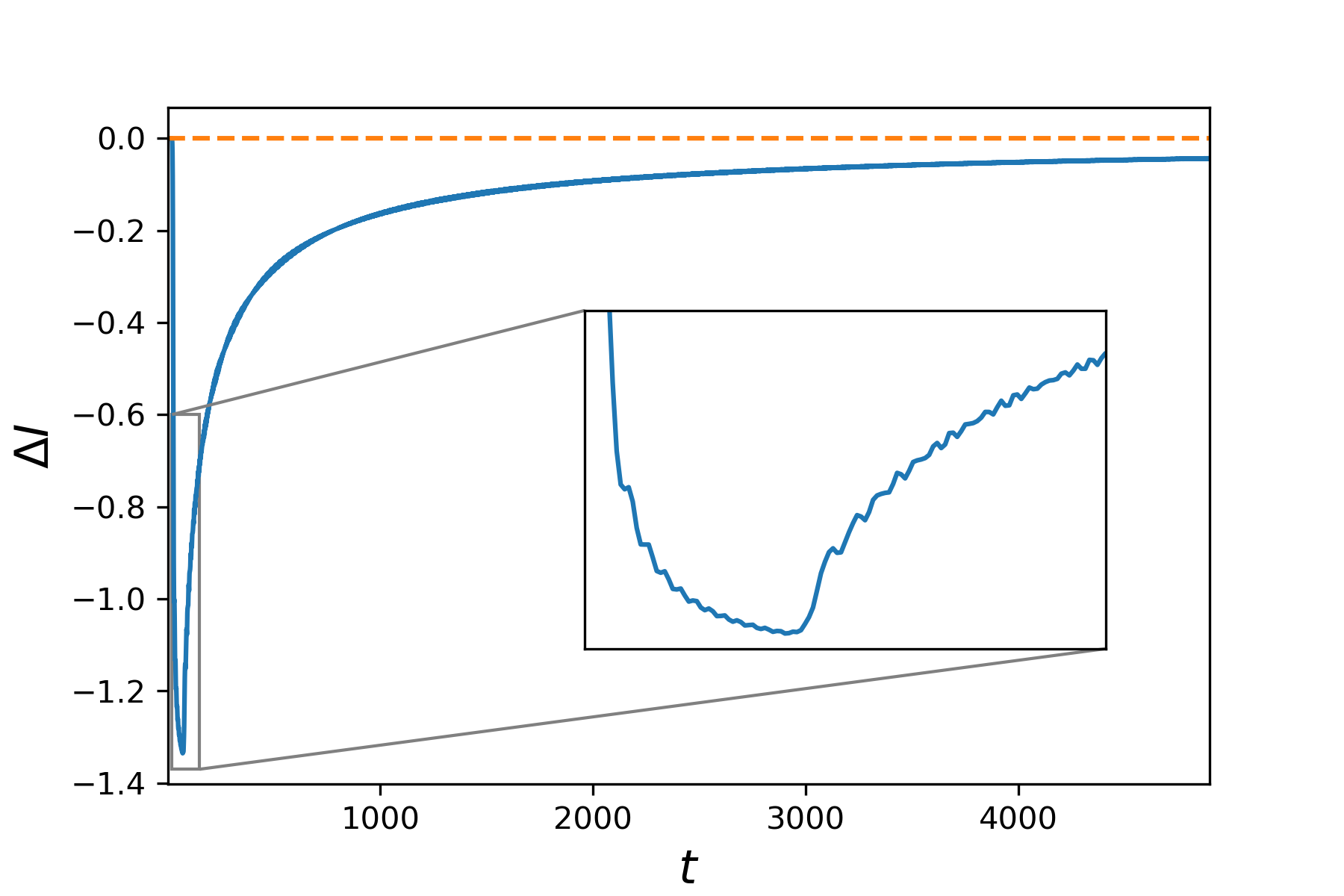}
    \includegraphics[width = .35\textwidth]{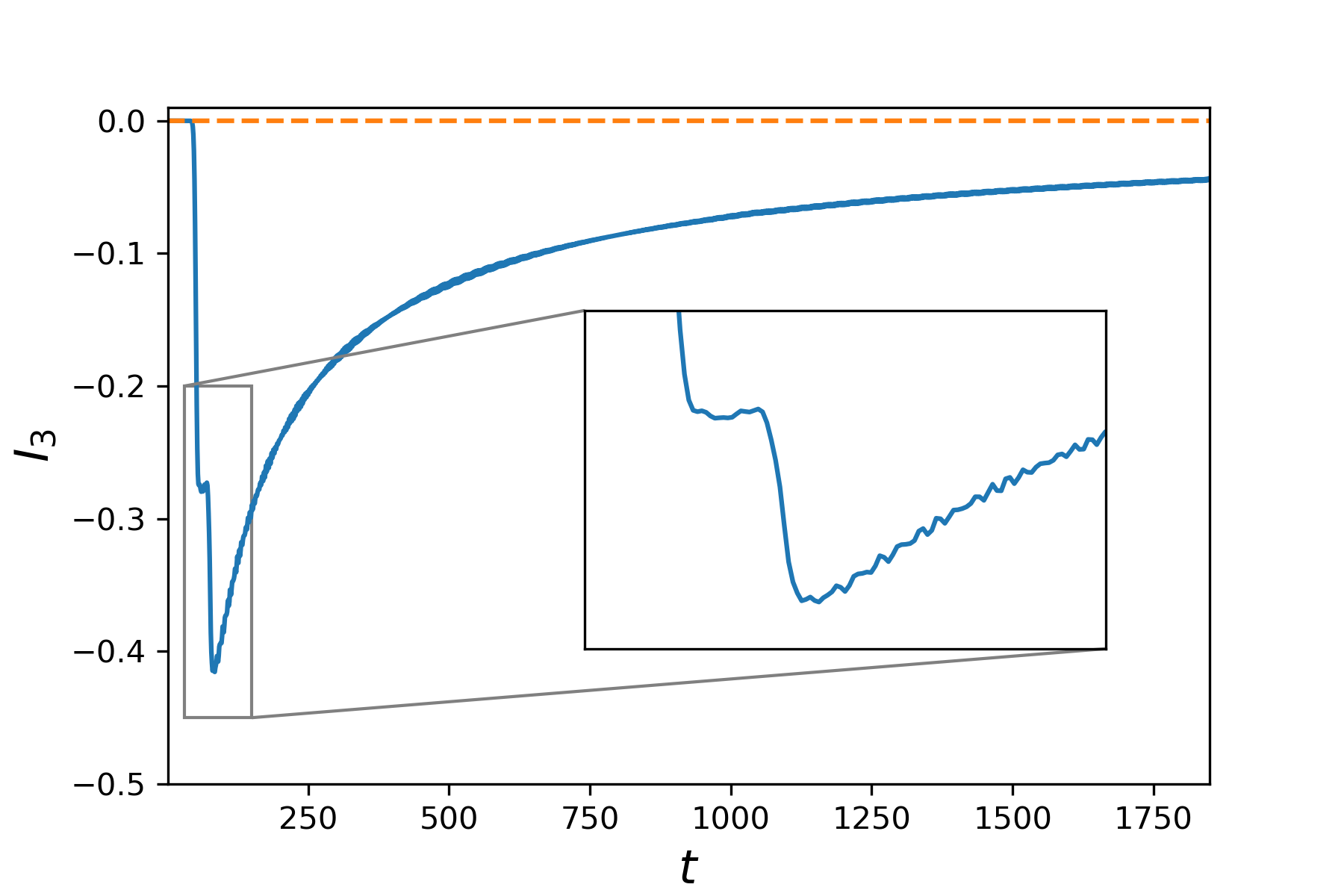}
    \caption{We show the evolution of the operator entanglement for the fermion parity operator at a single site $(-1)^{n_z}$. Left: the change in the operator mutual information is shown for symmetric intervals of length $50$ with the operator inserted $25$ sites away from the intervals. After a time corresponding to the distance from the operator to the intervals, the BOMI begins to drop. Once it passes through the intervals, it relaxes but asymptotes back to the value at which it started according to a power law with exponent between $-1/2$ and $-1$. This indicates that little information has been scattered. Right: the TOMI is shown for $l_A = 50$ and the partition of $B$ lying at the center of $A$. The local operator is again located $25$ sites away from $A$. We see minor delocalization of information when the operator has support in region $A$, but relaxes at late times back to zero.}
    \label{ff_plot}
\end{figure*}

We plot the results in Fig.~\ref{ff_plot} and find very different behavior than the random unitary circuits. In particular, the BOMI decreases from its initial value once the operator has time to enter the subregions. Then, it relaxes back once it has left the subregion. A similar analysis is made for the TOMI. It is presently unclear whether the values relax all the way back to their initial value because the lattice model has slow quasi-particle modes that take a very long time to travel through the intervals. This, however, 
is a moot point
because we clearly see that the operator scrambles very little (if any) information i.e.~the information in the free fermion channel is robust to perturbations, a quality we expect to be generic for integrable systems.

\section{Conformal field theory at large central charge}
\label{HolographicCFTs}

\subsection{Setup}
In this section, we compute the local operator entanglement for 2d conformal
field theories at large central charge.
These represent candidates theories possessing bulk gravitational duals well
described by semi-classical physics.
We now set up the path integral representation of operator entanglement.
We take a local operator situated at position $X$ in the Heisenberg picture and
expand in the energy eigen basis as
\begin{align}
    \mathcal{O}(X,t) &= e^{iH t}\mathcal{O}(X)e^{-iHt} 
    \nonumber
    \\
    &= \sum_{nm}e^{iE_n t}\mathcal{O}_{nm}(X)e^{-iE_mt}\ket{n}\bra{m}.
\end{align}
We then perform the state-operator map to create the local operator state in a doubled Hilbert space \footnote{It is understood that all states in the second Hilbert space are CPT conjugated.}
\begin{align}
    \ket{\mathcal{O}(X,t)}= \mathcal{N} \sum_{m,n}  e^{i (E_n-E_m) t -E_n \epsilon_1 - E_m \epsilon_2} \mathcal{O}_{nm}(X) \ket{n}_1\ket{m}_2
    \label{loc_op_state_eq}
\end{align}
where 
$\mathcal{N}$ is a normalization constant that ensures that the state has unit norm.
More specifically, the normalization squared is 
\begin{equation}
    |\mathcal{N}|^2 = \frac{1}{\text{Tr}\left[e^{-2(\epsilon_1+\epsilon_2)H}\mathcal{O}(X,2\epsilon_2)\mathcal{O}^\dagger(X)\right]}.
\end{equation}
Crucially, we have included regulators $\epsilon_1$ and $\epsilon_2$ in order to smear the operator and cut off the high-energy modes.

\begin{figure}
    \centering
    \includegraphics[width = .42 \textwidth]{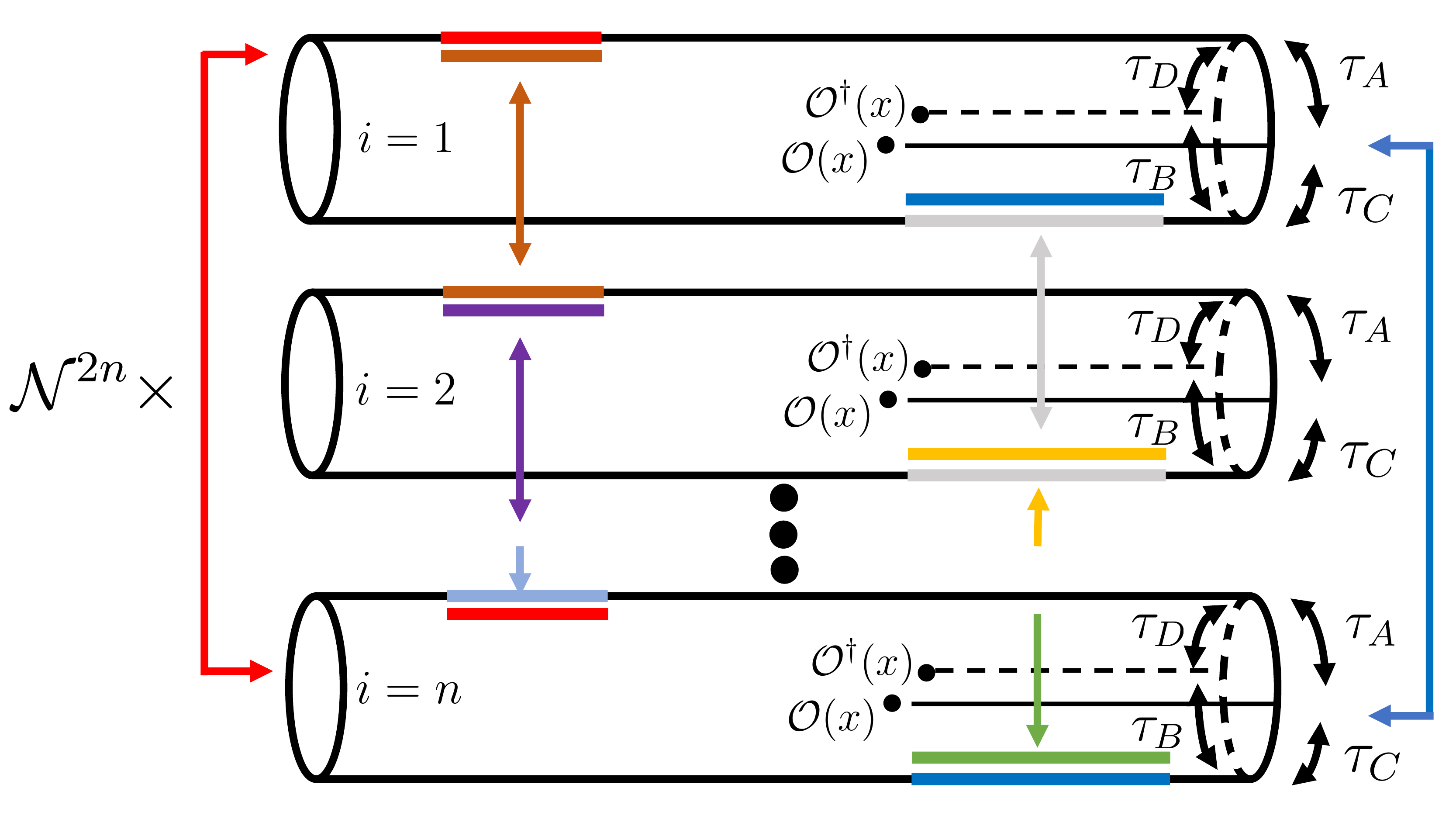}
    \caption{Displayed is the replica manifold for $S_{AB}$. The base manifolds are cylinders because the regulator sets the theory to finite temperature. The input and output intervals lie on opposite sides of the cylinder.}
    \label{pathintegral}
\end{figure}

Consider the density matrix corresponding to the state \eqref{loc_op_state_eq} in Euclidean signature
\begin{align}
    \rho_E &= \mathcal{N}^2 \sum_{n,m,A,B} \langle n| \mathcal{O}(X)| m\rangle\langle B|\mathcal{O}^\dagger(X)|A\rangle |n\rangle
    \nonumber
    \\
    &\times\langle A|_1\otimes|m\rangle\langle B|_2 e^{-\tau_A E_n}e^{-\tau_B E_A} e^{-\tau_C E_m} e^{-\tau_D E_B}.
\end{align}
After performing our computations of BOMI and TOMI in Euclidean space, we will perform the analytic continuation
\begin{equation}
  \tau_A \rightarrow \epsilon_1-i t,
  \quad \tau_B\rightarrow\epsilon_1+it,
  \quad \tau_C\rightarrow\epsilon_2+it,
  \quad\tau_D\rightarrow\epsilon_2-it.
\end{equation}
We bipartition the Hilbert spaces and write the density matrix elements in terms of the field configurations on these bipartitions
\begin{align}
    &\langle\Psi_{A_1},\Psi_1 |\langle\Phi_{B_1},\Phi_1|\rho_E|\Psi_{A_2},\Psi_2\rangle |\Phi_{B_2}\Phi_2\rangle 
     \nonumber 
     \\
    &
    = \mathcal{N}^2 \langle\Psi_{A_1},\Psi_1 |e^{-\tau_A H}\mathcal{O}(X) e^{-\tau_C H} |\bar{\Phi}_{B_1},\bar{\Phi}_1\rangle
    \nonumber
    \\
    &\quad \times\langle\bar{\Phi}_{B_2},\bar{\Phi}_2| e^{-\tau_D H}\mathcal{O}^\dagger(X) e^{-\tau_B H}|\Psi_{A_2},\Psi_2 \rangle
\end{align}
where the states corresponding to complex conjugated fields are defined by
$
\langle \bar{\Phi}_a,\bar{\Phi}_b|n\rangle = \langle n|\Phi_a,\Phi_b\rangle
$,
$
\langle n|\bar{\Phi}_a,\bar{\Phi}_b\rangle = \langle \Phi_a,\Phi_b |n\rangle.
$
Consider two subsystems $A$ and $B$ in the first and second Hilbert spaces respectively.
The reduced density matrix for the union of these two regions is
\begin{align}
    \rho_{A\cup B} 
  &=
    \mathcal{N}^2
    \int d\Psi_1 \int d\bar{\Phi}_1\,
    \nonumber \\
    &\quad 
    \times
      \langle\Psi_{A_1},\Psi_1 |e^{-\tau_A H}\mathcal{O}(X) e^{-\tau_C H} |\bar{\Phi}_{B_1},\bar{\Phi}_1\rangle
      \nonumber \\
    & \quad 
    \times 
      \langle\bar{\Phi}_{B_2},\bar{\Phi}_1| e^{-\tau_D H}\mathcal{O}^\dagger(X) e^{-\tau_B H}|\Psi_{A_2},\Psi_1 \rangle.
\end{align}
In order to compute the entropy and thence the mutual information, we must perform the replica trick
where we cyclically glue the path integrals defining the above state.
This replica manifold is shown in Fig.~\ref{pathintegral}.

Equivalently, we can consider a replicated theory on a single cylinder and introduce 
$\mathbb{Z}_n$ twist operators that implement the cyclic gluing, where $n$ is the number of replicas. 
The circumference of the cylinder is 
\begin{equation}
    \beta = 2(\epsilon_1+\epsilon_2).
\end{equation}
In the replicated theory, we consider the operator $\mathcal{O}_n$
which is the tensor product of the operators from each copy of the replica manifold
$
\mathcal{O}_n = \mathcal{O} \otimes \mathcal{O} \otimes \cdots \otimes \mathcal{O}.
$
If $\mathcal{O}$ has the conformal dimension $(h_\mathcal{O},\bar{h}_\mathcal{O})$,
then the corresponding operator in the replicated theory $\mathcal{O}_n$ has the conformal dimension $(n h_\mathcal{O},n \bar{h}_\mathcal{O})$.
The twist operators have conformal dimensions 
\begin{align}
    h_n = \bar{h}_n = \frac{c}{24}\left( n - \frac{1}{n}\right).
\end{align}
The operator entanglement entropy for two disjoint regions $A$ and $B$ is then computed by the following correlation functions
\begin{align}\label{SAB}
  &
    S_{A\cup B} 
    = \lim_{n\rightarrow 1}\frac{1}{1-n}
    \log\Bigg[
    \frac{\langle  
    \mathcal{O}_n^\dagger(w_1, \bar{w}_1)
    \mathcal{O}_n(w_2, \bar{w}_2)
    }
    {}
    \nonumber
    \\
    &\frac{\sigma_n(w_3, \bar{w}_3) 
    \bar{\sigma}_n(w_4, \bar{w}_4)
    \sigma_n(w_5, \bar{w}_5)
    \bar{\sigma}_n(w_6, \bar{w}_6) \rangle_{\beta}}{
    \langle \mathcal{O}^\dagger(w_1,\bar{w}_1)
    \mathcal{O}(w_2, \bar{w}_2)
    \rangle_{\beta}^n}\Bigg]
\end{align}
where we introduce the following coordinates on the cylinder: 
\begin{align}
    &w_1 = X+i\tau_B,
    &&w_2 = X+i(\tau_B+\tau_C+\tau_D),
    \nonumber \\
    &w_3=X_1, 
   &&w_4=X_2,
    \nonumber \\
    &w_5=Y_2+i(\tau_B+\tau_D),
    &&w_6=Y_1+i(\tau_B+\tau_D).
\end{align}
These coordinates are to be analytically continued at the end of the calculations:
\begin{align}
  &w_1  \xrightarrow{\text{a.c.}} X-t+i \epsilon_1,
    &&\bar{w}_1 \xrightarrow{\text{a.c.}} X+t-i \epsilon_1,
    \nonumber \\
  & w_2  \xrightarrow{\text{a.c.}}
    X-t+i(\beta-\epsilon_1),
    &&\bar{w}_2  \xrightarrow{\text{a.c.}} X+t-i(\beta-\epsilon_1),
    \nonumber \\
  &w_5 \xrightarrow{\text{a.c.}}Y_2+i\beta/2,
    &&\bar{w}_5\xrightarrow{\text{a.c.}}Y_2-i\beta/2,
    \nonumber \\
  & w_6\xrightarrow{\text{a.c.}}Y_1+i\beta/2,
    &&\bar{w}_6\xrightarrow{\text{a.c.}}Y_1-i\beta/2.
\end{align}
Similarly, the local operator entanglement entropy for the individual intervals
can be obtained by tracing out the other interval,
\begin{align}
S_A &= \lim_{n\rightarrow1}\frac{1}{1-n}
\nonumber
\\
&\log\left[\frac{
\langle \sigma_n(w_3,\bar{w}_3) 
\bar{\sigma}_n(w_4, \bar{w}_4) 
\mathcal{O}_n^\dagger(w_1, \bar{w}_1)
      \mathcal{O}_n(w_2, \bar{w}_2) \rangle_{\beta}}
      {\left(\langle \mathcal{O}^\dagger(w_1, \bar{w}_1) \mathcal{O}(w_2, \bar{w}_2)
      \rangle_{\beta}\right)^n} \right],
 \nonumber \\ 
S_B &= \lim_{n\rightarrow 1} \frac{1}{1-n}
\nonumber\\
&\log\left[\frac{
\langle 
\mathcal{O}_n^\dagger(w_1, \bar{w}_1)
\bar{\sigma}_n(w_6, \bar{w}_6)
\sigma_n(w_5, \bar{w}_5)
      \mathcal{O}_n(w_2, \bar{w}_2) \rangle_{\beta}}
      {\left(\langle \mathcal{O}^\dagger(w_1,\bar{w}_1) \mathcal{O}(w_2, \bar{w}_2)
      \rangle_{\beta}\right)^n}\right].
\end{align}
Performing the standard cylinder to plane map $z=e^{\frac{2\pi}{\beta}w}$, the
two-point function in the normalization is simply given by
\begin{equation}
    \langle \mathcal{O}^\dagger(w_1,\bar{w}_1) \mathcal{O}(w_2,\bar{w}_2)\rangle_\beta \nonumber
    \\
    =\left(\frac{2\pi}{\beta}\right)^{2h_\mathcal{O}+2\bar{h}_\mathcal{O}}\frac{(z_1z_2)^{h_\mathcal{O}} (\bar{z}_1\bar{z}_2)^{\bar{h}_\mathcal{O}}}{z_{12}^{2h_\mathcal{O}}\bar{z}_{12}^{2\bar{h}_\mathcal{O}}}.
\end{equation}

\subsection{Bipartite and tripartite information \label{section_BTOMI}}

The local operator R\'{e}nyi entropy for each individual set-up must be computed separately
as the monodromies of the conformal blocks are highly dependent on the spacetime
locations of the operators in the correlators.
The computations are thus rather repetitive so we leave them in Appendix \ref{HolographicCFTRenyiEntropyAppendix}.
In computing entropies, we use the four-point functions
given by the HHLL vacuum conformal block \cite{Fitzpatrick2015}:
\begin{align}
    \mathcal{F}^{\text{LL}}_{\text{HH}}(h_p|z)&=(1-z)^{h_L(\delta-1)}\left(\frac{1-(1-z)^\delta}{\delta}\right)^{h_p-2h_L} \nonumber
    \\
    &\times{}_{2}{F}_1(h_p,h_p,2h_p;1-(1-z)^\delta).
\end{align}
Here, 
in our case, 
the light operators are the twist operators,
$h_L = h_n$, 
$h_p = 0$ for the vacuum conformal block,
and 
\begin{align}
\delta=\bar{\delta} =\sqrt{1-\frac{24}{c}h_\mathcal{O}},
\end{align}
where we are considering scalar operators with $h_\mathcal{O} = \bar{h}_\mathcal{O}$.

The six-point function in \eqref{SAB} can be approximated by two four-point functions
using the OPE $\sigma_n(1)\times \bar{\sigma}_n(x,\bar{x}) \approx
\mathbb{I}\,+\,\mathcal{O}((1-x)^s)$ where $s\in \mathbb{Z}$.
Each four-point function will contain the local operators $\mathcal{O},\mathcal{O}^\dagger$ as well as the twist operators
$\sigma,\bar{\sigma}$.
If both twist operators in each four-point function correspond to the end-points of the same interval,
we say that the six-point function is computed in the disconnected channel.
On the other hand,
if both twist operator in each four-point function belongs to separate intervals,
we say that the six-point function is computed in the connected channel.
Holographically, the disconnected channel correponds to bulk geodesics starting and ending
on the endpoints of the same interval,
while the connected channel corresponds to geodesics beginning on an endpoint of
one interval and ending on an endpoint of the other interval.

Combining the various local operator entanglement entropies listed in Appendix
\ref{HolographicCFTRenyiEntropyAppendix},
we obtain the bipartite local operator mutual information.
In the following,
we list the mutual information (for the connected channel)
for various subsystem configurations.
Below, we set $\epsilon_1 = \epsilon_2$.
\begin{widetext}
\begin{description}
  \item[(i)]Symmetric intervals $X<X_2=Y_2<X_1=Y_1$:
\begin{align}
  I_{AB}^\text{con.}&=
  \frac{c}{3}\log\left[\sinh{\frac{\pi(X_1-X_2)}{\beta}}\sinh{\frac{\pi(Y_1-Y_2)}{\beta}}\right]
  +
    \frac{c}{6}
  \begin{cases}
    0,
    & t<X_2-X 
    \\ 
    \log\left[\frac{\tan\frac{\pi\bar{\delta}}{2}}{\bar{\delta}}e^{-\frac{2\pi}{\beta}(X+t-X_2)}\right],
    &
    X_2-X<t<X_1-X, \\
    -\log\left[\left(\frac{\sin{\pi \bar{\delta}}}{2\bar{\delta}}\right)^2e^{\frac{4\pi}{\beta}(X+t-\frac{X_1+X_2}{2})}\right],
    &
    t>X_1-X.
    \end{cases}.
\end{align}

\item[(ii)]Partially overlapping intervals I $X<X_2<Y_2<X_1<Y_1$:
\begin{align}\label{PartiallyOverlappingBOMI1}
  I_{AB}^\text{con.}
  &= 
    \frac{c}{3}\log\left[\frac{\sinh{\frac{\pi(X_1-X_2)}{\beta}}\sinh{\frac{\pi(Y_1-Y_2)}{\beta}}}{\cosh{\frac{\pi(X_1-Y_1)}{\beta}}\cosh{\frac{\pi(X_2-Y_2)}{\beta}}}\right]
    +
    \frac{c}{6}
    \begin{cases}
      0,
      &
    t<Y_2-X \\
    \log\left[
      \frac{\tan\frac{\pi\overline{\delta}}{2}}{2\overline{\delta}}
        e^{-\frac{2\pi}{\beta}(X+t-Y_2)}\right],
        &
    Y_2-X<t<X_1-X \\
    -
    \log\left[
      \frac
      {\sin{\pi\overline{\delta}}}
      {\overline{\delta}}
        e^{\frac{2\pi}{\beta}(X+t-Y_2)}
    \right],
    &
    X_1-X<t<Y_1-X\\
    -
    \log\left[
      \left(
        \frac {\sin{\pi\overline{\delta}}}{\overline{\delta}}
      \right)^2
      e^{\frac{4\pi}{\beta}(X+t-\frac{Y_1+Y_2}{2})}\right],
      &
    Y_1-X<t
    \end{cases}
\end{align}

\item[(iii)] Disjoint intervals $X<X_2<X_1<Y_2<Y_1$:
\begin{align}
  I_{AB}^\text{con.}
  &
    =
    \frac{c}{3}\log\left[\frac{\sinh{\frac{\pi(X_1-X_2)}{\beta}}\sinh{\frac{\pi(Y_1-Y_2)}{\beta}}}{\cosh{\frac{\pi(X_1-Y_1)}{\beta}}\cosh{\frac{\pi(X_2-Y_2)}{\beta}}}\right]
    +
    \frac{c}{6}
  \begin{cases}
    0,
    &
   t<X_1-X\\
   -\log
   \left[
   \frac{\sin{\frac{\pi\overline{\delta}}{2}}}{\overline{\delta}}
   \right]^2,
   &
   X_1-X<t<Y_2-X\\
   -\log\left[\frac{\sin{\pi\overline{\delta}}}{\overline{\delta}}e^{\frac{2\pi}{\beta}(X+t-Y_2)}\right],
   &
   Y_2-X<t<Y_1-X\\
   -\log\left[\left(\frac{\sin{\pi\overline{\delta}}}{\overline{\delta}}\right)^2
     e^{\frac{4\pi}{\beta}(X+t-\frac{Y_1+Y_2}{2})}
   \right],
   &
   t>Y_1-X
    \end{cases}
\end{align}

\item[(iv)]Partially overlapping intervals II $Y_2<X<X_2<Y_1<X_1$ with $X_2-X<Y_1-X<X_1-X<X-Y_2$:
\begin{align}\label{PartiallyOverlappingBOMI2}
  I_{AB}^\text{con.}
  &=
 \frac{c}{3}\log\left[\frac{\sinh{\frac{\pi(X_1-X_2)}{\beta}}\sinh{\frac{\pi(Y_1-Y_2)}{\beta}}}{\cosh{\frac{\pi(X_1-Y_1)}{\beta}}\cosh{\frac{\pi(X_2-Y_1)}{\beta}}}\right]
    +
    \frac{c}{6}
    \begin{cases}
      0,
      &
   0<t<X_2-X \\ 
   +\log\left[\frac{\sin{\frac{\pi\overline{\delta}}{2}}}{\overline{\delta}}\right]^2,
   &
   X_2-X<t<Y_1-X \\
   0,
   &
    Y_1-X<t<X_1-X  \\
    -\log\left[
      \frac{\sin{\pi\overline{\delta}}
      }{\overline{\delta}}
        e^{\frac{2\pi}{\beta}(X+t-X_1)}
    \right],
    &
    X_1-X<t
    \end{cases}
\end{align}

\item[(v)]Partially overlapping intervals III $Y_2<X<X_2<X_1<Y_1$ with $X_2-X<X_1-X<X-Y_2<Y_1-X$:
\begin{align}\label{PartiallyOverlappingBOMI3}
  I_{AB}^\text{con.}
  &=
    \frac{c}{3}\log\left[
    \frac{\sinh{\frac{\pi(X_1-X_2)} {\beta}}
    \sinh{\frac{\pi(Y_1-Y_2)}{\beta}}}
    {\cosh{\frac{\pi(X_1-Y_1)}{\beta}}
    \cosh{\frac{\pi(X_2-Y_2)}{\beta}}} \right]
    +
    \frac{c}{6}
    \begin{cases}
      0,
      &
      0<t<X_2-X\\
      +\log\left[\frac{\sin{\frac{\pi\overline{\delta}}{2}}}{\overline{\delta}}\right]^2,
      &
      X_2-X<t<X_1-X\\
      0,
      &
       X_1-X<t<Y_1-X\\
      -\log\left[\frac{\sin{\pi\overline{\delta}}}{\overline{\delta}}
            e^{\frac{2\pi}{\beta}(X+t-Y_1)}
      \right],
      &
       Y_1-X<t
    \end{cases}
\end{align}
\end{description}
\end{widetext}
Two comments are in order. (i) The bipartite local operator mutual information is constant until both intervals are within the light cone of the local operator.
(ii) We note that in these results,
the time-independent part of the bipartite local operator mutual information 
in the $\beta \rightarrow 0$ limit
is simply given by 
$\frac{2\pi c}{3\beta}l_{A\cap B}$, 
where $l_{A\cap B}$ is the length of the overlap of the two intervals $A$ and $B$.

\begin{figure*}
    \centering
    \includegraphics[width = .4\textwidth]{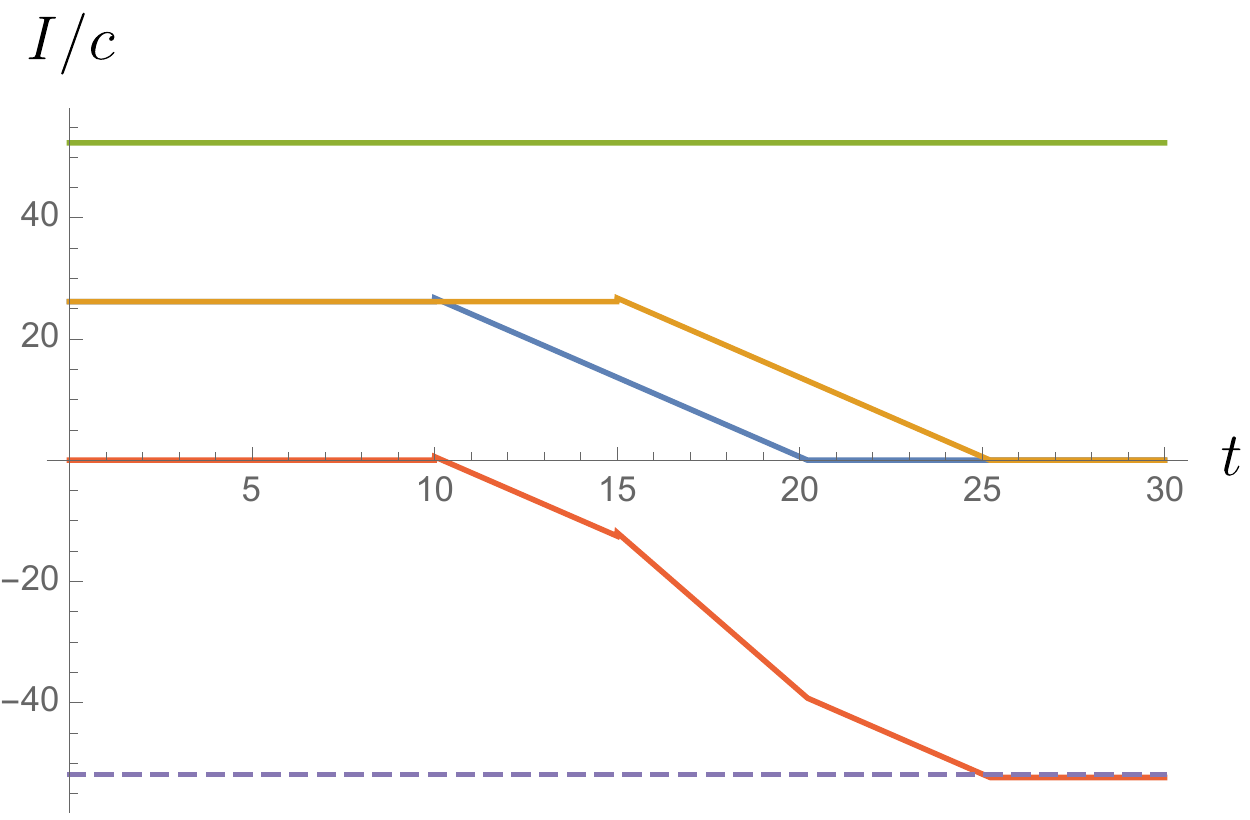}
    \includegraphics[width = .4\textwidth]{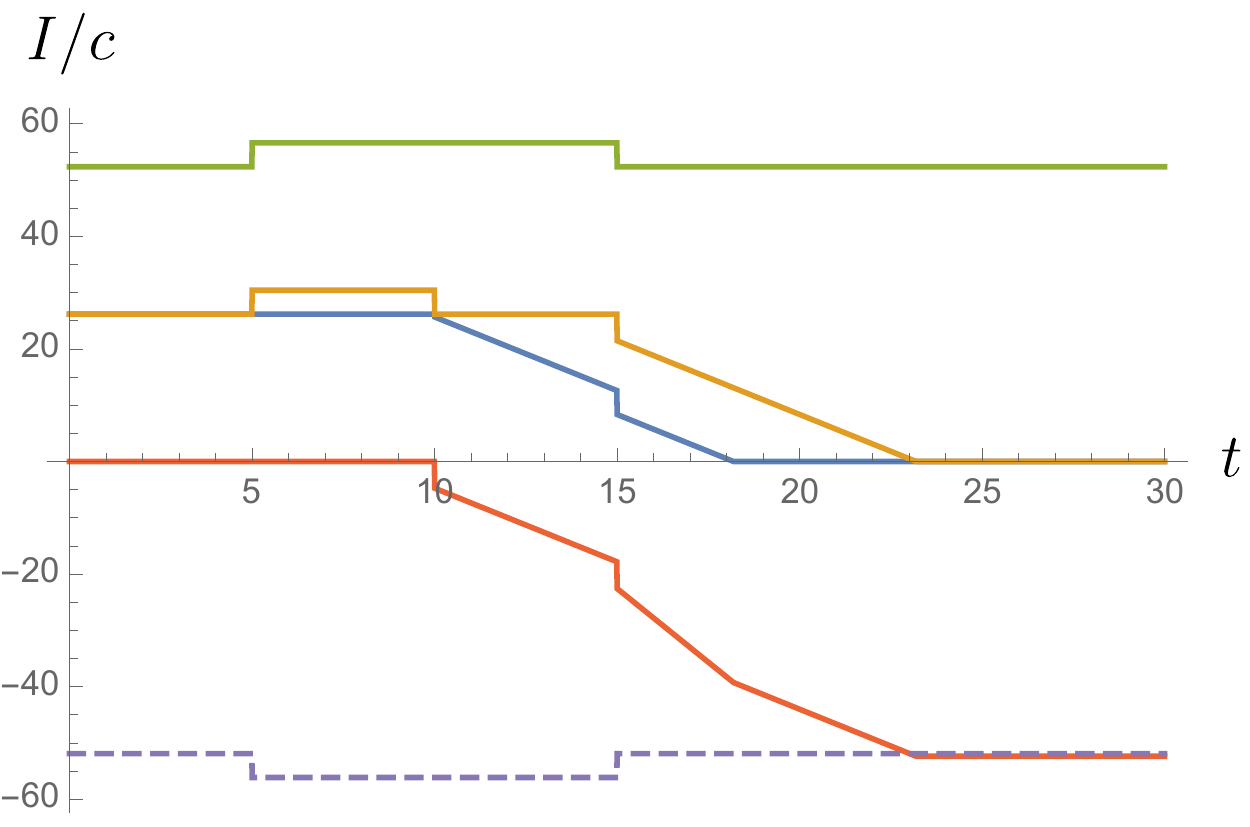}
    \caption{
      Bipartite local operator mutual information
      $I_{AB_1}$ (blue), $I_{AB_2}$ (orange) and $ I_{AB}$ (green),
      and
      tripartite local operator mutual information
      $I_3(A,B_1,B_2)$ (red) for large-$c$ CFTs.
      We are considering a light
      scalar ($\delta = \overline{\delta}= 0.99 $) on the left
      and a heavy
      scalar ($\delta = \overline{\delta}= 10 i$) on the right.
      The positions of the boundaries
      and local operators are $X_1=5,X_2=-5,Y_2=0$ and $X=-10$ respectively. The
      regulators are set to $\epsilon_1=\epsilon_2=0.1$.
      The dashed line at the bottom is given by $-2S_A^\text{reg.}$.
    }
    \label{CFT_holo_Is_light_heavy}
\end{figure*}

With the various bipartite local operator mutual information at hand,
we proceed at last to the tripartite local operator mutual information.
As a specific setup, we consider
subregions $A=[X_2,X_1]$, $B_1=[Y_2,Y_1]$ and $B_2=[Y_3,Y_2]$.
and insert the local operator to the left of subregion $A$
so that $Y_3<X<X_2<Y_2<X_1<Y_1$.
We also send $Y_1\rightarrow\infty$ and $Y_3\rightarrow -\infty$ so that $B_1$ and $B_2$ form a bipartition of the output.
The tripartite mutual information is obtained by taking the appropriate limits of
\eqref{PartiallyOverlappingBOMI1}, \eqref{PartiallyOverlappingBOMI2} and
\eqref{PartiallyOverlappingBOMI3}.
Those expressions are for the connected channel, and the actual bipartite local
operator mutual information is given by
$
I_{A\mathcal{R}} = \text{Max}\, (I_{A\mathcal{R}}^\text{con.},I_{A\mathcal{R}}^\text{discon.}) 
$
for $\mathcal{R}=B_1, B_2$ and $B=B_1 \cup B_2$.

We plot $I_{AB_1}, I_{AB_2},I_{AB}$ and $I(A,B_1,B_2)$ in Fig.~\ref{CFT_holo_Is_light_heavy}
both for light and heavy local operators.
The bipartite local operator mutual information for each semi-infinite interval
$I_{AB_1}$ and $I_{AB_2}$ vanish after a certain time while the bipartite local
operator mutual information $I_{AB}$
for $A$ and the entire output $B$ remains constant even at late times.
The tripartite local operator mutual information thus converges to $-I_{AB}$, 
\begin{equation}\label{HolographicTOMISaturation}
    \lim_{t\rightarrow\infty}I_3(A,B_1,B_2)=-\frac{2\pi c(X_1-X_2)}{12\epsilon_1}=-2S_A^\text{reg.}
\end{equation}
where $S_A^\text{reg.} \approx \frac{\pi c(X_1-X_2)}{12\epsilon_1}$  is the
regulated entanglement entropy for $A$
\cite{2018arXiv181200013N}.
This saturates the lower bound for $I_3$ as fast as is allowed by causality
just like the random unitary circuits in Section \ref{RandU_sec} and the tripartite unitary operator mutual information of holographic CFTs and random unitary circuits in Refs.~\cite{2018arXiv181200013N,2020JHEP...01..031K}.

Small discontinuities in the local operator mutual informations depend on the weight of the local operator. For a light operator, these discontinuities are small.
Here, 
the dependence of the bipartite and tripartite local operator mutual information
on the local operator comes
from their conformal dimension, 
$\delta = \bar{\delta} = \sqrt{1-\frac{24}{c}h_\mathcal{O}}$.
When the local operator is light, i.e.~$h_\mathcal{O}<\frac{c}{24}$,
$0<\delta<1$,
as it enters the expressions of bipartite local operator mutual information in the form of
$\frac{\text{Trig. Function}}{\delta}$,
its logarithm is much smaller than the kinematical factors that enter the
epxressions in terms of exponentials.
On the other hand, when the local operator is heavy,
$h_\mathcal{O}>\frac{c}{24}$,
$\delta$ is purely imaginary,
and hence the trigonometric functions become hyperbolic functions, and the
piecewise constant operator dependent terms give rise to more noticeable
discontinuities.
(Here, keep in mind that the bar does not refer to complex conjugation but instead refers to the anti-holomorphic conformal dimension.)
Since $I_{AB_1}^{\text{con.}}$ and $I_{AB_2}^{\text{con.}}$ decay to zero, and the late time value of $I_{AB}^{\text{con.}}$ and $S_A^{\text{reg.}}$ are operator independent, the tripartite local operator mutual information for heavy operators still satisfy the equality \eqref{HolographicTOMISaturation}. 

In Section \ref{RandU_sec}, we found precisely the same results as the large-$c$ calculations for \textit{light} operators i.e.~no discontinuities. This can be quantitatively verified once using the identification \eqref{cardy_bond_dim}.
It is interesting to consider if and how the discontinuities created by heavy operators can arise in the membrane theory. In Section \ref{RandU_sec}, we had neglected $O(1)$ contributions that can arise from the initial state. It is reasonable that by carefully accounting for these $O(1)$ contributions, one can find the discontinuities that depend on the specific operator.

\begin{figure}
    \centering
    \includegraphics[width = 0.2\textwidth]{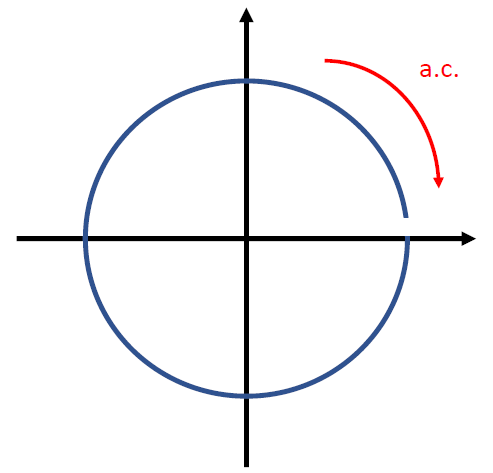}
    \caption{
    An example of the trajectory of a cross-ratio 
    on the complex plane
    during analytic continuation,
    relevant 
    for the calculations of the local operator entanglement.
    }
    \label{CrossRatioMonodromySchematicDrawing}
\end{figure}

Before concluding the section, let us comment on 
one of the key differences between 
the unitary operator entanglement and the local operator entanglement. 
During the analytic continuation,
the cross-ratios for the case of unitary operator entanglement are real and do not follow any non-trivial trajectories 
\cite{2018arXiv181200013N}.
On the other hand, the cross-ratios 
for local operator entanglement are complex and can encircle the branch point at the origin during analytic continuation as shown in figure \ref{CrossRatioMonodromySchematicDrawing}. 
The non-trivial time-dependent behaviour of the cross-ratios 
is a direct consequence of the insertion of local operators.
As a result, the dominant conformal block can acquire a monodromy,
which contributes to the late-time behavior.  
This behavior is essentially the same as 
the cross-ratios that appear in the computation of OTOCs
\cite{PhysRevLett.115.131603, 2016JHEP...08..129G, 2016PTEP.2016k3B06C}.
In some sense, 
one can think of the four-point functions in the computation of $S_{AB}$ as an OTOC with the operators $V$ and $W$ from \eqref{OTOC} being the twist field and local operator $\mathcal{O}$ respectively. 
The exponential decay of this OTOC (coming from the monodromy of the vacuum conformal block) 
manifests itself as  the linear decrease in bipartite local operator mutual information.

\subsection{Holographic description}
\label{holography_sec}

We conclude this section by
introducing the geometry that is holographically dual to \eqref{loc_op_state_eq}. 
It is the two-sided black hole with a massive object discussed in 
Refs.~\cite{2015JHEP...01..102C,2015JHEP...08..011C,2015JHEP...03..051R}.

In the holographic CFT, we study the time evolution of BOMI and TOMI on $\mathbb{R}^{1,1}$. 
Therefore, the gravity dual which we compute the operator entanglement
entropies is the geometry in the AdS-Schwarzschild patch 
\be \label{asymmetric}
ds^2=\frac{R^2}{z^2}\left[ -(1- M z^2)dt^2_{L,R}+\frac{dz^2}{1- M z^2}+dx^2\right],
\ee
where $\sqrt{M}={2\pi}/{\beta}$. 
In order to consider the gravitational dual to \eqref{loc_op_state_eq} 
(with $\epsilon_1=\epsilon_2=\epsilon$), 
we take the period to be 
$
\beta = 2(\epsilon_1+\epsilon_2)= 4\epsilon
$.
The Kruskal coordinates are related to  
AdS-Schwarzschild patch of two wedges of ${\it AdS}_3$ as:
\begin{align} \label{UVZT}
&
U=\pm \sqrt{\frac{1-\sqrt{M}z}{1+\sqrt{M}z}}e^{\sqrt{M}t_{L,R}},
\nonumber \\
&
 V=\mp \sqrt{\frac{1-\sqrt{M}z}{1+\sqrt{M}z}}e^{-\sqrt{M}t_{L,R}},  
 \nonumber \\
 &
 \pm R \frac{\sqrt{1-M z^2}}{\sqrt{M}z} \sinh{(\sqrt{M}t_{L,R})}=R \cdot \frac{U+V}{1+U V},
 \nonumber \\
 &
 \pm R \frac{\sqrt{1-M z^2}}{\sqrt{M}z} \cosh{(\sqrt{M}t_{L,R})}=R \cdot \frac{U-V}{1-U V}, 
 \nonumber \\
 &
 \frac{R}{\sqrt{M}z}\cosh{(\sqrt{M}x)}=R \cdot \frac{1-UV}{1+U V}\cosh{\psi},
   \nonumber \\
   &
 \frac{R}{\sqrt{M}z}\sinh{(\sqrt{M}x)}=R \cdot \frac{1-UV}{1+U V}\sinh{\psi}.
\end{align}
The resuling metric is
\be \label{krskl}
ds^2 = R^2 \frac{-4dVdU+(-1+UV)^2d \psi^2}{(1+UV)^2}
\ee 
where $U$ and $V$ are defined in the region $-1<UV<1$.
The conformal boundaries, where the two copies of the CFT live, is at $UV=-1$,
the horizons are at $UV=0$, and the singularities are at $UV=1$.
The regions which correspond to the left and right CFTs are defined by
\begin{align}
&\text{Left}: \left\{ 0 \le U, - 1 \le UV \le 0\right\},\nonumber \\
  &\text{Right}: \left\{U\le 0, -1 \le UV \le 0\right\}.
\end{align}

We now place a massive object located at
\begin{align}
    (z,x) = (\alpha,0),
    \quad
    \forall t_{L,R}
\end{align}
in the coordinate of \eqref{asymmetric}.
The geometry back-reacted by the massive object can be constructed by first considering 
the metric of AdS$_3$ black hole in the global coordinate,
\be \label{gbc}
ds^2=-(r^2+R^2-\mu)d\tau^2+\frac{R^2 dr^2}{r^2+R^2-\mu} +r^2 d \psi^2,
\ee 
where the black hole (of mass $m = \mu /(8G_N R^2)$) is located at the center of the cylinder.  
The parameter $\mu$ is related to the conformal dimension of local operator $\mathcal{O}$: 
\be 
\delta =\sqrt{1-\frac{\mu}{R^2}}=\sqrt{1-\frac{24h_{\mathcal{O}}}{c}}.
\ee
\begin{widetext}
This metric can then be mapped by the following boost and coordinate transformation
so that the resulting metric describes
the massive object at the origin of the coordinates in the AdS-Schwarzschild patch
\begin{align}
&\sqrt{R^2+r^2}\sin{\tau}= R \frac{e^{\Lambda_1}U+e^{-\Lambda_1}V}{1+UV}, \\
&\sqrt{R^2+r^2}\cos{\tau}=\frac{R \cosh{\Lambda_2}(1-UV)}{1+UV}
 \left(\cosh{\psi}-\tanh{\Lambda_2}\frac{e^{\Lambda_1}U-e^{-\Lambda_1}V}{1-UV}\right),\\
&r \sin{\psi}= R \frac{1-UV}{1+UV} \sinh{\psi}, \\
& r\cos{\psi}
=\frac{R \cosh{\Lambda_2}(1-UV)}{1+UV}
 \left(\frac{e^{\Lambda_1}U-e^{-\Lambda_1}V}{1-UV}-\tanh{\Lambda_2} \cosh{\psi}\right),
\end{align}
where $(U,V,\psi)$ 
is the Kruskal coordinate.
In terms of $(U,V)$ coordinate, $r$ is given by
\begin{align}
&r=\left|\frac{R(1-UV)\cosh{\Lambda_2}}{1+UV}\right|
\sqrt{\left(\frac{\sinh{\psi}}{\cosh{\Lambda_2}}\right)^2+\left(\frac{e^{\Lambda_1}U-e^{-\Lambda_1}V}{1-UV}-\tanh{\Lambda_2} \cosh{\psi}\right)^2}.
\end{align}
Since $(U,V ,\psi)$ are transformed to $(z, t_{L,R}, x)$ as in (\ref{UVZT}),
the boost parameters $\Lambda_1$ and $\Lambda_2 $ 
can be determined by requiring the massive object at $(z,t,x)=(\alpha, 0, 0)$ 
in AdS-Schwarzschild patch corresponds to $r=0$ in global coordinate:
\be
\Lambda_1=0, ~\tanh{\Lambda_2}= \sqrt{1-M \alpha^2}.
\ee

The above transformation gives the metric in Kruskal coordinates
that takes into account the back reaction from the massive object.
For our purpose of computing entanglement entropies, let us write
the global coordinate for the two wedges
$(r^L, \tau, \psi_L)$ and $(r^R, \tau, \psi_R)$ in terms of AdS-Schwarzschild coordinate:
\begin{align}
\label{rpkl}
  &
  r^p
  = \frac{R}{\alpha M z}\Big[M\alpha^2 \sinh^2{(\sqrt{M}x)}
    + \Big( \sqrt{1-M z^2}\cosh{(\sqrt{M}t_p)}
    \pm \sqrt{1-M\alpha^2}\cosh{(\sqrt{M}x)}\Big)^2\Big]^{1/2},
  \nonumber \\
  &
  \sqrt{R^2+(r^p)^2}\sin{\tau}
  = \mp \frac{R\sqrt{1-Mz^2}}{\sqrt{M}z}
  \sinh{(\sqrt{M}t_p)},
  \nonumber \\
  &
  \sqrt{R^2+(r^p)^2}\cos{\tau}
  =\frac{R}{\alpha M z}\cosh{(\sqrt{M}x)}
  \pm
  \frac{\sqrt{1-M\alpha^2}}
  {\alpha M}\cdot\frac{R\sqrt{1-Mz^2}}{z} \cosh{(\sqrt{M}t_p)},
  \nonumber \\
  &
  r^p \sin{\left(\psi_p\right)}
  =\frac{R}{\sqrt{M}z} \sinh{(\sqrt{M}x)},
  \nonumber \\
  & r^p \cos{\left(\psi_p\right)}
  =\mp \frac{R}{M\alpha z}\Big[\sqrt{1-Mz^2}\cosh{(\sqrt{M}t_p)}
  \pm \sqrt{1-M\alpha^2}\cosh{(\sqrt{M}x)}\Big],
\end{align}
for the left ($p=L$) and right boundaries ($p=R$), respectively.
The left and right regions in Kruskal coordinate corresponds to the geometries, which are back-reacted by the massive object and their asymptotic regions are AdS-Schwarzschild.   

\begin{figure*}
    \centering
    \includegraphics[width = .33\textwidth]{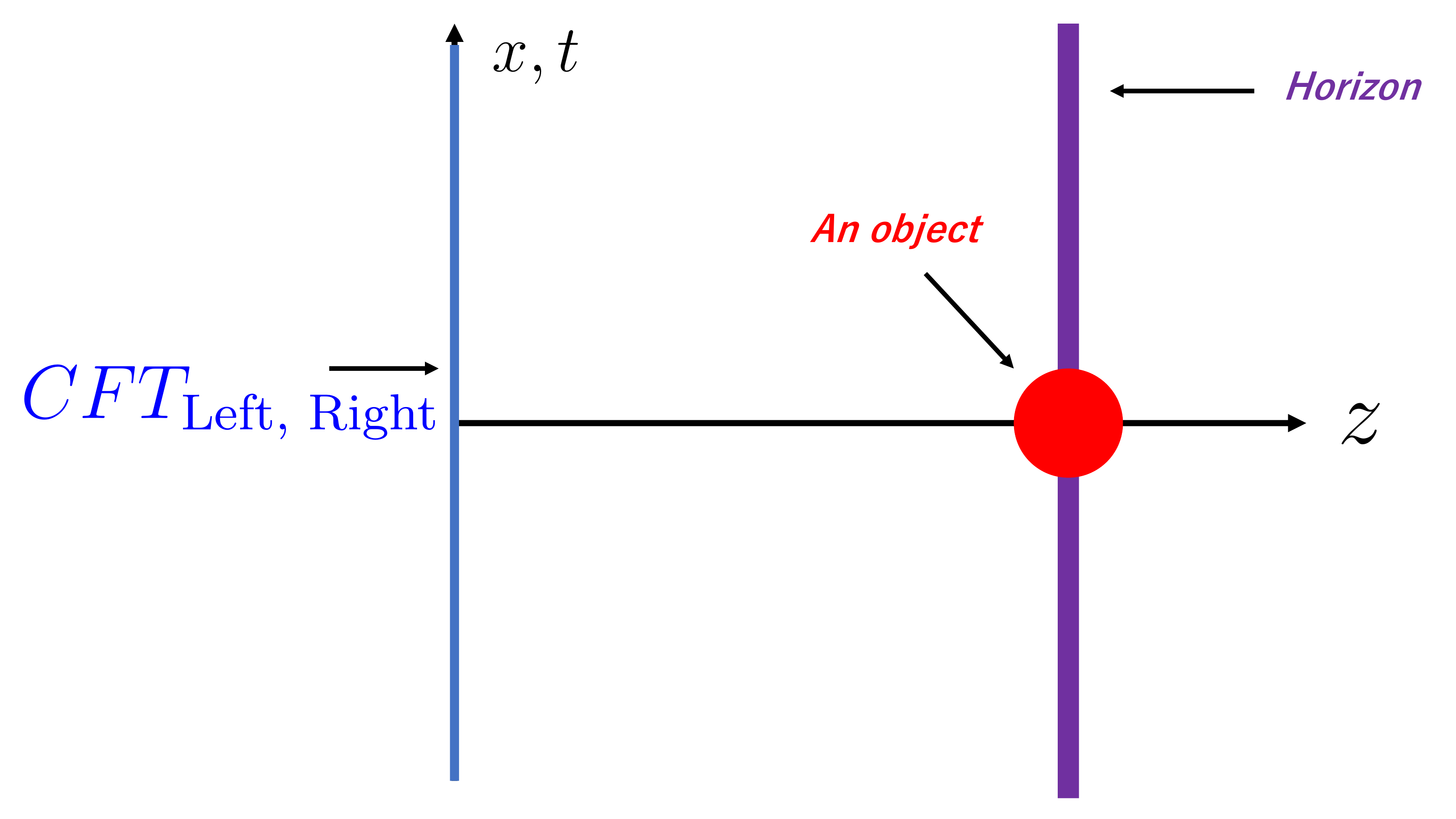}
    \qquad
    \includegraphics[width = .33\textwidth]{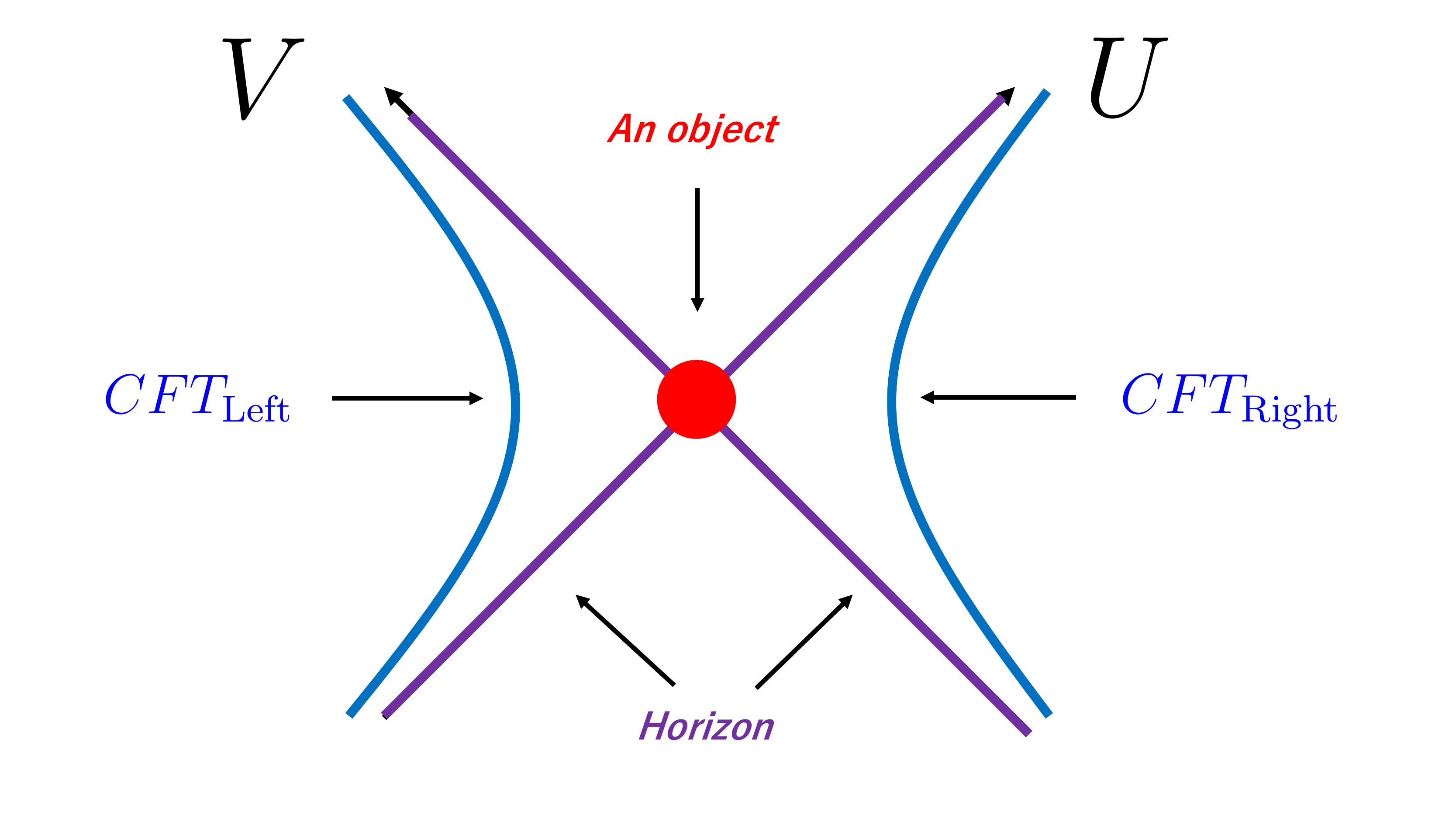}
    \caption{
      The location of the massive object
      in the AdS-Schwarzschild and Kruskal coordinate. 
      The left panel is for the AdS-Schwarzschild patch, and the right one for Kruskal coordinate.   
      \label{infalling}}
\end{figure*}

Now, in terms of global coordinates
the holographic entanglement entropy $S_A$ 
is given by \cite{2013JHEP...05..080N}
\be \label{EEformula}
S_A = \frac{c}{6} \log{\left[ \frac{2 r_{1}\cdot r_{2}}{R^2}\cdot\frac{\cos{\left(| \tau_1-\tau_2|\cdot \delta\right)}-\cos{\left(| \psi_1-\psi_2|\cdot\delta\right)}}{\delta^2}\right]},
\ee
where the boundaries of subsystem $A$ are at $(r_{2}, \tau_2, \psi_2 )$ and $(r_{1}, \tau_1, \psi_1 )$. 
Since the holographic entanglement entropy is diffeomorphism invariant,
we are able to compute holographic entanglement entropy in the two-sided black hole with the massive object by using (\ref{EEformula}).

As the final step, let us determine the parameter $\alpha$. 
The expectation value of energy momentum tensor for the state in
(\ref{dual_to_local_operator})
is equal to the three point function which is universal in $2d$ CFTs. 
This can be done by requiring the energy density in the gravity side should be
equal to that for the state in (\ref{dual_to_local_operator}) with $\epsilon_1=\epsilon_2=\epsilon$:
\be
\langle T^{L,R}_{00}\rangle
=\frac{\bra{\mathcal{O}(x,t)}T^{L,R}_{00}\ket{\mathcal{O}(x,t)}}{ \langle\mathcal{O}(x,t) |\mathcal{O}(x,t)   \rangle}\Big{|}_{\epsilon_1=\epsilon_2=\epsilon}.
\ee
This is related to 
its holographic counterpart $\langle T^{L,R}_{tt}\rangle_{\text{hol}}$ 
as 
$
\langle T^{L,R}_{tt}\rangle_{\text{hol}}=\langle T^{L,R}_{xx}\rangle_{\text{hol}}
=\frac{1}{2\pi} \cdot \langle T^{L,R}_{00} \rangle
$
\cite{1999CMaPh.208..413B, 2001CMaPh.217..595D}. 
Using this dictionary, $\alpha$ is determined as 
\begin{equation}
\alpha^2 M =1.
\end{equation}
Then, the massive object is pinned to the horizon of AdS-Schwarzschild coordinate
and the origin of Kruskal coordinate as in Fig.~\ref{infalling}.

In AdS-Schwarzschild coordinate, the subsystems $A$ and $B$ are defined as
$A=\{z,t_L,x|z=a,t_L=t, X_2<x<X_1\}$
and
$
B=\{z,t_R,x|z=a, t_R=t, Y_2<x<Y_1\}$,
respectively,
where $a \ll 1$ is the inverse UV cutoff. For $M \alpha^2=1$,
the small $a$ expansions of the variables in (\ref{EEformula}) for $A$, $B$ and
$A \cup B$ are given to leading order by 
\begin{align}
  \label{eqAB}
  A:
  &\left|\tau^L_1-\tau^L_2\right|
    =\arccos{\Bigg[\frac{\sinh^2{(\sqrt{M} t)}+D^L_1D^L_2}
    {\sqrt{
    N_{LL1} N_{LL2}}}
    \Bigg]},
    \qquad
  r_{i}^{L}=\frac{R}{a} \sqrt{M\cdot N_{LLi}},
  \nonumber \\
  &
    \left|\psi^L_1-\psi^L_2\right|
    =\arccos\Bigg[\frac{-\cosh{(\sqrt{M}(X_1-X_2))}}
    {\sqrt{
    N_{LL1} N_{LL2}}}
     +\frac{\cosh^2{(\sqrt{M}t)}+D^L_1D^L_2}{\sqrt{
    N_{LL1} N_{LL2}}}
    \Bigg],
  \nonumber \\
  B: 
  &\left|\tau^R_1-\tau^R_2\right|
    =\arccos\Bigg[\frac{\sinh^2{(\sqrt{M} t)}+D^R_1D^R_2}
    {\sqrt{
    N_{RR1} N_{RR2} }}
    \Bigg],
    \qquad
  r_{i}^{R}=\frac{R}{a} \sqrt{M \cdot  N_{RRi}},
  \nonumber \\
  &\left|\psi^R_1-\psi^R_2\right|
    =\arccos\Bigg[\frac{-\cosh{(\sqrt{M}(Y_1-Y_2))}}
    {\sqrt{
    N_{RR1} N_{RR2} }}
     +\frac{+\cosh^2{(\sqrt{M}t)}+D^R_1D^R_2}{\sqrt{
    N_{RR1} N_{RR2} }}
    \Bigg],
  \nonumber \\
  A\cup B:
  &\left|\tau^R_i-\tau^L_i\right|
    =\arccos\Bigg[\frac{-\sinh^2{(\sqrt{M} t)}+D^R_iD^L_i}
    {\sqrt{
    N_{RRi} N_{LLi} }}
    \Bigg],
  \nonumber \\
  &\left|\psi^R_i-\psi^L_i\right|
    =\arccos\Bigg[\frac{-\cosh{(\sqrt{M}(X_i-Y_i))}
    }
    {\sqrt{
    N_{RRi} N_{LLi} }}
    +\frac{-\cosh^2{(\sqrt{M}t)}
   +D^R_iD^L_i}{\sqrt{
    N_{RRi} N_{LLi} }}
    \Bigg],
\end{align}
where $i= 1, 2$.
Here, we assume $|X_2|<|X_1|$ and $|Y_2|<|Y_1|$,
and
we introduced
\begin{align}\label{eqABpara}
  &
  D^L_{i}=\cosh{(\sqrt{M}X_i)},
  \quad
  D^R_{i}=\cosh{(\sqrt{M}Y_i)},
  \nonumber \\
  &
  N_{pq1}
  = \sinh^2{(\sqrt{M}{t}_p)}+(D_1^{q})^2, 
  \quad
  N_{pq2}
  =
    \sinh^2{(\sqrt{M}{t}_p)}+(D_2^{q})^2,
\end{align}
for $p,q=L,R$.
The above equations
can be used to compute
the operator entanglement entropies in terms of Poincare coordinate.
By choosing the minimum one of (\ref{EEformula}) to be the operator
entanglement entropy,
we verify that 
the time evolution of holographic BOMI and TOMI match precisely with those in Section \ref{section_BTOMI}.
\end{widetext}

\section{Discussion}

In this work, we have studied a strong version of the butterfly effect in quantum many-body systems from an information theoretic perspective. We have found that local operators in chaotic theories entirely delocalize information, regardless of the details of the operator. In certain ``large-$N$'' theories such as holographic CFTs and random unitary circuits with large local Hilbert space dimension, we have found this delocalization process to occur at the fastest possible rate allowed by causality. In contrast, we have found that integrable theories are robust against these perturbations.

There are several interesting avenues for further study of local operator entanglement. These include higher-dimensional calculations which may be made possible through the holographic membrane theory \cite{2018PhRvD..98j6025M,2019arXiv191211024M}. Holographically, it should also be tractable to compute the entanglement wedge cross-section in the massive-particle geometry of Section \ref{holography_sec}. 
It will be important to understand if the reflected entropy remains parametrically larger than the mutual information as this is a novel phenomenon never seen for simpler quantum systems and is hinting at the fundamental role of multipartite entanglement.
Moving beyond holography and maximally chaotic systems, it would be fascinating to understand this notion of the butterfly effect in more generic quantum systems. In particular, it is important to understand the universal features in generic interacting RCFTs and irrational CFTs as has been previously done for OTOC, local quenches, global quenches, and unitary operator entanglement \cite{2014arXiv1403.0702H,2015JHEP...09..110A,2019JHEP...08..063K,2020JHEP...01..175K,2020arXiv200811266K,2019arXiv190706646K,2019arXiv190906790K,2020arXiv200105501K,2016PTEP.2016k3B06C,2016JHEP...08..129G}. Similarly, it is desirable to understand non-conformal theories that are not maximally scrambling such as spin chains and random unitary circuits with finite onsite Hilbert space dimension \cite{2019PhRvL.122y0603A,2019arXiv190907407B,2019arXiv190907410B,2020PhRvB.101i4304P,2020arXiv200413697B,op_ent_spinchain_draft,2019PhRvB..99q4205Z,2018arXiv180300089J,2018PhRvX...8b1014N,2019arXiv191212311Z} and interacting integrable systems that exhibit diffusion, a tractable example being the Rule 54 chain \cite{2019PhRvL.122y0603A, 2020arXiv200602788A}.

\acknowledgments

The authors would like to acknowledge insightful discussions with Yuya Kusuki, Mark Mezei, Xiaoliang Qi, and Tadashi Takayanagi.
This  work  was  supported  by  
the National Science Foundation under award number DMR-2001181,  
and by a Simons Investigator Grant
from the Simons Foundation 
(Award Number:  566116).
MN is supported by JSPS Grant-in-Aid for Scientific Research (Wakate) No.\ 19K14724, RIKEN iTHEMS Program, and the RIKEN Special Postdoctoral Researcher program.

{\it Note added:}
While the revision of this 
manuscript  was at the final stage,
a preprint appeared on arXiv
\cite{dong2021holographic}.
Our analysis 
in Appendix \ref{app_stat_mech}
is consistent with 
\cite{dong2021holographic}.
For further discussion and 
more detailed 
version of the analysis of 
Appendix \ref{app_stat_mech}
will be published elsewhere
\cite{unpublished}.

\appendix
\begin{widetext}

\section{Membrane theory for negativity and reflected entropy
-- Mapping to classical spin model}
\label{app_stat_mech}

In this appendix, we make progress in the derivation of
the membrane theory for logarithmic negativity and reflected entropy conjectured
in Refs.~\cite{2020JHEP...01..031K,2020arXiv200105501K} and used in the main text\footnote{We emphasize that this derivation is valid for the global quench and
  operator entanglement circuits.
  However, for the local operator entanglement considered in this paper,
  the future and past light cones are correlated which requires extra care.
  In Section \ref{RandU_sec}, we treated them as independent.
  Technically, one must average over the future and past light cones together.
  In this case, the resultant geometry is just the future light cone and the
  effective spins live in the symmetric group with $S_N\rightarrow S_{2N}$.
  In the large-$q$ limit, we wind up with identical results to the heuristics
  shown in this section,
  so, for simplicity, we omit the subtlety.
  We note that there also may be $O(1)$ effects from the operator choice similar to the initial state choice contribution in Ref.~\cite{2018arXiv180300089J}. 
}.
We do this using the formalism developed in
Refs.~\cite{2018PhRvX...8b1013V,2019PhRvB..99q4205Z} for R\'enyi entropies.
We will work in the $q\rightarrow \infty$ limit where significant simplifications may be made. It will be interesting to understand finite-$q$ effects in the future. The $q\rightarrow \infty$ limit is relevant to irrational CFTs where the effective bond dimension is determined by the Cardy density of states
\eqref{cardy_bond_dim},
$q = e^{\frac{\pi c}{3\beta}}.$

Progress can be made because of analytic formulas known for averaging over an arbitrary numbers of Haar random unitary matrices
\begin{align}
    \includegraphics[width = 6cm]{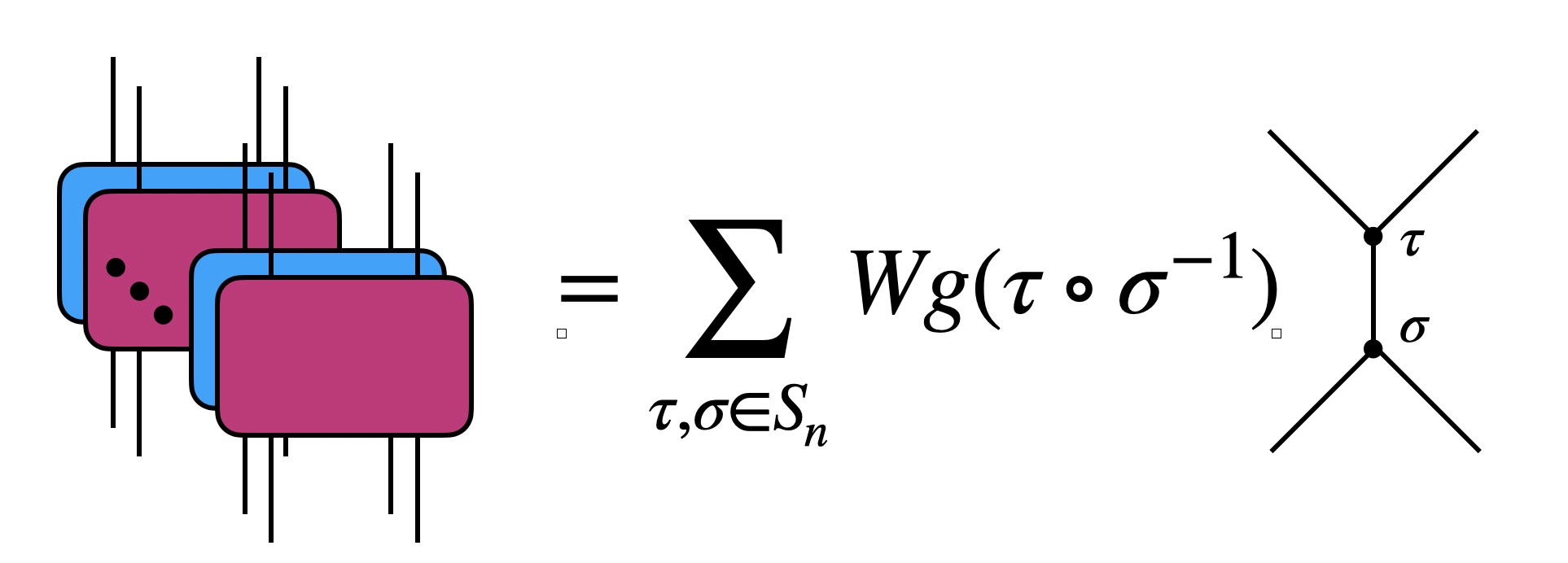}
\end{align}
where $\sigma$ and $\tau$ are elements of the permutation group $S_n$ and $n$ is the
number of unitaries (and duals) averaged over.
Once applying this independent averaging on every unitary matrix, we end up with
an effective hexagonal lattice of classical $S_n$ spins.
We are thus instructed to compute the partition function on this lattice.
It has been shown that the partition function simplifies by summing over the $\tau$ variables and we end up with a triangular lattice with \textit{positive} three-spin interactions involving the Weingarten function \cite{2018PhRvX...8b1014N}
\begin{align}
    \includegraphics[width = 6cm]{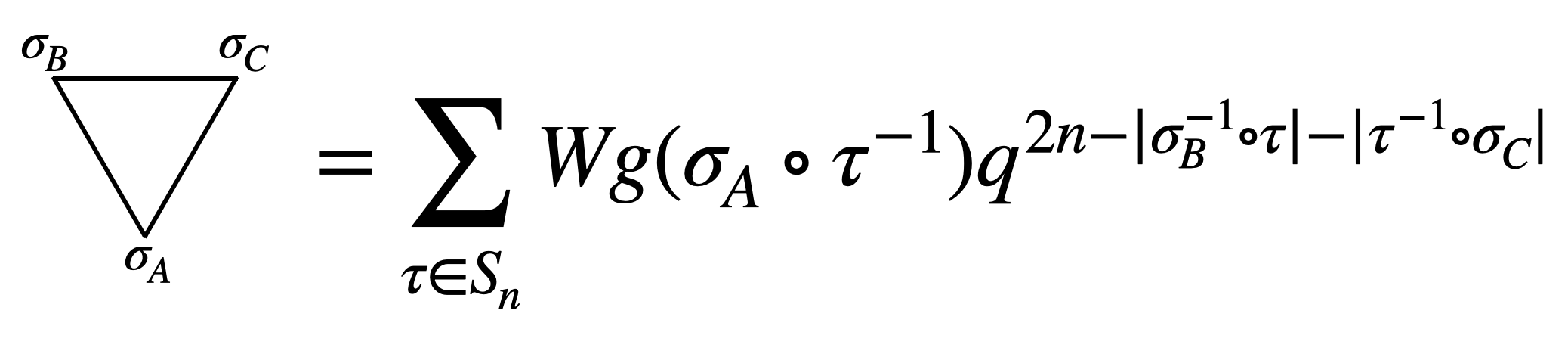}.
\end{align}
These interactions are still quite complicated but they simplify greatly in the $q\rightarrow \infty$ limit where they equal $q^{|\sigma_B^{-1}\circ \sigma_c|-n}$. $|\cdot|$ for a permutation element denotes the total number of cycles in that permutation.

Rather than considering spin configurations in the partition function, it is more convenient to consider $S_n$ domain wall configurations. 
With these simplifications, we can reduce the problem of computing the partition function to finding the dominant ``bulk saddle'' completely analogous to the story for conformal field theories in the large-$N$ limit \cite{2013JHEP...08..090L}, though, so far, the discussion has been quite general and we have not specified to the entanglement entropy. In order to apply this general framework to the specific quantities that we are interested in studying, we must apply appropriate boundary conditions.

\subsection{Mutual information}

\begin{figure}
    \centering
    \includegraphics[width = .325\textwidth]{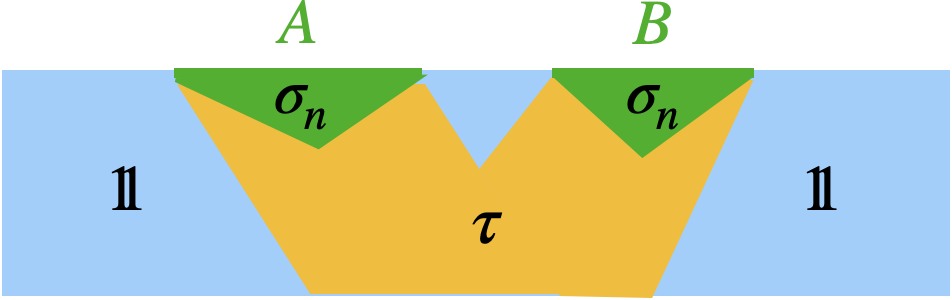}
    \includegraphics[width = .325\textwidth]{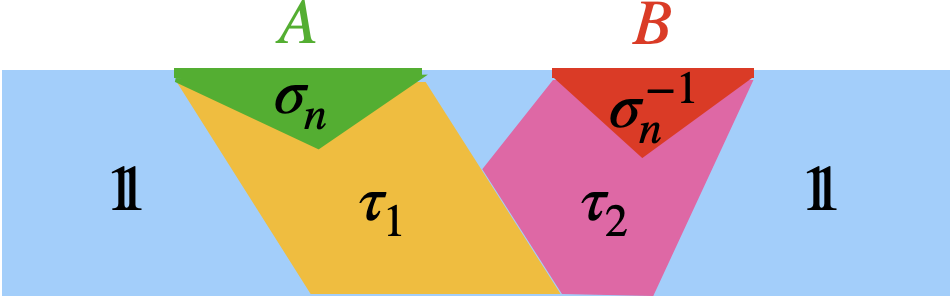}
    \includegraphics[width = .325\textwidth]{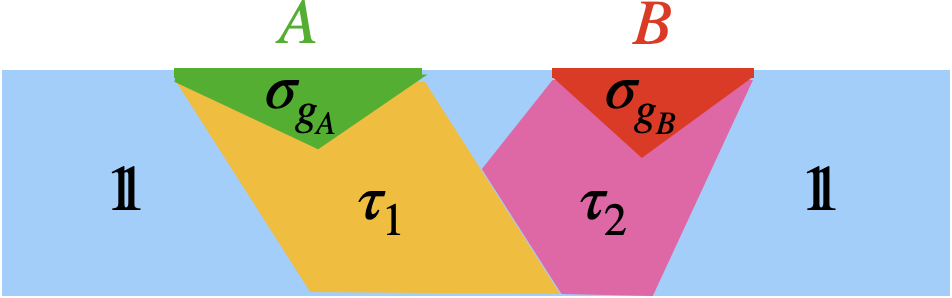}
    \caption{Cartoons of the Ising-like partition functions are shown in the ``ordered phase.'' From left to right, these are the partition functions for the R\'enyi entropy, negativity, and reflected entropy. The differences are present in the boundary conditions for the spins at the top boundary. The labels in the bulk represent the permutation elements of the spins in the dominant phases. All $\tau$'s are summed over. Focusing on the middle figure, $\gamma_{dis,A}$ ($\gamma_{dis,B}$) is the length of the domain wall between the green (red) sections on the yellow (pink) section, $\gamma_{con,A}$ ($\gamma_{con,B}$) is the length of the domain wall between the blue section on the yellow (pink) sections, and $E_W$ is the length of the domain wall between the yellow and pink sections.}
    \label{EW_linetension_configs}
\end{figure}

For the R\'enyi entropies, one must impose $\mathbb{Z}_n$ permutations on the boundaries within the regions of interest while imposing identity elements everywhere else (see Fig.~\ref{EW_linetension_configs}). This imposes the correct trace structure. The $\mathbb{Z}_n$ permutation is represented as
\begin{align}
    \sigma_n = (1,2,\dots,n)
\end{align}
where we are using cycle notation to label the elements of $S_n$. To leading order in $q$, the replica partition function is
\begin{align}
    \Tr \rho_{AB}^n = \frac{Z_n(A \cup B)}{Z_1(A\cup B)^n} = \sum_{\tau \in S_n} q^{\gamma_{con}(|\tau|-n) +\gamma_{dis}(|\sigma_n^{-1}\circ\tau|-n)}.
\end{align}
Here, $\gamma_{con}$ and $\gamma_{dis}$ represent the areas of the extremal surfaces in the circuit of different topologies. In the large $q$ limit, when $\gamma_{con} > \gamma_{dis}$, we need to maximize $|\tau|$. This is achieved when $\tau$ is the identity element, $e$, because $|e| = n$. Analogously, when $\gamma_{dis} > \gamma_{con}$, we need to maximize $|\sigma_n^{-1}\circ\tau|$ which occurs when $\tau = \sigma_n$. Thus, to leading order, the R\'enyi entropies are all equal
\begin{align}
    S_n(A) =  \min[\gamma_{con},\gamma_{dis}]
    \log q.
\end{align}
This may be described by a membrane theory because it is equivalent to finding the minimal membrane in the circuit with membrane tension $\log q$.


\subsection{Negativity}
For the logarithmic negativity, we must compute even powers of the partial transposed density matrix
\begin{align}
     \Tr\left( \rho_{AB}^{T_B}\right)^{n}=\frac{{Z}^{(PT)}_{n}}{{Z}_1^{n}} .
\end{align}
For this partition function, we must modify the boundary conditions of the $S_{n_e}$ spin model to incorporate the simultaneous cyclic and anticyclic gluing of the replica manifold. In cycle notation, the anti-cyclic permutation is
\begin{align}
    \sigma_n^{-1} = (n,n-1,\dots,1)
\end{align}
Spins in region $A$ have $\mathbb{Z}_{n_e}$ permutations while spins in $B$ have $\mathbb{Z}_{n_e}$ permutations in the opposite directions. All spins on the boundary outside of $A\cup B$ are set to the identity. Because regions $A$ and $B$ have different boundary conditions, there can now be a domain wall between them in the bulk. This complicates the computation by necessitating a double sum over the permutation group (see Fig.~\ref{EW_linetension_configs})
\begin{align}
    \frac{{Z}^{(PT)}_{n}}{{Z}_1^{n}} = \sum_{\tau_1\in S_n}\sum_{\tau_2 \in S_n}q^{\gamma_{A,dis}(|\sigma_n^{-1}\circ\tau_1|-n)+\gamma_{B,dis}(|\sigma_n\circ\tau_2|-n)+E_W(|\tau_1^{-1}\circ\tau_2|-n) + \gamma_{A,con}(|\tau_1| - n) + \gamma_{B,con}(|\tau_2| - n)}.
\end{align}

We will now perform a $0^{th}$ order analysis of this double sum and show various subtleties that arise that are not present in the case of R\'enyi entropies. The simplest regime is when $A$ and $B$ are sufficiently distant. In this case, we must maximize both $|\tau_1|$ and $|\tau_2|$ which means that we take both of them to be the identity. 
This contributes to the partition function as 
\begin{align}
    \frac{{Z}^{(PT)}_{n}}{{Z}_1^{n}} \supset q^{(1-n)(\gamma_{A,dis}+\gamma_{B,dis})}.
\end{align}

In the opposite regime where we expect correlations to be present between $A$ and $B$, ignoring the term with $E_W$, we must maximize both $|\sigma_n^{-1}\circ\tau_1|$ and $|\sigma_n\circ\tau_2|$ which means that we take $\tau_1 = \sigma_n$ and $\tau_2 = \sigma_n^{-1}$. This contributes to the partition function as 
\begin{align}
    \frac{{Z}^{(PT)}_{n}}{{Z}_1^{n}} \supset q^{E_W(2-n) + (1- n)(\gamma_{A,con}+ \gamma_{B,con})}.
\end{align}

While these two choices for permutations seem most natural, due to the term proportional to $E_W$, there exists another permutation (with degeneracy) that can become important. These permutation elements can be thought of as being halfway between $\sigma_n$ and $\sigma_n^{-1}$, leading to the following contribution
\begin{align}
    \frac{{Z}^{(PT)}_{n}}{{Z}_1^{n}} \supset q^{(1-n/2)(\gamma_{A,dis}+\gamma_{B,dis})- (n/2)(\gamma_{A,con} + \gamma_{B,con})}.
\end{align}

The three contributions described above are the most clear leading contributions to the partition function. However, we have not proven that other permutations are not also important or even that it is valid to take a single dominant contribution. If we make the assumption that these are the only important saddles and that it is sufficient to throw away the rest, we find as $q\rightarrow \infty$
%
\begin{align}
    \mathcal{E}^{(n)} &\equiv \log \frac{{Z}^{(PT)}_{n}}{{Z}_1^{n}} = - \min \Big[(n -1)\gamma_{con} +\left({n} -2\right)E_W
,(n -1)\gamma_{dis}
,
   \frac{n}{2}\gamma_{con}+\left(\frac{n}{2} -1\right)\gamma_{dis})\Big]\log q ,
\end{align}
where $\gamma_{con} \equiv \gamma_{con,A} + \gamma_{con,B}$ and $\gamma_{dis} \equiv \gamma_{dis,A} + \gamma_{dis,B}$. Here, we have approximated the logarithm of the sum of the three contributions as a ``min'' function due to the $q\rightarrow \infty$ limit. In reality, 
we are summing the many terms of the form $q^{\#}$.
For the disconnected regime ($\gamma_{con} \geq \gamma_{dis}$), the middle term will dominate for $n > 1$, leading to 
\begin{align}
    \mathcal{E}^{(n)} = (1-n)\gamma_{dis}\log q.
\end{align}
Naively taking the replica limit, this vanishes, implying that the logarithmic negativity is zero. This makes sense because it corresponds to the regime where the intervals are either sufficiently distant or we are at sufficiently late times when thermalization has occurred.

The connected regime ($\gamma_{dis} \geq \gamma_{con}$) is more subtle. First note that the first and third terms are identical when $n = 2$. Being linear in $n$, this means that one of the terms is minimal for all $n > 2$ and the other is minimal for $n < 2$ i.e.~there is a \textit{replica transition}. This is a novel phenomenon that we did not see for R\'enyi entropies. To determine which term is dominant in which regime, take $n = 1$ for simplicity. In this case, we are comparing the size of $E_W$ versus $(\gamma_{dis} - \gamma_{con})/2$. It is a simple geometric exercise to show that $E_W$ is always the greater of these two, thus the first term would appear to be dominant for $n < 2$ while the third term is dominant for $n > 2$. If we are to take the analytic continuation seriously, we would determine that the logarithmic negativity is given by
\begin{align}
    \mathcal{E} = E_W \log q.
    \label{LN_EW_eq}
\end{align}
This is precisely the membrane theory used in Refs.~\cite{2020JHEP...01..031K,2020arXiv200105501K}. Moreover, it shows how the entanglement wedge cross-section in the context of negativity can emerge outside of holographic conformal field theories. However, this analytic continuation was too naive.

We warn the reader that the above derivation was not rigorous both due to dropping all subleading terms in the double sum and in the analytic continuation to one. We have not proven that other terms are not important. We leave a more rigorous proof (or disproof) that involves computing the full negativity spectrum to future work \cite{unpublished}. One peculiarity in this result that must be resolved is the following: Recently, the logarithmic negativity has been analytically computed for arbitrary tripartitions of Haar random states \cite{2020arXiv201101277S}. Haar random states are expected to accurately describe the late time states after random local quantum circuit evolution, an intuitive idea that has been made precise in Ref.~\cite{2019arXiv190512053H}. However, if one compares the results from Ref.~\cite{2020arXiv201101277S} with those predicted by \eqref{LN_EW_eq} at late-times for a finite size system, one finds disagreements. Rather than $E_W$, Ref.~\cite{2020arXiv201101277S} found an answer that is more reminiscent of a R\'enyi mutual information. Resolving this tension is an important future direction\footnote{We note here that this tension may indeed have been resolved by the calculations of Ref.~\cite{dong2021holographic} which generalized the computation of the negativity spectrum of a single random tensor \cite{2020arXiv201101277S} to arbitrarily large random tensor networks. This suggests that \eqref{LN_EW_eq} is not correct in the regime we are probing.}.

\subsection{Reflected entropy}


We are able to run through similar analysis for the reflected entropy by imposing yet another boundary condition on the effective spin system. The partition function is indexed by two replica numbers
\begin{align}
    S_R^{(n)} = \frac{1}{n-1} \log \frac{\mathcal{Z}_{n,m}}{(\mathcal{Z}_{1,m})^n}.
\end{align}
The computation of these partition functions has an associated replica trick
(see Ref.~\cite{2019arXiv190500577D} for details).
For our purposes, we simply need to recall the cycles defined for the relevant twist operators so that we may set appropriate boundary conditions on the $S_{mn}$ spin model. In region $B$, the permutation element is
\begin{align}
    \sigma_{g_B} = \prod_{k = 1}^n (k, k+n, \dots, k+n(m-1)).
\end{align}
Each factor consists of $m-1$ elementary swaps, so in total, the domain wall between this element and the identity is composed of $n(m-1)$ elementary domain walls. Similarly, the permutation elements acting on region $A$ give $n(m-1)$ elementary domain walls, but the cycles are of a different form
\begin{align}
    \sigma_{g_A} &= \prod_{k = 1}^n (k, k+n, \dots, k+n(m/2-1),k+1+nm/2,\dots,k+1+n(m-1)).
\end{align}
The product of these permutation elements is
\begin{align}
    \sigma_{g^{\ }_B}\circ\sigma_{g_A}^{-1}&= (1,2,\dots n)(n(m/2+1),n(m/2+1)-1,\dots, nm/2+1),
\end{align}
which is composed of $2(n-1)$ elementary domain walls. The reason why we are concerned with this product is it may dominate the replica partition function in the sums over the permutation group.
The total replica partition function at large $q$ is
\begin{align}
    {{Z}_{n,m}} = \sum_{\tau_1\in S_{nm}}\sum_{\tau_2 \in S_{nm}}q^{\gamma_{A,dis}(|\sigma_{g_A}^{-1}\circ\tau_1|-nm)+\gamma_{B,dis}(|\sigma_{g_B}^{-1}\circ\tau_2|-nm)+E_W(|\tau_1^{-1}\circ\tau_2|-nm)+ \gamma_{A,con}(|\tau_1|-nm)  + \gamma_{B,con})|\tau_2|-nm) }.
\end{align}

We consider just two contributions that can dominate the sum. The first is most likely dominant when $A$ and $B$ are sufficiently separated or we are at late times. That is, $\tau_1,\tau_2 = e$, the identity. In this case, we have
\begin{align}
    {{Z}_{n,m}} \supset q^{(\gamma_{A,dis}+\gamma_{B,dis})n(1-m) }.
\end{align}
The other regime is when $\tau_1 = \sigma_{g_A}$ and $\tau_2 = \sigma_{g_B}$, in which case
\begin{align}
    {{Z}_{n,m}} \supset q^{2(1-m)E_W+ (\gamma_{A,con}  + \gamma_{B,con})n(1-m) }.
\end{align}
Assuming these are the only important saddles, we have
\begin{align}
    \log Z_{n,m} = -\log q \min\Big[ (\gamma_{A,dis}+\gamma_{B,dis})n(m-1),2(n-1)E_W+ (\gamma_{A,con}  + \gamma_{B,con})n(m-1)\Big]
\end{align}
The $n\rightarrow 1$ limit is
\begin{align}
    \log Z_{1,m}^n =-n\log q \min\Big[ (\gamma_{A,dis}+\gamma_{B,dis})(m-1),(\gamma_{A,con}  + \gamma_{B,con})(m-1)\Big]
\end{align}
When $\gamma_{con } > \gamma_{dis}$, we therefore have
\begin{align}
    \log \frac{Z_{n,m}}{Z_{1,m}^n} = 0.
\end{align}
Thus, the reflected entropy is zero in the disconnected regime. 
In the connected regime when $\gamma_{con} < \gamma_{dis}$ , we have
\begin{align}
    \log \frac{Z_{n,m}}{Z_{1,m}^n} =\max \left[ 2(1-n)E_W, n(1-m)(\gamma_{dis} -\gamma_{con}) \right].
\end{align}
The second term is undesirable so we wish to take $n \rightarrow 1$ before $m \rightarrow 1$. This was previously noted in Refs.~\cite{2019arXiv190906790K,2019arXiv190706646K} as necessary for picking the correct ``entanglement wedge.'' However, this order of limits subtlety is only present because we have taken $q\rightarrow \infty$. At finite $q$, there should be no ambiguity. A better understanding of the necessity of this order of limits deserves further attention. Taking the proper order of limits, the reflected entropy becomes
\begin{align}
    S_R = 2E_W
\end{align}
as advertised. Again, we stress that we have made serious assumptions about the dominating terms in the sums over the permutation group. A more thorough analysis of this issue is being pursued \cite{SR_RTN}.

\section{R\'enyi entropies of the Holographic CFTs}
\label{HolographicCFTRenyiEntropyAppendix}

In this appendix, we compute the local operator R\'{e}nyi entropy for the holographic CFTs discussed in section \ref{HolographicCFTs}.
Consider the local operator entanglement entropy for a single subregion
$\mathcal{R}=A$ or $B$ residing on either the first or second Hilbert space:
\begin{align}
  &S_\mathcal{R}^{(n)} = \frac{1}{1-n}
  \log\left[\frac{\langle\mathcal{O}_n^\dagger(w_1,\bar{w}_1) \mathcal{O}_n(w_2,\bar{w}_2)\sigma_n(w_k,\bar{w}_k)\bar{\sigma}_n(w_l,\bar{w}_l)\rangle_\beta}{\langle\mathcal{O}^\dagger(w_1,\bar{w}_1)\mathcal{O}(w_2,\bar{w}_2)\rangle_\beta}\right].
\end{align}
By conformal transformation
$
  \chi =\frac{(z-z_k)z_{21}}{(z-z_1)z_{2k}},
$
$S^{(n)}_{\mathcal{R}}$ can be expressed in terms of
correlation functions on the complex plane as
\begin{align}\label{SingleIntervalRenyiEntropy}
  &S_\mathcal{R}^{(n)}
    =\frac{1}{1-n}\log\Bigg[
    \left(\frac{2\pi}{\beta}\right)^{4h_n}\frac{|z_kz_l|^{2h_n}}{|z_{kl}|^{4h_n}} |\chi_l|^{4h_n}
    \lim_{\chi_1,\bar{\chi}_1\rightarrow\infty} \chi_1^{2nh_\mathcal{O}} \bar{\chi}_1^{2n\bar{h}_\mathcal{O}}\langle\mathcal{O}_n^\dagger(\chi_1,\bar{\chi}_1) \mathcal{O}_n(1)\bar{\sigma}_n(\chi_l,\bar{\chi}_l)\sigma_n(0)\rangle_\mathbb{C}
    \Bigg].
\end{align}
This expression is completely general. Let us now specialize to the case where subregion $\mathcal{R}$ is subregion $A$ or $B$.
As in the main text, we always take $\epsilon_1=\epsilon_2$
in this Appendix.

\subsection{$S_A$}
Setting $w_k = w_3$ and $w_l = w_4$,
the R\'{e}nyi entropy is
\begin{align}
  S_A^{(n)}
  &=\frac{1}{1-n}\log\Bigg[
    \left(\frac{\pi}{\beta}\right)^{4h_n}\frac{|\chi_l^A|^{4h_n}}{\left(\sinh{\frac{\pi(X_1-X_2)}{\beta}}\right)^{4h_n}}
     \lim_{\chi_1,\bar{\chi}_1\rightarrow\infty} \chi_1^{2nh_\mathcal{O}} \bar{\chi}_1^{2n\bar{h}_\mathcal{O}}\langle\mathcal{O}_n^\dagger(\chi_1,\bar{\chi}_1) \mathcal{O}_n(1)\bar{\sigma}_n(\chi_l^A,\bar{\chi}_l^A)\sigma_n(0)\rangle_\mathbb{C}
    \Bigg].
\end{align}
The holomorphic and anti-holomorphic cross-ratios are
\begin{align}
  \chi_l^A &\xrightarrow{\text{a.c.}} \frac{-i\sin{\frac{2\pi \epsilon_1}{\beta}}\sinh{\frac{\pi(X_1-X_2)}{\beta}}}{\sinh{\frac{\pi(X-t-X_2+i\epsilon_1)}{\beta}} \sinh{\frac{\pi(X-t-X_1-i\epsilon_1)}{\beta}}},
  \quad
  \bar{\chi}_l^A \xrightarrow{\text{a.c.}} \frac{i\sin{\frac{2\pi \epsilon_1}{\beta}}\sinh{\frac{\pi(X_1-X_2)}{\beta}}}{\sinh{\frac{\pi(X+t-X_2-i\epsilon_1)}{\beta}} \sinh{\frac{\pi(X+t-X_1+i\epsilon_1)}{\beta}}}    
\end{align}
\paragraph{Configuration 1: Local operator left of subregion.}
For our purposes,
the only relevant configuration for $S_A$ is
The cross ratios have the following behaviour as we send the regulators to zero.
\begin{align}
  &\chi_l^A  \xrightarrow{\epsilon_1=\epsilon_2\rightarrow 0}0,
  \quad
  \bar{\chi}_l^A\xrightarrow{\epsilon_1=\epsilon_2\rightarrow 0}
  \begin{cases}
    0,&t<|X_2-X|\quad\text{ or }\quad t>X_1-X\\
    2,&|X_2-X|<t<X_1-X
  \end{cases}
\end{align}
The four-point functions can be approximated by the HHLL vacuum conformal block \cite{Fitzpatrick2015}.
The R\'{e}nyi entropy becomes
\begin{align}
  S_A^{(n)}&=\frac{c}{6}\frac{n+1}{n}\log\left[\frac{\beta}{\pi}\sinh{\frac{\pi(X_1-X_2)}{\beta}}\right]
  +\frac{c}{12}\frac{n+1}{n}\log\left[\frac{1-(1-\chi_l^A)^\delta}{\delta\, \chi_l^A (1-\chi_l^A)^{\frac{\delta-1}{2}}}\frac{1-(1-\bar{\chi}_l^A)^{\bar{\delta}}}{\bar{\delta}\, \bar{\chi}_l^A (1-\bar{\chi}_l^A)^{\frac{\bar{\delta}-1}{2}}}    \right]
\end{align}
The term that depends on the cross-ratio takes the following simple forms in the two relevant limits
\begin{align}
  &\lim_{\chi\rightarrow 0}\frac{1-(1-\chi)^\delta}{\delta\,\chi(1-\chi)^{\frac{\delta-1}{2}}} = 1,
  \quad 
  \lim_{\chi\rightarrow 2}\frac{1-(1-\chi)^\delta}{\delta\,\chi(1-\chi)^{\frac{\delta-1}{2}}}=\frac{\sin{\frac{\pi \delta}{2}}}{\delta}
\end{align}
The von Neumann entropy for subsystem $A$ when $X<X_2$ is
\begin{align}\label{SAHolographicCFT}
  &S_A =
  \frac{c}{3}\log\left(\frac{\beta}{\pi}\sinh{\frac{\pi(X_1-X_2)}{\beta}}\right)
  +
  \frac{c}{6}
  \begin{cases}
    0
    & t<X_2-X\,\text{ or }\,t>X_1-X \\
    \log\frac{\sin{\frac{\pi\bar{\delta}}{2}}}{\bar{\delta}} &X_2-X<t<X_1-X
  \end{cases}
\end{align}

\subsection{$S_B$}
This corresponds to $w_k = w_5$ and $w_l = w_6$ in \eqref{SingleIntervalRenyiEntropy}. The holomorphic and anti-holomorphic cross-ratios are
\begin{align}
  &\chi_l^B \xrightarrow{\text{a.c.}} \frac{-i\sin{\frac{2\pi\epsilon_1}{\beta}} \sinh{\frac{\pi(Y_1-Y_2)}{\beta}}}{\cosh{\frac{\pi(X-t+i\epsilon_1-Y_1)}{\beta}}\cosh{\frac{\pi(X-t-i\epsilon_1-Y_2)}{\beta}}},
  \quad
\bar{\chi}_l^B \xrightarrow{\text{a.c.}} \frac{i\sin{\frac{2\pi\epsilon_1}{\beta}} \sinh{\frac{\pi(Y_1-Y_2)}{\beta}}}{\cosh{\frac{\pi(X+t-i\epsilon_1-Y_1)}{\beta}}\cosh{\frac{\pi(X+t+i\epsilon_1-Y_2)}{\beta}}}.
\end{align}

\paragraph{Configuration 1: Local operator left of subregion.}
$X<Y_2<Y_1$
The chiral and anti-chiral cross-ratios have the following limits
\begin{align}
  &\chi_l^B\xrightarrow{\epsilon_1=\epsilon_2\rightarrow 0}0,
  \quad
  \bar{\chi}_l^B\xrightarrow{\epsilon_1=\epsilon_2\rightarrow 0}
  \begin{cases}
    0& t<Y_2-X\,\text{ or }\,t>Y_1-X \\
    2& Y_2-X<t<Y_1-X
  \end{cases}
\end{align}
Repeating a computation similar to that for $S_A$, we find
\begin{align}
  S_B &=
  \frac{c}{3}\log\left(\frac{\beta}{\pi}\sinh{\frac{\pi(Y_1-Y_2)}{\beta}}\right)
  +
  \frac{c}{6}
  \begin{cases}
    0
    & t<Y_2-X\, \text{ or }\,t>Y_1-X \\
   \log\left(\frac{\sin{\frac{\pi \bar{\delta}}{2}}}{\bar{\delta}}\right)& Y_2-X<t<Y_1-X
  \end{cases}
\end{align}

\paragraph{Configuration 2: Local operator within subregion and closer to right boundary.}
When the local operator is within the subregion, i.e.~$Y_2<X<Y_1$, the cross-ratios have the following trajectories.
\begin{align}\label{SBLocalOperatorWithinSubregion}
  &\chi_l^B\xrightarrow{\epsilon_1=\epsilon_2\rightarrow 0}\begin{cases}
    2& t<X-Y_2\\
    0& t>X-Y_2
  \end{cases}
  ,
  \quad
  \bar{\chi}_l^B\xrightarrow{\epsilon_1=\epsilon_2\rightarrow 0}
  \begin{cases}
    2& t<Y_1-X \\
    0& t>Y_1-X
  \end{cases}.
\end{align}
Since both cross-ratios vanish at different times for generic set-ups, we have to consider whether the local operator is closer to the right or left boundary separately. First, consider the former, where $X-Y_2>Y_1-X$, with $\epsilon_1=\epsilon_2$.
The von Neumann entropy for holographic CFTs is
\begin{align}
  S_B&=
  \frac{c}{3}\log\left(\frac{\beta}{\pi}\sinh{\frac{\pi(Y_1-Y_2)}{\beta}}\right)
  +
  \frac{c}{6}
  \begin{cases}\label{SBCloserToRight}
    \log\left(\frac{\sin{\frac{\pi\delta}{2}} \sin{\frac{\pi\overline{\delta}}{2}}}{\delta\overline{\delta}}\right)& t<Y_1-X
    \\ 
    \log\left(\frac{\sin{\frac{\pi\delta}{2}} }{\delta}\right)& Y_1-X<t<X-Y_2\\
    0
    &t>X-Y_2
  \end{cases}
\end{align}

\paragraph{Configuration 3: Local operator within subregion and closer to left boundary.}
Finally, we consider the case where $Y_2<X<Y_1$ but $Y_1-X>X-Y_2$. Since the cross-ratios $\chi_l$ and $\overline{\chi}_l$ each depend on either $Y_2$ or $Y_1$ but not both, their individual trajectories are unchanged from the previous configuration and are given by \eqref{SBLocalOperatorWithinSubregion}.
For a holographic CFT, the von Neumann entropy is
\begin{align}
  &S_B=
  \frac{c}{3}\log\left(\frac{\beta}{\pi}\sinh{\frac{\pi(Y_1-Y_2)}{\beta}}\right)
  +
  \frac{c}{6}
  \begin{cases}\label{SBCloserToLeft}
    \log\left(\frac{\sin{\frac{\pi\delta}{2}} \sin{\frac{\pi\overline{\delta}}{2}}}{\delta\overline{\delta}}\right)& t<X-Y_2
    \\ 
   \log\left(\frac{\sin{\frac{\pi\overline{\delta}}{2}} }{\overline{\delta}}\right)& X-Y_2<t<Y_1-X
    \\
    0&t>Y_1-X
  \end{cases}
\end{align}
The expressions for the von Neumann entropy when the local operator is within the subregion \eqref{SBCloserToRight} and \eqref{SBCloserToLeft} are symmetrical, as they should be.

\subsection{$S_{AB}$}
The R\'{e}nyi entropy for two regions \eqref{SAB} can be written in terms of the complex plane coordinates. Let the coordinates of the twist operators be arbitrary for now so that we can specialize to either the connected or disconnected case later.
\begin{align}\label{SAB03}
  &S_{AB}^{(n)} 
  =\frac{1}{1-n}\log\Bigg[\left(\frac{2\pi}{\beta}\right)^{8h_n}z_{12}^{2nh_\mathcal{O}} \bar{z}_{12}^{2n\bar{h}_\mathcal{O}} |z_az_bz_cz_d|^{2h_n}
    \langle\mathcal{O}_n^\dagger(z_1,\bar{z}_1) \mathcal{O}_n(z_2,\bar{z}_2)\sigma_n(z_a,\bar{z}_a)\bar{\sigma}_n(z_b,\bar{z}_b)
    \sigma_n(z_c,\bar{z}_c)\bar{\sigma}_n(z_d,\bar{z}_d) \rangle_\mathbb{C}
    \Bigg].
\end{align}
By  the conformal transformation
$\zeta =\frac{(z-z_2)z_{a1}}{(z-z_1)z_{a2}}$,
introducing a resolution of identity,
the six-point correlation function can be written as
\begin{align}
  &\langle\mathcal{O}_n^\dagger(z_1,\bar{z}_1) \mathcal{O}_n(z_2,\bar{z}_2)\sigma_n(z_a,\bar{z}_a)\bar{\sigma}_n(z_b,\bar{z}_b)  
  \sigma_n(z_c,\bar{z}_c)\bar{\sigma}_n(z_d,\bar{z}_d) \rangle_\mathbb{C} 
  \nonumber \\
  &
  =
    \prod_{i=1,2,a,b,c,d}\left[\left(\frac{\partial\zeta}{\partial z}\right)_{z_i}^{h_i}\left(\frac{\partial\bar{\zeta}}{\partial \bar{z}}\right)_{\bar{z}_i}^{\bar{h}_i}\right]    
    \sum_p\sum_{l=|M|=|N|}
\sum_{\bar{l}=|\bar{M}|=|\bar{N}|} \left[G_p^{(l)}\right]_{MN}^{-1}\left[G_p^{(\bar{l})}\right]_{\bar{M}\bar{N}}^{-1}
\nonumber \\
&\quad \times
    \langle\mathcal{O}_n^\dagger(\infty) \sigma_n(1)\bar{\sigma}_n(\zeta_b,\bar{\zeta}_b)|\nu_{p,M},\nu_{p,\bar{M}}\rangle  
  \langle \nu_{p,N}\nu_{p,\bar{N}}| \sigma_n(\zeta_c,\bar{\zeta}_c)\bar{\sigma}_n(\zeta_d,\bar{\zeta}_d)\mathcal{O}_n(0) \rangle_\mathbb{C}
\end{align}
where $G$ is the Gram matrix and the sums are over primaries fields and
descendants. For the second four-point function,
we perform
the conformal transformation
$
  \eta = \frac{\zeta}{\zeta_c}
$.
Noting 
\begin{align}
  \prod_{i=1,2,a,b,c,d}
  \left(\frac{\partial\zeta}{\partial z}\right)_{z_i}^{h_i}
  &= \frac{\zeta_1^{2nh_\mathcal{O}}}{z_{21}^{2nh_\mathcal{O}}}
  \left(\frac{z_{12}z_{a1}}{z_{a2}}\right)^{4h_n}
  \frac{1}{(z_{a1}z_{b1}z_{c1}z_{d1})^{2h_n}},
  \quad
  \prod_{j=c,d}
  \left(\frac{\partial\eta}{\partial\zeta}\right)_{\zeta_j}^{h_j}
  =\left(\frac{z_{c1}z_{a2}}{z_{c2}z_{a1}}\right)^{2h_n},
\end{align}
the R\'{e}nyi entropy for subsystem $A\cup B$ is given by
\begin{align}\label{SAB04}
  S_{AB}^{(n)}
  &=
    \frac{1}{1-n}\log \bigg[\left(\frac{2\pi}{\beta}\right)^{8h_n}\frac{|z_az_bz_cz_d|^{2h_n}}{|z_{ab}z_{cd}|^{4h_n}}|1-\zeta_b|^{4h_n}|1-\eta_d|^{4h_n}
    \nonumber \\
  &\qquad 
    \times \sum_p\sum_{l=|M|=|N|}\sum_{\bar{l}=|\bar{M}|=|\bar{N}|}
    \left[G_p^{(l)}\right]_{MN}^{-1}
    \left[G_p^{(\bar{l})}\right]_{\bar{M}\bar{N}}^{-1}
    \nonumber \\
  &\qquad  
    \times \lim_{\zeta_1,\bar{\zeta}_1\rightarrow\infty}\zeta_1^{2nh_\mathcal{O}} 
    \bar{\zeta}_1^{2n\bar{h}_\mathcal{O}}
    \langle\mathcal{O}_n^\dagger(\zeta_1,\bar{\zeta}_1) \sigma_n(1)\bar{\sigma}_n(\zeta_b,\bar{\zeta}_b)|\nu_{p,M},\nu_{p,\bar{M}}\rangle 
    \langle \nu_{p,N}\nu_{p,\bar{N}}| \sigma_n(1)\bar{\sigma}_n(\eta_d,\bar{\eta}_d)\mathcal{O}_n(0) \rangle_\mathbb{C}\bigg]
\end{align}
The holomorphic cross ratios are
\begin{align}
  &\zeta_b = \frac{z_{1a}z_{b2}}{z_{1b}z_{a2}} = \frac{\sinh{\frac{\pi w_{1a}}{\beta}}\sinh\frac{\pi w_{b2}}{\beta}}{\sinh{\frac{\pi w_{1b}}{\beta}}\sinh\frac{\pi w_{a2}}{\beta}},
  \quad
            \eta_d = \frac{z_{1c}z_{d2}}{z_{1d}z_{c2}} =  \frac{\sinh{\frac{\pi w_{1c}}{\beta}}\sinh\frac{\pi w_{d2}}{\beta}}{\sinh{\frac{\pi w_{1d}}{\beta}}\sinh\frac{\pi w_{c2}}{\beta}}
\end{align}

The calculation up to this point is completely general as we have neither specified a theory nor a channel, nor have we performed any analytic continuation. Note that the expression is symmetric in terms of the operator coordinates.

\subsubsection{Connected Channel}
To obtain the connected channel, set
\begin{equation}
  w_a = w_3,\quad w_b = w_6,\quad w_c=w_5,\quad w_6=w_4.
\end{equation}
The cross-ratios are
\begin{align}\label{SABConnectedCrossRatios}
  \zeta^{\text{con}}_{b} &\xrightarrow{\text{a.c.}}-\frac{\sinh{\frac{\pi(X-t-X_1+i\epsilon_1)}{\beta}}\cosh{\frac{\pi(Y_1-X+t+i\epsilon_1)}{\beta}}}{\sinh{\frac{\pi(X_1-X+t+i\epsilon_1)}{\beta}}\cosh{\frac{\pi(X-t-Y_1+i\epsilon_1)}{\beta}}},
  \quad
  \bar{\zeta}^{\text{con}}_{b} \xrightarrow{\text{a.c.}}-\frac{\sinh{\frac{\pi(X+t-X_1-i\epsilon_1)}{\beta}}\cosh{\frac{\pi(Y_1-X-t-i\epsilon_1)}{\beta}}}{\sinh{\frac{\pi(X_1-X-t-i\epsilon_1)}{\beta}}\cosh{\frac{\pi(X+t-Y_1-i\epsilon_1)}{\beta}}},
  \nonumber\\
  \eta^{\text{con}}_{d} &\xrightarrow{\text{a.c.}}-\frac{\cosh{\frac{\pi(X-t-Y_1+i\epsilon_1)}{\beta}} \sinh{\frac{\pi(X_2-X+t+i\epsilon_1)}{\beta}}}{\sinh{\frac{\pi(X-t-X_2+i\epsilon_1)}{\beta}} \cosh{\frac{\pi(Y_2-X+t+i\epsilon_1)}{\beta}}},
  \quad
  \bar{\eta}^{\text{con}}_{d} \xrightarrow{\text{a.c.}}-\frac{\cosh{\frac{\pi(X+t-Y_2-i\epsilon_1)}{\beta}} \sinh{\frac{\pi(X_2-X-t-i\epsilon_1)}{\beta}}}{\sinh{\frac{\pi(X+t-X_2-i\epsilon_1)}{\beta}} \cosh{\frac{\pi(Y_2-X-t-i\epsilon_1)}{\beta}}}.
\end{align}

Let us compute $S_{AB}$ for various configurations.
\paragraph{Configuration 1: Symmetric intervals $X<X_2=Y_2<X_1=Y_1$}
Sending the regulators to zero, the cross-ratios go to
\begin{align}
  \zeta^{\text{con}}_{b}
  &\xrightarrow{\epsilon_1 = \epsilon_2 \rightarrow 0}1, 
  \quad
  \bar{\zeta}^{\text{con}}_{b}
  \xrightarrow{\epsilon_1 = \epsilon_2 \rightarrow 0}
                        \begin{cases}
                          1,&t<X_1-X\\
                          e^{2\pi i},& t>X_1-X
                        \end{cases} 
                        \\ \nonumber
  \eta^{\text{con}}_{d}
  &\xrightarrow{\epsilon_1 = \epsilon_2 \rightarrow 0}1, 
  \quad
  \bar{\eta}^{\text{con}}_{d} \xrightarrow{\epsilon_1 = \epsilon_2 \rightarrow 0}\begin{cases}
                          1,&t<X_2-X \\
                          e^{-2\pi i},&t>X_2-X
                        \end{cases}
\end{align}
At early time $t<|X_2-X|$, before any monodromy can be acquired, the twist fields have the following OPE 
\begin{align}
  &\sigma_n(1)\times \bar{\sigma}_n(
  \zeta^{\text{con}}_{b},
  \bar{\zeta}^{\text{con}}_{b}) 
  \approx \mathbb{I}\,+\,
  \mathcal{O}((1-\zeta^{\text{con}}_{b})^r) \nonumber \\ 
  &\sigma_n(1)\times \bar{\sigma}_n(\eta^{\text{con}}_{d},
  \bar{\eta}^{\text{con}}_{d}) 
  \approx 
  \mathbb{I}\,+\,\mathcal{O}((1-\eta^{\text{con}}_{d})^s) ,\qquad r,s\in \mathbb{Z}
\end{align}
This implies that the six-point conformal functions factorizes into two four-point conformal functions\footnote{This OPE simplification is not rigorous but is justified for large $c$ theories in the $n\rightarrow 1$ limit. It will generally be incorrect when $n\neq 1$. See e.g.~Ref.~\cite{2019arXiv190906790K} for an example where a different operator dominates this OPE.}
\begin{align}
  &\sum_p\sum_{l=|M|=|N|}\sum_{\bar{l}=|\bar{M}|=|\bar{N}|}
    \left[G_p^{(l)}\right]_{MN}^{-1}
    \left[G_p^{(\bar{l})}\right]_{\bar{M}\bar{N}}^{-1}
  \lim_{\zeta_1,\bar{\zeta}_1\rightarrow\infty}
  \zeta_1^{2nh_\mathcal{O}} \bar{\zeta}_1^{2n\bar{h}_\mathcal{O}}
  \nonumber \\
  &\quad 
  \times
    \langle\mathcal{O}_n^\dagger(\zeta_1,\bar{\zeta}_1) 
    \sigma_n(1)
    \bar{\sigma}_n(\zeta^{\text{con}}_{b},
    \bar{\zeta}^{\text{con}}_{b})
    |\nu_{p,M},\nu_{p,\bar{M}}\rangle    
    \langle \nu_{p,N}\nu_{p,\bar{N}}| \sigma_n(1)\bar{\sigma}_n(
    \eta^{\text{con}}_{d},
    \bar{\eta}^{\text{con}}_{d})
    \mathcal{O}_n(0) \rangle_\mathbb{C}
    \nonumber \\
  &= \lim_{\zeta_1,\bar{\zeta}_1\rightarrow\infty}\zeta_1^{2nh_\mathcal{O}} 
    \bar{\zeta}_1^{2n\bar{h}_\mathcal{O}}
    \langle\mathcal{O}_n^\dagger(\zeta_1,\bar{\zeta}_1)\mathcal{O}_n(1)   
     \bar{\sigma}_n(1-\zeta^{\text{con}}_{b},
    1-\bar{\zeta}^{\text{con}}_{b})
    \sigma_n(0)\rangle_\mathbb{C}
 \nonumber \\
  &\qquad \times 
  \lim_{\zeta_1,\bar{\zeta}_1\rightarrow\infty}\zeta_1^{2nh_\mathcal{O}} 
    \bar{\zeta}_1^{2n\bar{h}_\mathcal{O}}
    \langle\mathcal{O}_n^\dagger(\zeta_1,\bar{\zeta}_1)\mathcal{O}_n(1)    
   \bar{\sigma}_n(1-\eta^{\text{con}}_{d},
    1-\bar{\eta}^{\text{con}}_{d})
    \sigma_n(0)\rangle_\mathbb{C}.
\end{align}
Using the HHLL vacuum conformal block
for each four-point function,
\begin{align}
  S_{AB}^{\text{con.}(n)}
  &
    = \frac{1}{1-n}\log\bigg[\left(\frac{2\pi}{\beta}\right)^{8h_n}\frac{|z_3z_6z_5z_4|^{2h_n}}{|z_{36}z_{54}|^{4h_n}}
    \nonumber \\
    &\quad 
    \times
    \bigg[
    \frac{\delta\, 
    \zeta_{b}^{{\rm con}\frac{\delta-1}{2}} (1-\zeta^{\text{con}}_{b}) }
    {1-\zeta^{\text{con}\delta}_{b}}
    \frac{\bar{\delta}\, 
    \bar{\zeta}_{b}^{{\rm con} \frac{\bar{\bar{\delta}}-1}{2}}
    (1-\bar{\zeta}^{\text{con}}_{b}) }{1-\bar{\zeta}_{b}^{{\rm con} \bar{\delta}}}
    \frac{\delta\, 
    \eta_{d}^{{\rm con}\frac{\delta-1}{2}} 
    (1-\eta^{\text{con}}_{d}) }
    {1-\eta_{d}^{\text{con}\delta} }
    \frac{\bar{\delta}\, 
    \bar{\eta}_{d}^{{\rm con} 
    \frac{\bar{\bar{\delta}}-1}{2}}
    (1-\bar{\eta}^{\text{con}}_{d}) }{1-\bar{\eta}_{d}^{{\rm con}\bar{\delta}}}   \bigg]^{2h_n}        
    \bigg]. 
\end{align}
We take the $\epsilon_1=\epsilon_2\rightarrow 0$ limit before taking the $n\rightarrow 1$ limit to obtain the von Neumann entropy at various times
\begin{align}
  &S_{AB}^{\text{con.}} =
  \frac{c}{3}\log\left(\frac{\beta}{\pi}\right)^2
  +
  \frac{c}{6}
  \begin{cases}
    0
    & t<X_2-X \\
    \log\left[\frac{\sin{\pi \overline{\delta}}}{2\overline{\delta}}e^{\frac{2\pi}{\beta}(X+t-X_2)}\right],&X_2-X<t<X_1-X \\
   \log\left[\left(\frac{\sin{\pi\overline{\delta}}}{2\overline{\delta}}\right)^2e^{\frac{4\pi}{\beta}(X+t-\frac{X_1+X_2}{2})}\right], &t>X_1-X
  \end{cases}
\end{align}

\paragraph{Configuration 2: Partially overlapping intervals
    I $X<X_2<Y_2<X_1<Y_1$}
    
Consider the configuration of two partially overlapping intervals where neither the ends
of the intervals are aligned nor are the intervals disjoint.
 More precisely, let $X<X_2<Y_2<X_1<Y_1$ with $\epsilon_1=\epsilon_2$.
The chiral and anti-chiral cross ratios follow the following trajectories:
\begin{align}\label{SABAsymmetricCrossRatioTrajectories}
  \zeta^{\text{con.}}_{b} &\xrightarrow{\epsilon_1=\epsilon_2\rightarrow 0}1\quad\forall t,
  \quad
  \bar{\zeta}^{\text{con}}_{b}
  \xrightarrow{\epsilon_1=\epsilon_2\rightarrow 0}\begin{cases}
    1,&t<X_1-X \\ 
    -1,&X_1-X<t<Y_1-X \\
    e^{2\pi i},& t>Y_1-X
  \end{cases} \\ \nonumber
  \eta^{\text{con}}_{d} &\xrightarrow{\epsilon_1=\epsilon_2\rightarrow 0}1\quad\forall t,
  \quad
  \bar{\eta}^{\text{con}}_{d}
  \xrightarrow{\epsilon_1=\epsilon_2\rightarrow 0}\begin{cases}
    1,&t<X_2-X \\ 
    -1,&X_2-X<t<Y_2-X \\
    e^{-2\pi i},& t>Y_2-X
  \end{cases}
\end{align}
The six-point function factorizes as before.
Again, we use the HHLL vacuum conformal block
to obtain the von Neumann entropy.
\begin{align}
  &S_{AB}^{\text{con.}}=
                        \frac{c}{3}\log\left[\left(\frac{\beta}{\pi}\right)^2\cosh{\frac{\pi(X_1-Y_1)}{\beta}}\cosh{\frac{\pi(X_2-Y_2)}{\beta}} \right]
                        +
                        \frac{c}{6}
                        \begin{cases}
                         0,
                         &
                          t<X_2-X \\                         \log\frac{\sin{\frac{\pi\bar{\delta}}{2}}}{\overline{\delta}},
                          &
                          X_2-X<t<Y_2-X \\
                        \log\left(\frac{\sin{\pi\overline{\delta}}}{\overline{\delta}}e^{\frac{2\pi}{\beta}(X+t-Y_2)}\right),
                        &
                          Y_2-X<t<X_1-X \\                       \log\left[\frac{\sin{\pi\overline{\delta}}\sin{\frac{\pi\overline{\delta}}{2}}}{\overline{\delta}^2}e^{\frac{2\pi}{\beta}(X+t-Y_2)}\right],
                          &
                          X_1-X<t<Y_1-X\\ 
 \log\left[\left(\frac{\sin{\pi\overline{\delta}}}{\overline{\delta}}\right)^2e^{\frac{2\pi}{\beta}(2X+2t-Y_1-Y_2)}\right],
 &
                          Y_1-X<t
                        \end{cases}.
\end{align}

\paragraph{Configuration 3: Disjoint intervals $X<X_2<X_1<Y_2<Y_1$}
The holomorphic and anti-holomorphic cross ratios follow the same trajectory
under analytic continuation as in the case of partially overlapping intervals,
\eqref{SABAsymmetricCrossRatioTrajectories}.
Taking the $\epsilon_1=\epsilon_2 \rightarrow 0$ limit before taking the $n\rightarrow 1$ limit gives
\begin{align}
  &S_{AB}^{\text{con.}}
    =
    \frac{c}{3}\log\left[\left(\frac{\beta}{\pi}\right)^2\cosh{\frac{\pi(X_1-Y_1)}{\beta}}\cosh{\frac{\pi(X_2-Y_2)}{\beta}} \right]
    +
    \frac{c}{6}
    \begin{cases}
     0,
     &
      t<X_2-X\\
      \log
      \left(
        \frac{\sin{\frac{\pi\overline{\delta}}{2}}}{\overline{\delta}} \right)^2,
        &
      X_2-X<t<Y_2-X\\
      \log\left[\frac{\sin{\frac{\pi\overline{\delta}}{2}}\sin{\pi\overline{\delta}} e^{\frac{2\pi}{\beta}(X+t-Y_2)}}{\overline{\delta}^2} \right],
      &
      Y_2-X<t<Y_1-X \\
      \log\left[\left(\frac{\sin{\pi\overline{\delta}}}{\overline{\delta}}\right)^2e^{\frac{2\pi}{\beta}(2X+2t-Y_1-Y_2)}\right],
      &
      Y_1-X<t
    \end{cases}
\end{align}

\paragraph{Configuration 4: Partially Overlapping intervals II}
Consider again the situation where the intervals have a non-trivial intersection but the local operator is now contained within subregion $B$. More precisely, let $Y_2<X<X_2$ and $Y_1,X_1>X$. The cross-ratios \eqref{SABConnectedCrossRatios} have the following limits.
\begin{align}
\zeta^{\text{con}}_{b}
&\xrightarrow{\epsilon_1=\epsilon_2
\rightarrow 0} 1,
\quad
\overline{\zeta}^{\text{con}}_{b} \xrightarrow{\epsilon_1=\epsilon_2
\rightarrow 0}\begin{cases}
1,
&t<\text{Min}\{X_1-X,Y_1-X\} \\
-1,
&\text{Min}\{X_1-X,Y_1-X\}<t <\text{Max}\{X_1-X,Y_1-X\} \\
e^{2\pi i},& t>\text{Max}\{X_1-X,Y_1-X\}
\end{cases} 
\\
\eta^{\text{con}}_{d}
  &\xrightarrow{\epsilon_1=\epsilon_2
                        \rightarrow 0}\begin{cases}
                        -1,&t<X-Y_2 \\ 
                        1,&t>X-Y_2
                      \end{cases},
                      \quad
                      \overline{\eta}^{\text{con}}_{d}
                      \xrightarrow{\epsilon_1=\epsilon_2
                            \rightarrow 0}\begin{cases}
                            -1,&t<X_2-X \\
                            1,&t>X_2-X
                          \end{cases}
\end{align}
Since $\zeta_b\rightarrow 1$, the six-point function factorizes as before.
Let us further restrict ourselves to the case where $X-Y_2>X_1-X>Y_1-X>X_2-X$ with $\epsilon_1=\epsilon_2$ as usual. The vacuum conformal block gives
\begin{align}\label{SABHolographicCFTCloserToRight}
  &S_{AB}^{\text{con.}}
    =
    \frac{c}{3}\log\left[\left(\frac{\beta}{\pi}\right)^2\cosh{\frac{\pi(X_1-Y_1)}{\beta}}\cosh{\frac{\pi(X_2-Y_2)}{\beta}}\right]
    +
    \frac{c}{6}
    \begin{cases}
      \log\left(\frac{\sin{\frac{\pi\delta}{2}}\sin{\frac{\pi\overline{\delta}}{2}}}{\delta\overline{\delta}}\right),
      &
      0<t<X_2-X \\
     \log\left(\frac{\sin{\frac{\pi\delta}{2}}}{\delta}\right),
     &
      X_2-X<t<Y_1-X \\
    \log\left(\frac{\sin{\frac{\pi\delta}{2}}\sin{\frac{\pi\overline{\delta}}{2}}}{\delta\overline{\delta}}\right),
    &
      Y_1-X<t<X_1-X \\ 
      \log\left(\frac{\sin{\frac{\pi\delta}{2}}\sin{(\pi\overline{\delta})}e^{\frac{2\pi}{\beta}(X+t-X_1)}}{\delta\overline{\delta}}\right),
      &
      X_1-X<t<X-Y_2 \\
 \log\left(\frac{\sin{(\pi\overline{\delta})}e^{\frac{2\pi}{\beta}(X+t-X_1)}}{\overline{\delta}}\right),
 &
      X-Y_2<t 
    \end{cases}
\end{align}

\paragraph{Configuration 5: Partially Overlapping intervals III $Y_2<X<X_2<X_1<Y_1$ and $X_2-X<X_1-X<X-Y_2<Y_1-X$}
\begin{align}\label{SABHolographicCFTCloserToLeft}
  &S_{AB}^{\text{con.}}
    =
    \frac{c}{3}\log\left[\left(\frac{\beta}{\pi}\right)^2\cosh{\frac{\pi(X_1-Y_1)}{\beta}}\cosh{\frac{\pi(X_2-Y_2)}{\beta}}\right]
    +
    \frac{c}{6}
    \begin{cases}
      \log\left(\frac{\sin{\frac{\pi\delta}{2}}\sin{\frac{\pi\overline{\delta}}{2}}}{\delta\overline{\delta}}\right),
      &
      0<t<X_2-X\\
      \log\left(\frac{\sin{\frac{\pi\delta}{2}}}{\delta}\right),
      &
      X_2-X<t<X_1-X\\
      \log\left(\frac{\sin{\frac{\pi\delta}{2}}\sin{\frac{\pi\overline{\delta}}{2}}}{\delta\overline{\delta}}\right),
      &
      X_1-X<t<X-Y_2\\
      \log\left(\frac{\sin{\frac{\pi\overline{\delta}}{2}}}{\overline{\delta}}\right),
      &
       X-Y_2<t<Y_1-X\\
     \log\left(\frac{\sin{\pi\overline{\delta}}e^{\frac{2\pi}{\beta}(X+t-Y_1)}}{\overline{\delta}}\right),
     &
       t>Y_1-X
    \end{cases}
\end{align}

\subsubsection{Disconnected Channel}
Let us now consider the bipartite local operator mutual information for $A\cup B$ in the disconnected channel. This corresponds to geodesics beginning and ending on the same interval for holographic theories.
The R\'{e}nyi entropy is given by \eqref{SAB04} with $w_a=w_3,w_b=w_4,w_c=w_5,w_d=w_6$. The chiral and anti-chiral cross-ratios are
\begin{align}
  \zeta^{\text{discon}}_{b}
  &= \frac{z_{13}z_{42}}{z_{14}z_{32}} 
  \xrightarrow{\text{a.c.}} \frac{\sinh{\frac{\pi(X-t+i\epsilon_1-X_1)}{\beta}} \sinh{\frac{\pi(X_2-X+t+i\epsilon_1)}{\beta}}}{\sinh{\frac{\pi(X-t+i\epsilon_1-X_2)}{\beta}} \sinh{\frac{\pi(X_1-X+t+i\epsilon_1)}{\beta}}},
    \nonumber \\
  \bar{\zeta}^{\text{discon}}_{b}
  &= \frac{\bar{z}_{13}\bar{z}_{42}}{\bar{z}_{14}\bar{z}_{32}} 
  \xrightarrow{\text{a.c.}} \frac{\sinh{\frac{\pi(X+t-i\epsilon_1-X_1)}{\beta}} \sinh{\frac{\pi(X_2-X-t-i\epsilon_1)}{\beta}}}{\sinh{\frac{\pi(X+t-i\epsilon_1-X_2)}{\beta}} \sinh{\frac{\pi(X_1-X-t-i\epsilon_1)}{\beta}}},
    \nonumber \\
  \eta^{\text{discon}}_{d}
  &=\frac{z_{15}z_{62}}{z_{16}z_{52}}
  \xrightarrow{\text{a.c.}}\frac{\cosh{\frac{\pi(X-t+i\epsilon_1-Y_2)}{\beta}} \cosh{\frac{\pi(Y_1-X+t+i\epsilon_1)}{\beta}}}{\cosh{\frac{\pi(X-t+i\epsilon_1-Y_1)}{\beta}}\cosh{\frac{\pi(Y_2-X+t+i\epsilon_1)}{\beta}}},
    \nonumber \\
  \bar{\eta}^{\text{discon}}_{d}
  &=\frac{\bar{z}_{15}\bar{z}_{62}}{\bar{z}_{16}\bar{z}_{52}}
  \xrightarrow{\text{a.c.}}\frac{\cosh{\frac{\pi(X+t-i\epsilon_1-Y_2)}{\beta}} \cosh{\frac{\pi(Y_1-X-t-i\epsilon_1)}{\beta}}}{\cosh{\frac{\pi(X+t-i\epsilon_1-Y_1)}{\beta}}\cosh{\frac{\pi(Y_2-X-t-i\epsilon_1)}{\beta}}}.
\end{align}
Since $X<X_2$, $\zeta^{\text{discon}}_{b}\xrightarrow{\epsilon_1=\epsilon_2\rightarrow 0}1$ for all time, the six-point function factorizes
\begin{align}\label{SAB05}
  S_{AB}^{(n)}
  &=
    \lim_{\zeta_1,\bar{\zeta}_1\rightarrow\infty}
    \frac{1}{1-n}\log \bigg[
    \left(\frac{2\pi}{\beta}\right)^{8h_n}\frac{|z_az_bz_cz_d|^{2h_n}}{|z_{ab}z_{cd}|^{4h_n}}
    |1-\zeta^{\text{discon}}_{b}|^{4h_n}
    |1-\eta^{\text{discon}}_{d}|^{4h_n}
    \nonumber \\
  &\quad
    \times
    \zeta_1^{2nh_\mathcal{O}} 
    \bar{\zeta}_1^{2n\bar{h}_\mathcal{O}}
    \langle
    \mathcal{O}_n^\dagger(\zeta_1,\bar{\zeta}_1) \sigma_n(1)\bar{\sigma}_n(\zeta^{\text{discon}}_{b},
    \bar{\zeta}^{\text{discon}}_{b})|\mathcal{O}_n\rangle
    \langle \mathcal{O}_n| \sigma_n(1)\bar{\sigma}_n(\eta^{\text{discon}}_{d},
    \bar{\eta}^{\text{discon}}_{d})
    \mathcal{O}_n(0) \rangle_\mathbb{C}\bigg].
\end{align}
The cross-ratios for the disconnected channel are related to the cross-ratios for the single interval entropies as follows:
\begin{align}
  &1-\zeta^{\text{discon}}_{b} = \chi_l^A,
\quad
  1-\bar{\zeta}^{\text{discon}}_{b} = \bar{\chi}_l^A,
  \nonumber
  \\
  &1-\eta^{\text{discon}}_{d} = \chi_l^B,
  \quad
  1-\bar{\eta}^{\text{discon}}_{d} = \bar{\chi}_l^B.
\end{align}
Conformal invariance implies the following identity for four-point functions.
\begin{equation}
  \langle \phi_i(\infty)\phi_j(1)\phi_k(u,\bar{u})\phi_l(0)\rangle_\mathbb{C}=\langle \phi_i(\infty)\phi_l(1)\phi_k(1-u,1-\bar{u})\phi_j(0)\rangle_\mathbb{C}
\end{equation}
Applying this to the bipartite local operator entanglement in the disconnected channel, we find that
\begin{equation}
  S_{AB}^{\text{discon.}(n)}=S_A^{(n)}+S_B^{(n)}
\end{equation}
Thus, the bipartite local operator mutual information in the disconnected channel vanishes
\begin{equation}
  I_{AB}^{\text{discon.}(n)}=0.
\end{equation}

One obtains a simple step function for the entropy when the subregion is a single interval on either Hilbert space. When the subregion is composed of two intervals, one on each Hilbert space, there is no contribution to the local operator entanglement entropy when the operator is within the spatial intersection of both intervals. The local operator entanglement for two intervals $S_{AB}$ also begins a linear increase due to the acquisition of a monodromy by the conformal blocks when both left or right boundaries enter either the holomorphic or anti-holomorphic light cone. This will lead to a linear decrease in the bipartite operator mutual information.

\end{widetext}
%

\end{document}